\documentclass[fleqn,usenatbib]{mnras}
\usepackage[utf8]{inputenc}
\usepackage{graphicx,amsmath,amssymb,deluxetable}
\usepackage{appendix}

\newcommand\msun{\hbox{\ensuremath{{\rm M}_{\odot}}}}
\newcommand{\scinot}[1]{\ensuremath{\times 10^{#1}}}
\newcommand{\jwst}{{\it JWST}}

\title[XMDs in IllustrisTNG]{Extreme Metallicity Dwarf Galaxies in IllustrisTNG}
\author[Carleton \& Monkiewicz]{Timothy Carleton$^1$\thanks{tmcarlet@asu.edu}, Jacqueline Monkiewicz$^1$\\$^1$ Arizona State University}

\begin{document}
\label{firstpage}
\pagerange{\pageref{firstpage}--\pageref{lastpage}}
\maketitle

\begin{abstract}
The use of extremely metal-deficient dwarf galaxies (XMDs) as nearby analogs for high-redshift protogalaxies is generating renewed interest due to recent JWST observations studying these protogalaxies. However, the existence of a population of unenriched galaxies at $z\sim0$ raises fundamental questions about how galaxies with such pristine gas reservoirs could be formed. To address these questions we study XMDs in the IllustrisTNG simulation. We find that XMDs at $z=0$ are not relics of the first galaxies, but dwarf galaxies that experience a dramatic $\sim0.3$ dex drop in their gas-phase metallicity in the past few Gyr. We investigate possible causes of this drop in metallicity including high gas fractions, outflow efficiency or inflow/outflow rates, unique environments, pristine inflow metallicity, and inflow/SFR interactions. Of these, we find that inflow/outflow interactions, parameterized by the cumulative regional SFR experienced by inflows, has the strongest correlation with dwarf galaxy metallicity and XMD formation. In other words, inefficient gas enrichment during the short time between its accretion from the CGM and the initiation of star formation is the most important cause of XMD formation in the simulation. We identify differences in star formation history between XMDs and non-XMDs (with XMDs having significantly decreased star formation rates on $1-5$ Gyr timescales) and differences in galaxy size (with XMDs having a more extended young stellar population) as the primary observable differences between the two populations. These results highlight the importance of inflow enrichment efficiency as a possible driver of dwarf galaxy metallicities.
\end{abstract}

\section{Introduction}
Extremely metal-deficient (XMD) galaxies, with nebular oxygen abundances of less than 1/10th the Solar abundance \citep{Izotov1997,Madden2006,Izotov2004,Papaderos2008,Morales-Luis2011,Kojima2020}, are a rare class of nearby dwarf galaxies commonly treated as Local analogs for early protogalaxies \citep[e.g.~][]{Berg2019,Papaderos2012}. 
With the James Webb Space Telescope (\jwst{}) beginning to observe protogalaxies directly \citep[e.g.~][]{Nakajima2023,Isobe2023,Sanders2023}, interest in XMDs has recently rekindled.
But XMDs themselves remain somewhat of a mystery: How have these galaxies retained such pristine, metal-poor gas all the way to the present epoch? Are these galaxies truly high-$z$ analogs, or do their properties differ significantly from the first galaxies? Early \jwst{} observations suggest that high-$z$ XMDs appear similar to their low-$z$ counterparts in terms of their nebular emission properties \citep{Schaerer2022,Taylor2022}, but these studies are still limited to a relatively small number of objects.

Systematic studies of XMDs have proven to be difficult given their low luminosities, but some general properties have been established. The brightest parts of the galaxies are star-forming regions with very young stellar populations \citep{Papaderos2008}, suggesting some overlap with ``green-pea" or ``blueberry galaxies" \citep{Cardamone2009,Yang2017}, which in turn has led to speculation that XMDs might candidate leakers of ionizing Lyman continuum photons. Their morphology appears particularly clumpy, with high surface brightness knots of star formation against a backdrop of a lower surface brightness older stellar population \citep{Morales-Luis2011}. The gas-phase metallicities associated with the active star-forming clumps are quite low, while gas associated with the underlying disk seems to be quite a bit higher \citep{SanchezAlmeida2014,James2020,Fernandez-Arenas2023}. The H$\alpha$-derived star formation rates of XMDs appear significantly higher than non-XMDs of similar mass \citep{Isobe2021}.

Despite all of these observations, the formation of XMDs is not well understood. In particular, it is not clear if these objects are an extension of the metallicity scaling relations for dwarf galaxies, or if some unique circumstances lead to their extreme metallicities. Large outflows \citep{Dekel1986,Ferrara2000} or inflows \citep{Ekta2010,Rupke2010,Torrey2012, SanchezAlmeida2018} have both been invoked to explain their low metallicities. While many simulation and observational studies suggest that outflows are generally not strong enough to drive substantial evolution in the metallicities of dwarf galaxies \citep{Dalcanton2007,Ma2016,McQuinn2019,Xu2022}, it is possible that in certain cases outflows may influence a galaxy's metallicity.

The general scaling relations describing galaxy metallicity are more well-studied. The theory of gas enrichment proceeding in parallel with stellar mass buildup has had success in describing the mass-metallicity relation and how it evolves with redshift \citep[e.g.~][]{Tremonti2004, Zahid2014}. 
While secondary parameters such as star formation rate \citep{Mannucci2010,Lara-Lopez2010}, gas fraction \citep{Hughes2013,Bothwell2013,Brown2018,Zu2020,Chen2022}, or environment \citep{Cooper2008,Ellison2009,Alpaslan2015} have been identified as affecting galaxy metallicity, they are not as well understood. The successful extension of the mass-metallicity relation to the dwarf galaxy regime \citep{Kirby2011} demonstrates that dwarf galaxies still largely follow this formation-enrichment paradigm.

These secondary parameters are often understood in terms of the gas cycle within galaxies. In an accreting, leaky box galaxy evolution model, metallicity ($Z$) evolution can be written as \citep[following][]{Lilly2013}:
\begin{equation}
    \frac{dZ}{dt}={\rm SFE} y (1-R) - {\rm SFE}(Z-Z_0)(1-R+\eta)- (Z-Z_0) \frac{1}{m_{\rm gas}}\frac{dm_{\rm gas}}{dt},
    \label{eqn:zmodel0}
\end{equation}
which reaches an equilibrium of
\begin{equation}
    Z=Z_0+\frac{y}{1+\eta(1-R)^{-1}+M_{\rm gas} M_*^{-1}},
    \label{eqn:zmodel}
\end{equation}
with $Z_0$ representing the metallicity of inflowing gas, $R$ representing the fraction of gas retained in stars, $\eta$ representing the ratio of the mass outflow rate to the star formation rate, $M_{\rm gas}$ representing the available gas mass, SFR describing the star formation rate, SFE describing the efficiency of converting gas into stars (SFR/M$_{\rm gas}$), $M_*$ representing the formed stellar mass, and $y$ representing the average metal yield per stellar mass formed. Based on a direct (or ``naive") interpretation of Eqn.~\ref{eqn:zmodel}), one would expect secondary variables to the mass-metallicity relation to include environment (affecting $Z_0$), outflow efficiency, or gas fraction.

The observed anti-correlation between metallicity and star formation rate or gas fraction can be understood either by associating the influence of outflows ejecting newly-enriched gas \citep[increasing $\eta$;][]{Dave2011,Forbes2014} and/or assuming a constant star formation efficiency (which decreases the gas fraction over time). This can also be interpreted in the context of competing gas accretion and consumption, as recently accreted metal-poor gas initially lowers the overall metallicity of the galaxy. It subsequently drives star formation until the gas is depleted, resulting in a metal-enriched ISM \citep[e.g.~][]{Dayal2013,DeRossi2015,Torrey2018}. The impact of environment can be understood as a consequence of a higher metallicity gas reservoir in denser environments \citep[][but see \citealt{Hughes2013,Genel2016}]{Oppenheimer2006,Dave2011,Peng2014}. A natural expectation is that these trends should continue for low-mass galaxies, in which case XMDs may simply reflect extremes of environment, SFR, or gas fraction. However, there appears to be a larger degree of scatter in the mass-metallicity relation of dwarf galaxies compared to more massive galaxies \citep{Tremonti2004}, suggesting these secondary parameters may be more important for dwarf galaxies than their more massive counterparts. This is consistent with the fact that dwarf galaxies have other notable differences compared to massive galaxies, such as a relatively larger neutral gas reservoir and lower gravitational potential. The physics of gas enrichment may be different for dwarf galaxies than more massive galaxies, and the nature of XMDs may be different still.

To gain insight into the formation and nature of XMDs, we turn to the IllustrisTNG cosmological hydrodynamic simulation. The IllustrisTNG simulation incorporates all the effects described above, and allows us to study the formation of XMDs in detail. In this paper, we present an analysis of the formation, evolution, and properties of XMDs identified in the IllustrisTNG simulation compared to other similar-mass dwarf galaxies. Section~\ref{sec:data} describes the identification of XMDs in the IllustrisTNG simulation. Section~\ref{sec:formation} analyzes the applicability of previously proposed XMD formation scenarios within the IllustrisTNG model, and Sec.~\ref{sec:nomix} details the actual XMD formation mechanism identified in IllustrisTNG. Section~\ref{sec:properties} describes observable properties of XMDs in the simulation. Lastly, Section~\ref{sec:conculsions} summarizes our conclusions. When applicable, we use the \cite{PlanckCollaboration2016} cosmological parameters, with $H_0=67.74$ km~s$^{-1}$~Mpc$^{-1}$, $\Omega_m=0.3089$, and $\Omega_\Lambda=0.6911$.

\begin{figure}
    \centering
    \includegraphics[width=1\linewidth]{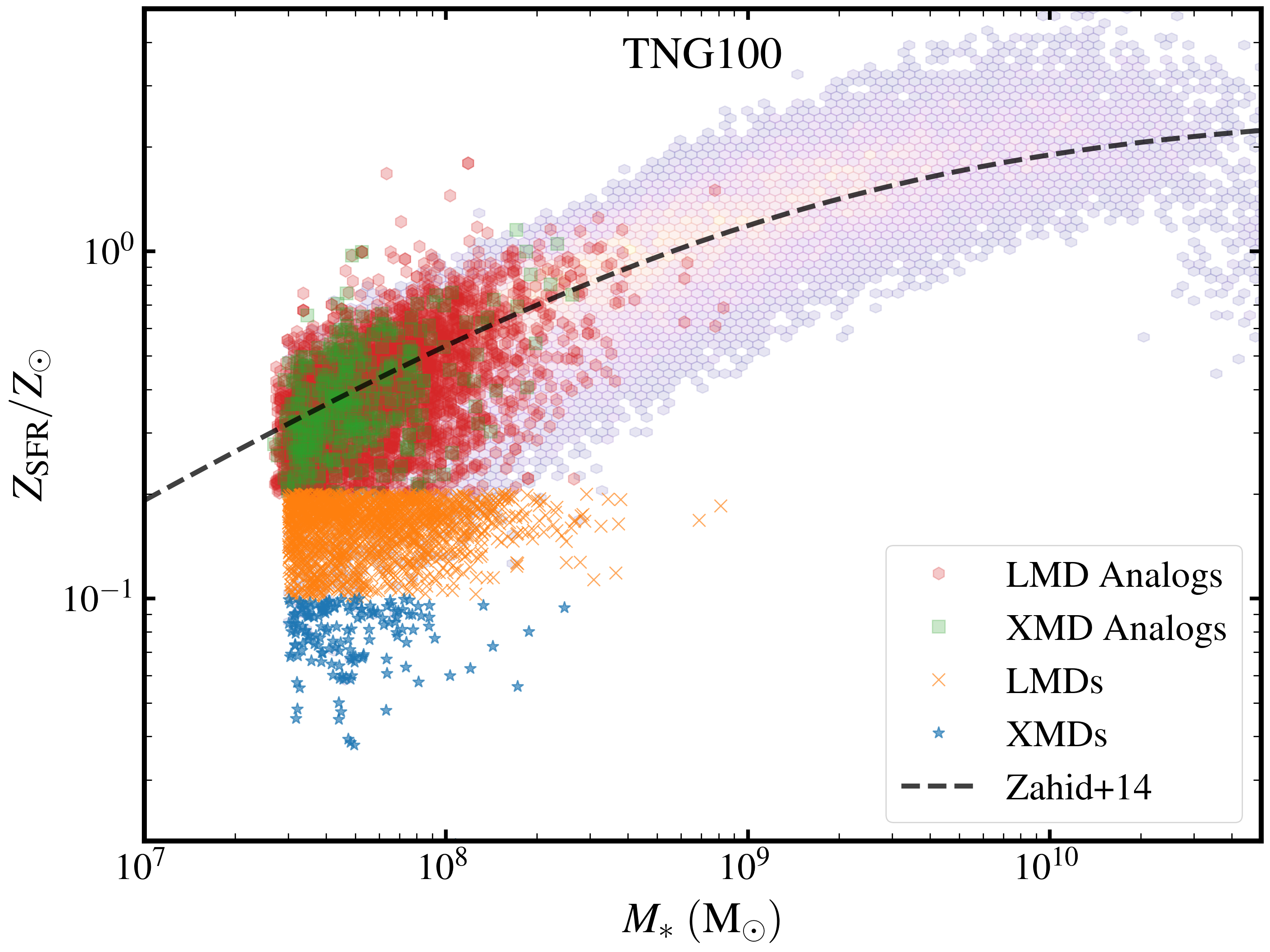} \\ \includegraphics[width=1\linewidth]{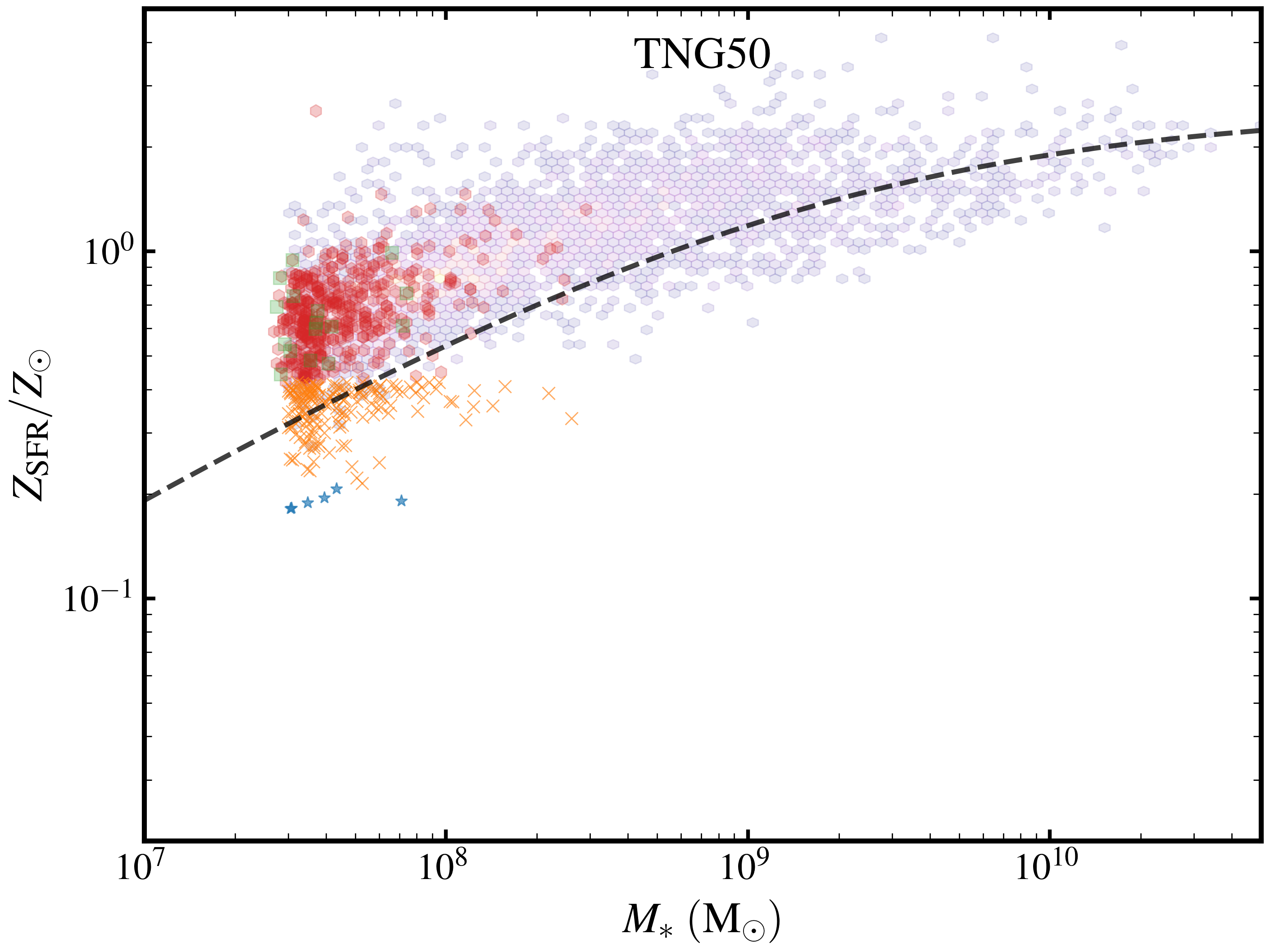}
    \caption{The mass-metallicity relation of objects in our sample at $z=0$, taken from TNG100 (top) and TNG50 (bottom). The coloured shading corresponds to the density of star-forming objects in this space, and points correspond to XMDs (blue stars), XMD analogs (green boxes), LMDs (orange ``x"s), and LMD analogs (red hexagons). The dashed black line corresponds to the observed mass-metallicity relation taken from \protect \cite{Zahid2014}. Overall, objects in TNG100 match the observations well, but objects in TNG50 overpredict gas-phase metallicities. Regardless, dwarf galaxies in both simulations have a range of metallicities, allowing us to compare the formation of metal-deficient dwarf galaxies with normal metallicity dwarf galaxies.}
    \label{fig:mzrplot}
\end{figure}

\section{Data}
\label{sec:data}
\subsection{XMDs in Illustris}
\label{sec:xmdsample}
To study the formation and evolution of XMDs, we identify extreme metallicity dwarfs in the IllustrisTNG100-1 Simulation \citep{Springel2018, Pillepich2018, Naiman2018, Marinacci2018}. IllustrisTNG100 was run starting with the Planck 2015 cosmology \citep{PlanckCollaboration2016}, and has a comoving volume of $(75$ Mpc$/h)^3$ and a baryon particle mass of $9.4\scinot{5}~\msun{}/h$, ideally suited to study rare, low mass objects like XMDs.
In addition, to examine the properties of XMDs in more detail, we investigate objects in the higher-resolution TNG50 simulation \citep{Nelson2019,Pillepich2019}. TNG50 has a volume of $(35$ Mpc$/h)^3$ and a mass resolution a factor of 16 times smaller than TNG100, such that galaxies with our $M_*>3\scinot{7}~\msun{}$ selection are well resolved with approximately $350$ stellar particles.

The IllustrisTNG simulation was run with the moving-mesh {\sc Arepo} hydrodynamics code \citep{Springel2010}, and uses a galaxy formation model \citep[described in][]{Weinberger2017,Pillepich2018mod} largely based on the original Illustris simulation, described in \cite{Vogelsberger2013}. The most relevant details for our study are the star formation and chemical enrichment model. Star formation is modeled in a subgrid fashion following \cite{Springel2003}, with stars forming stochastically above a density threshold (taken to be $0.13$ cm$^{-3}$) with a characteristic timescale ($2.2$~Gyr). Stars are formed with a universal \cite{Chabrier2003} initial mass function. Stellar evolution then returns mass and metals to the ISM, with relevant yields taken from \cite{Karakas2010} for AGB stars, \cite{Portinari1998} for core-collapse supernovae, and \cite{Thielemann2003} and \cite{Travaglio2004} for Type Ia supernovae.

The stellar mass-gas phase metallicity relation (or MZR) in IllustrisTNG has been shown to match observations well \citep{Torrey2019}, assuming that the star-formation-rate-weighted metal fraction ($Z_{\rm SFR}$) is taken as the gas-phase metallicity. This makes sense, given that observational metallicity tracers are primarily sensitive to active star-forming regions, not the large (lower metallicity) gas reservoir in galaxies' outskirts. Given this agreement, we can make an observationally-motivated selection of XMDs. First, we select galaxies in TNG100 with stellar masses ($M_*$) greater than 3\scinot{7}~$M_\odot$(corresponding to $\sim 20$ stellar particles) and a current star formation rate above 0. The mass cut is to ensure our galaxies are resolved, and the SFR cut is designed to ensure that gas-phase metallicity measurements in the simulation are comparable to observations (i.e. gas-phase metallicity measurements are typically conducted using the ratios of emission-line strengths, so ionizing emission from young stars (i.e. SFR$>0$) is necessary). We also discard any objects with dark-matter masses less than $10^8$~\msun{}. These objects represent tidal debris picked up by the halo finder and are not resolved enough to study in detail. Lastly, there are some objects in IllustrisTNG with low metallicities, but high stellar masses ($M_*>10^9$~\msun{}). These are predominantly very low SFR galaxies where a recent accretion of gas results in an overall low metallicity, and while interesting in their own right, are beyond the scope of this paper. From this parent sample, we select XMDs as objects with gas-phase abundances ([O/H]) less than $1/10^{\rm th}$ solar\footnote{\label{foot:part}In this paper, $Z/Z_\odot$ refers to the gas-phase oxygen abundance relative to solar, which we use as a proxy for total metallicity given the prevalence of Oxygen-based metallicity measurements among observations. However, individual particle data does not track individual species, so we refer to the {\sc GFM\_Metallicity} property when utilizing that particle data, such as Fig.~\ref{fig:zflowstack}. We also divide it by $0.0127$ to adjust to solar metallicity and denote it as $Z_p/Z_\odot$ to differentiate it from and facilitate comparisons to $Z/Z_\odot$.}. Additionally, we select a sample of low-metallicity dwarfs (LMDs) with $Z_{\rm SFR}/Z_\odot$ between $0.1$ and $0.2$. We also select a mass-matched sample of XMD and LMD analogs with metallicities above the LMD threshold to study how XMDs and LMDs compare with the overall dwarf galaxy population in IllustrisTNG. Following this selection, there are $160$ ($1557$) XMDs (LMDs) in TNG100 at $z=0$, corresponding to an abundance of $1.2$\scinot{-4} Mpc$^{-3}$ ($1.19$\scinot{-3} Mpc$^{-3}$) and $0.47\%$ ($4.6\%$) of star-forming objects in our mass range. While this abundance is lower than the abundance of XMDs in the local group \citep{Monkiewicz2019}, it matches the abundance of XMDs among a sample of objects with similar stellar masses in SDSS \citep{Morales-Luis2011} remarkably well, suggesting that the XMDs selected in TNG100 can be treated as analogs of the observed XMD population and that the difference in abundance relative to the Local Group can be attributed to differences in the stellar mass range of our samples, although the limited resolution means IllustrisTNG probably overestimates XMD abundance. Tables of the $z=0$ halo IDs and basic properties of our XMD, LMD, and normal-metallicity sample are included as Table~\ref{tab:objecttable} for TNG100 and Table~\ref{tab:objecttable50} for TNG50. The top panel of Figure~\ref{fig:mzrplot} illustrates the mass-metallicity relation in TNG100 and highlights the selection of galaxy samples we use in this study. 

Before selecting objects from TNG50, we note that the overall mass-metallicity relation in TNG50 is $0.22$ dex offset from the measured relation (taken for galaxies with stellar masses between $10^7$ and $10^9$~\msun{}). The mass-metallicity relation also has a $28\%$ lower scatter in TNG50 than TNG100 (0.19 dex vs 0.23 dex). For these comparisons, we define XMDs in TNG50 to be objects with gas-phase metallicities $21\%$ of solar or lower and LMDs to be objects with gas-phase metallicities $21-42\%$ of solar. Aside from this difference, we select objects in the same manner as TNG100.
This results in 6 XMDs and 166 LMDs, such that the ratio of XMDs to star-forming galaxies in our mass range is 9\scinot{-4} and the ratio of LMDs to star-forming galaxies in our mass range is 2.5\scinot{-2}.
This lower density of XMDs is likely a consequence of differences in gas cooling, star formation, and metal enrichment achieved with the higher resolution. We note that the effect of resolution on stellar metallicity in IllustrisTNG has been observed before \citep{Nelson2018}, and the effect of resolution on the IllustrisTNG model is discussed in the appendices of \cite{Pillepich2018} and \cite{Pillepich2019}. Regardless, as we will see below, this sample offers valuable insight into the drivers of gas-phase metallicity in low-mass galaxies. Except when noted, particularly as part of the discussion of inflow gas enrichment, the results of the TNG50 and TNG100 simulations are consistent.

\subsection{Derived Quantities}
\label{sec:measurements}
To track the evolution of XMDs, LMDs, and their analogs over time, we use the Sublink trees \citep{Rodriguez-Gomez2015} published as part of the IllustrisTNG data release, taking the massive-progenitor branch. When discussing inflow and outflow rates, we consider the halo cutout of the most massive progenitor at times earlier than $z=0$. Any gas cells that are within $5r_h$ at time $t_i$ that were not within $5r_h$ at time $t_{i-1}$ are considered inflow. Likewise, any gas cells that are not within $5r_h$ at time $t_i$ that were within $5r_h$ at time $t_{i-1}$ are considered outflow. When considering the evolution of gas particles (Sec.~\ref{sec:nomix}), we identify tracer particles that were hosted by the relevant gas particles at $z=0$. We track the tracer particles back in time, taking the relevant properties of gas particles hosting the tracer particles \citep{Genel2013} \footnote{https://www.tng-project.org/data/forum/topic/490/sfh-for-individual-gas-particles/}. 

\begin{table*}
    \centering
\begin{tabular}{c|c|c|c|c|c|c|c|c}
Type & TNG100 ID & 12+log(O/H) & $\log M_*/\msun{}$ & $\log$ SFR ($\msun{}$~yr$^{-1}$) & $\log M_{\rm gas}/\msun{}$ & $\log M_{\rm halo}/\msun{}$ & $r_h$ (kpc) & $Z_*/Z_\odot$ \\
\hline
XMD & 69629 & 7.39 & 7.64 & -3.15 & 9.09 & 10.54 & 3.44 & -1.13 \\
XMD & 83366 & 7.65 & 7.84 & -2.83 & 9.42 & 10.64 & 3.28 & -1.17 \\
XMD & 177769 & 7.67 & 7.70 & -1.83 & 9.20 & 10.37 & 1.93 & -1.08 \\
XMD & 272337 & 7.64 & 7.49 & -2.86 & 9.10 & 10.48 & 4.89 & -1.61 \\
XMD & 432493 & 7.47 & 7.69 & -3.58 & 9.20 & 10.31 & 3.78 & -1.23 \\
XMD & 447769 & 7.69 & 7.88 & -2.34 & 9.37 & 10.39 & 3.23 & -1.13 \\
\end{tabular}
    \caption{Information about all objects in our IllustrisTNG100 galaxy sample, including XMDs, LMDs, and their analogs. A subset is shown here, see the electronic manuscript for the complete table.}
    \label{tab:objecttable}
\end{table*}

\begin{table*}
    \centering
\begin{tabular}{c|c|c|c|c|c|c|c|c}
Type & TNG50 ID & 12+log(O/H) & $\log M_*/\msun{}$ & $\log$ SFR ($\msun{}$~yr$^{-1}$) & $\log M_{\rm gas}/\msun{}$ & $\log M_{\rm halo}/\msun{}$ & $r_h$ (kpc) & $Z_*/Z_\odot$ \\
\hline
XMD & 567383 & 8.01 & 7.64 & -3.62 & 8.16 & 10.04 & 1.60 & -0.83 \\
XMD & 802456 & 7.98 & 7.60 & -2.06 & 9.42 & 10.46 & 1.85 & -0.95 \\
XMD & 839242 & 7.97 & 7.85 & -2.88 & 9.16 & 10.29 & 1.41 & -0.67 \\
XMD & 844914 & 7.95 & 7.48 & -2.92 & 8.64 & 10.34 & 1.18 & -0.87 \\
XMD & 856894 & 7.97 & 7.54 & -3.95 & 8.69 & 10.25 & 0.97 & -0.82 \\
XMD & 891332 & 7.95 & 7.49 & -3.11 & 8.63 & 10.03 & 1.37 & -0.86 \\
\end{tabular}
    \caption{Information about all objects in our IllustrisTNG50 dwarf galaxy sample, including XMDs, LMDs, and their analogs. A subset is shown here, see the electronic manuscript for the complete table.}
    \label{tab:objecttable50}
\end{table*}

\begin{figure}
    \centering
    \includegraphics[width=1\linewidth]{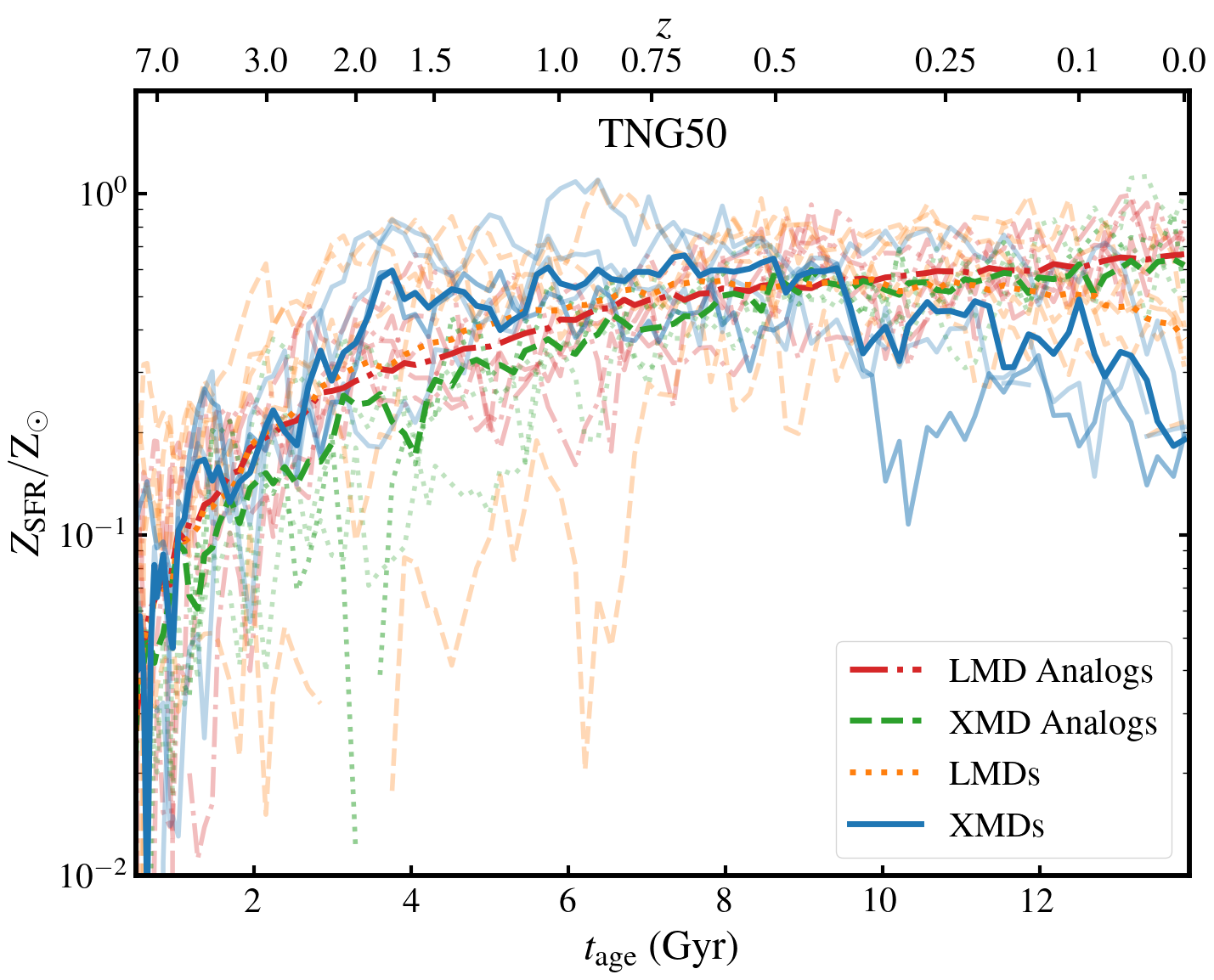} \\
    \includegraphics[width=1\linewidth]{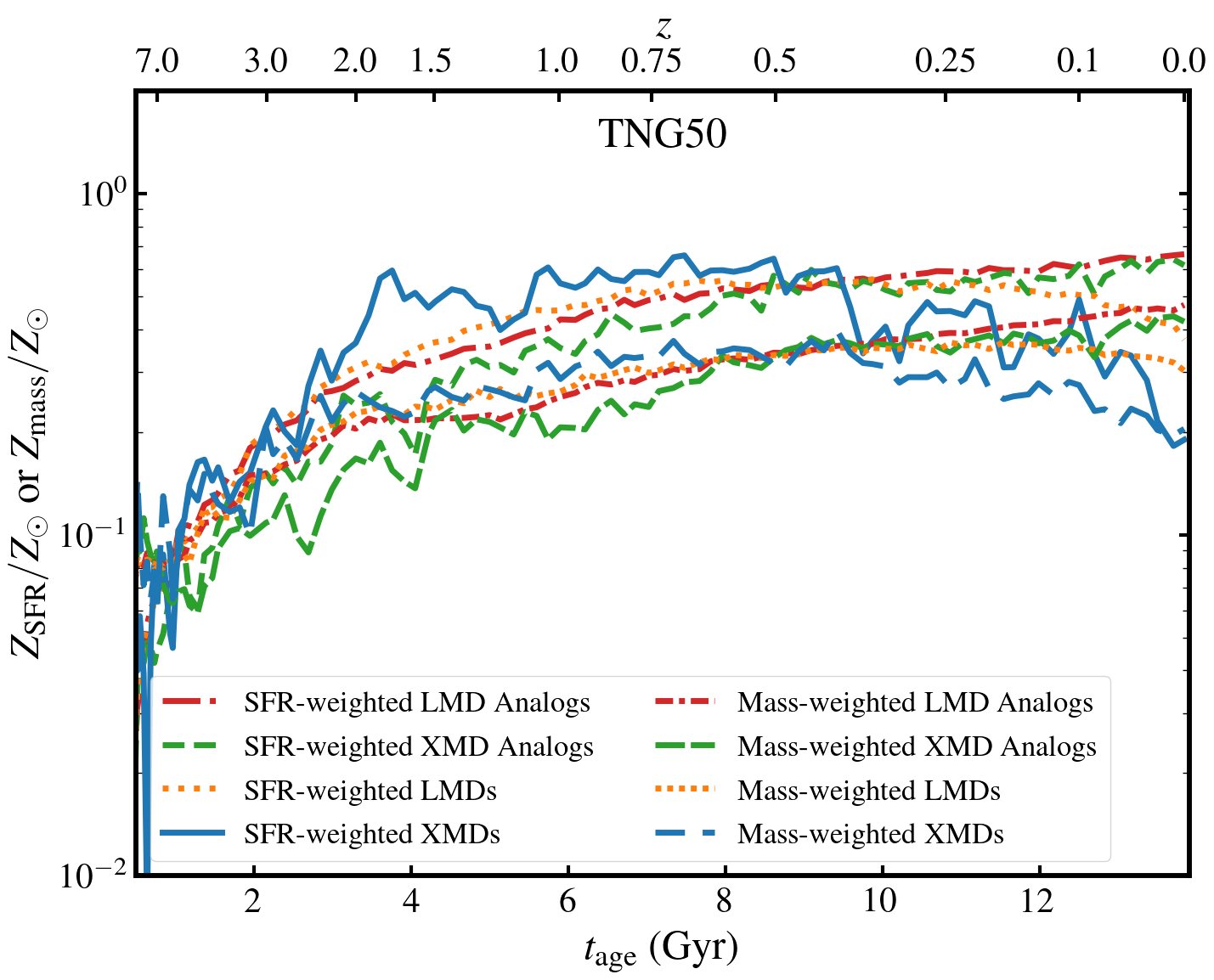}
    \caption{{\bf Top:} The metallicity evolution of our sample including XMDs (shown as blue lines), mass-matched analogs of XMDs (orange dotted lines), LMDs (green dashed lines), and mass-matched analogs of LMDs (red dot-dashed lines). The thick lines represent the median of all objects, whereas the thin lines represent a randomly-selected sample of 5 objects in each subsample in order to represent the scatter. Although there is a large scatter in metallicity evolution for both subsamples, XMDs show a substantial downturn toward in metallicity at redshift $0$ not seen in the analog subsample.
    {\bf Bottom:} The metallicity evolution of objects in our sample, as in Fig.~\ref{fig:metvstime}, but including mass-weighted metallicity in addition to SFR-weighted metallicity. For LMDs, XMD analogs, and LMD analogs, the more-densely packed lines correspond to the mass-weighted measurements, whereas the long dashed blue line corresponds to the mass-weighted metallicity of XMDs and the solid blue line corresponds to the SFR-weighted metallicity of XMDs. While the metallicities of individual objects vary significantly over time, the metallicity evolution of XMDs is not driven exclusively by low-metallicity gas suddenly participating in star formation --- the entire gas reservoir is decreasing in metallicity at late times.
    }
    \label{fig:metvstime}
\end{figure}

\begin{figure*}
    \centering
    \begin{tabular}{ccc}
    \includegraphics[width=.3\linewidth]{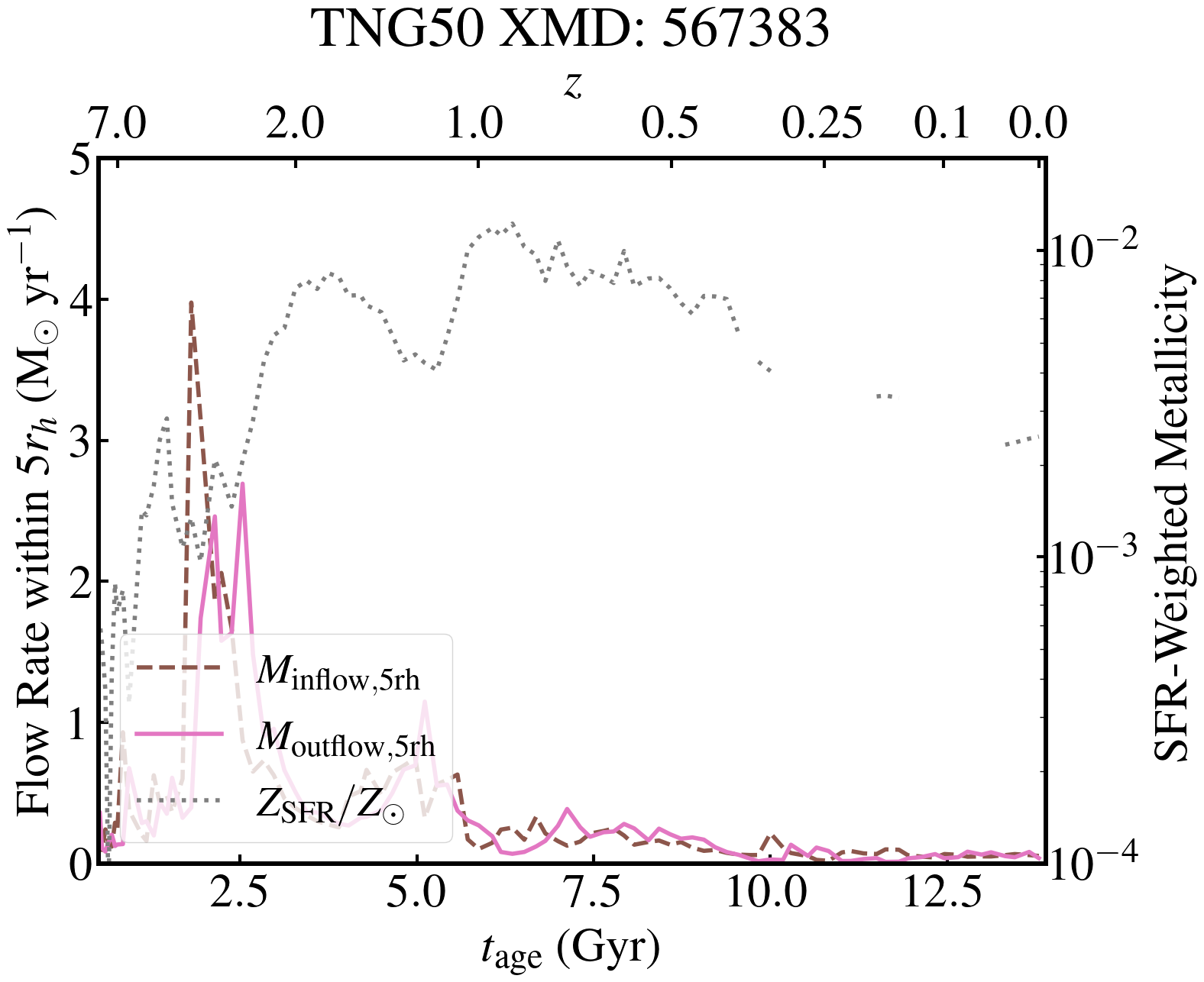} & 
    \includegraphics[width=.3\linewidth]{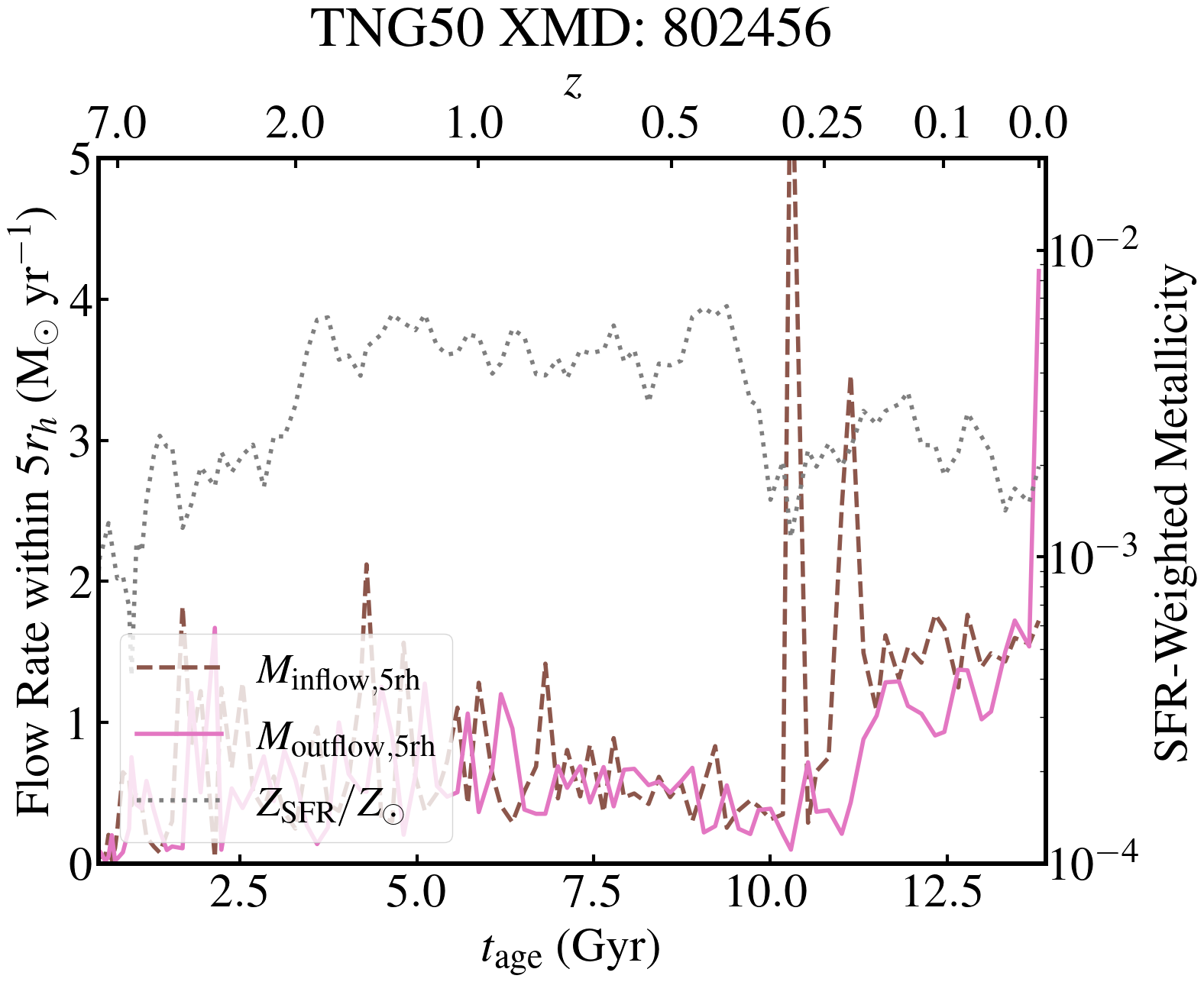} & 
    \includegraphics[width=.3\linewidth]{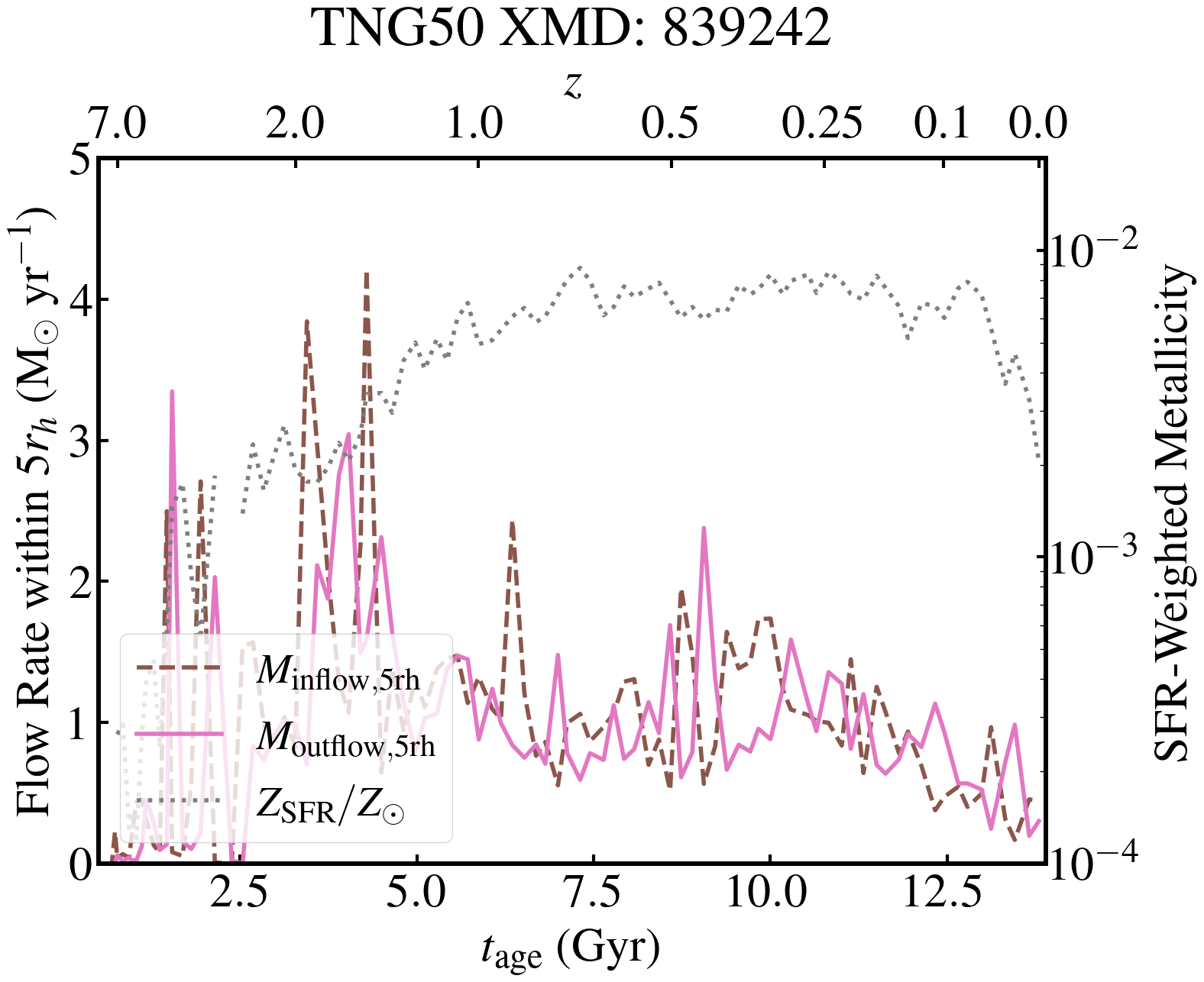}
    \end{tabular}
    \caption{An illustration of the inflow and outflow properties of XMDs in our sample over time. The brown dashed line illustrates the inflow rate of all gas to within $5r_h$. The pink line illustrates the same quantity, but measuring outflowing gas rather than inflowing gas. For comparison, the grey dotted line illustrates the SFR-weighted metallicity of the gas (not formally restricted to within $5r_h$, although most star formation does primarily occur there). Gaps in that line correspond to instances when the SFR=0~\msun{}~yr$^{-1}$.}
    \label{fig:flowxmd}
\end{figure*}

\begin{figure*}
    \centering
    \begin{tabular}{ccc}
    \includegraphics[width=.3\linewidth]{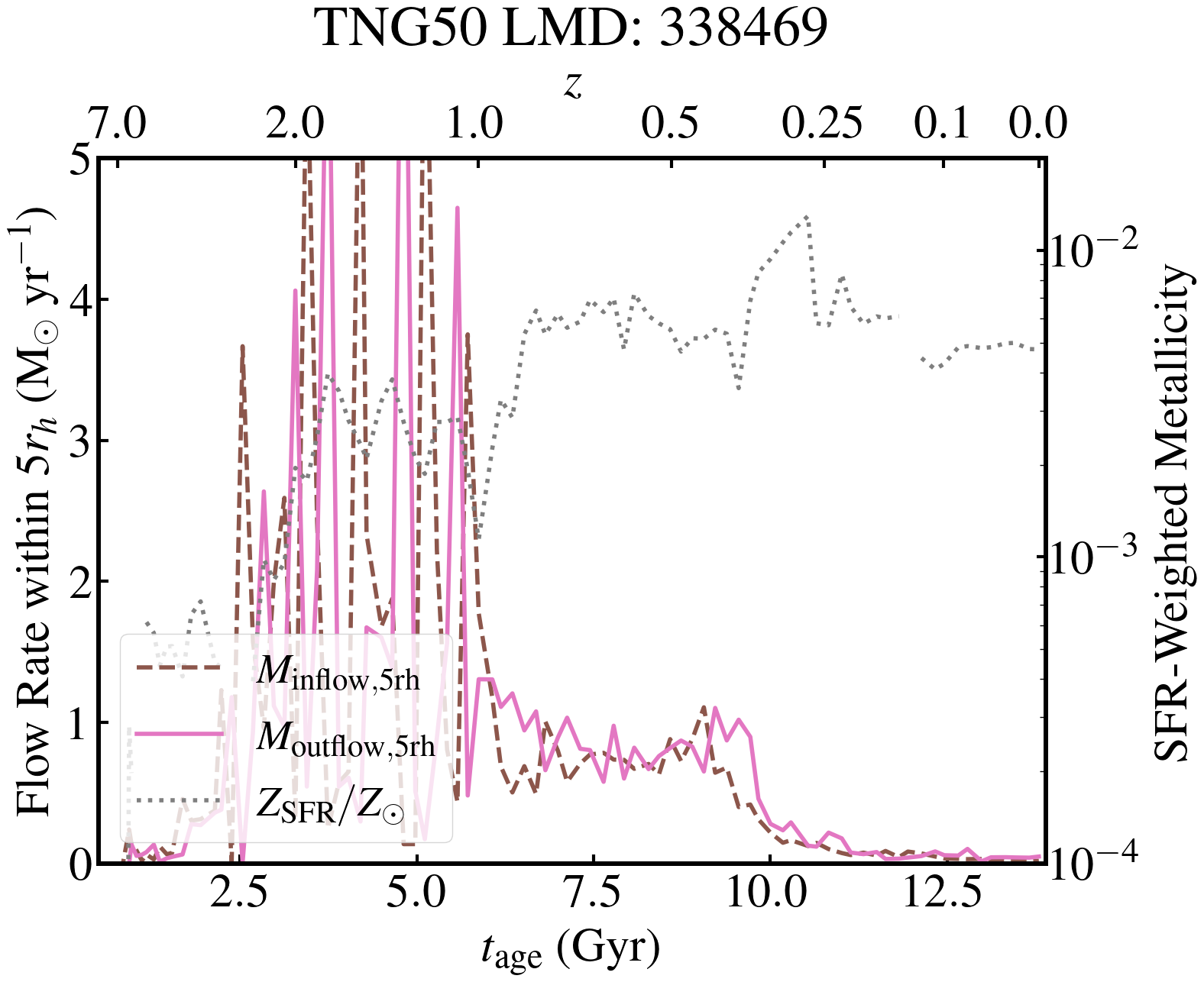} & 
    \includegraphics[width=.3\linewidth]{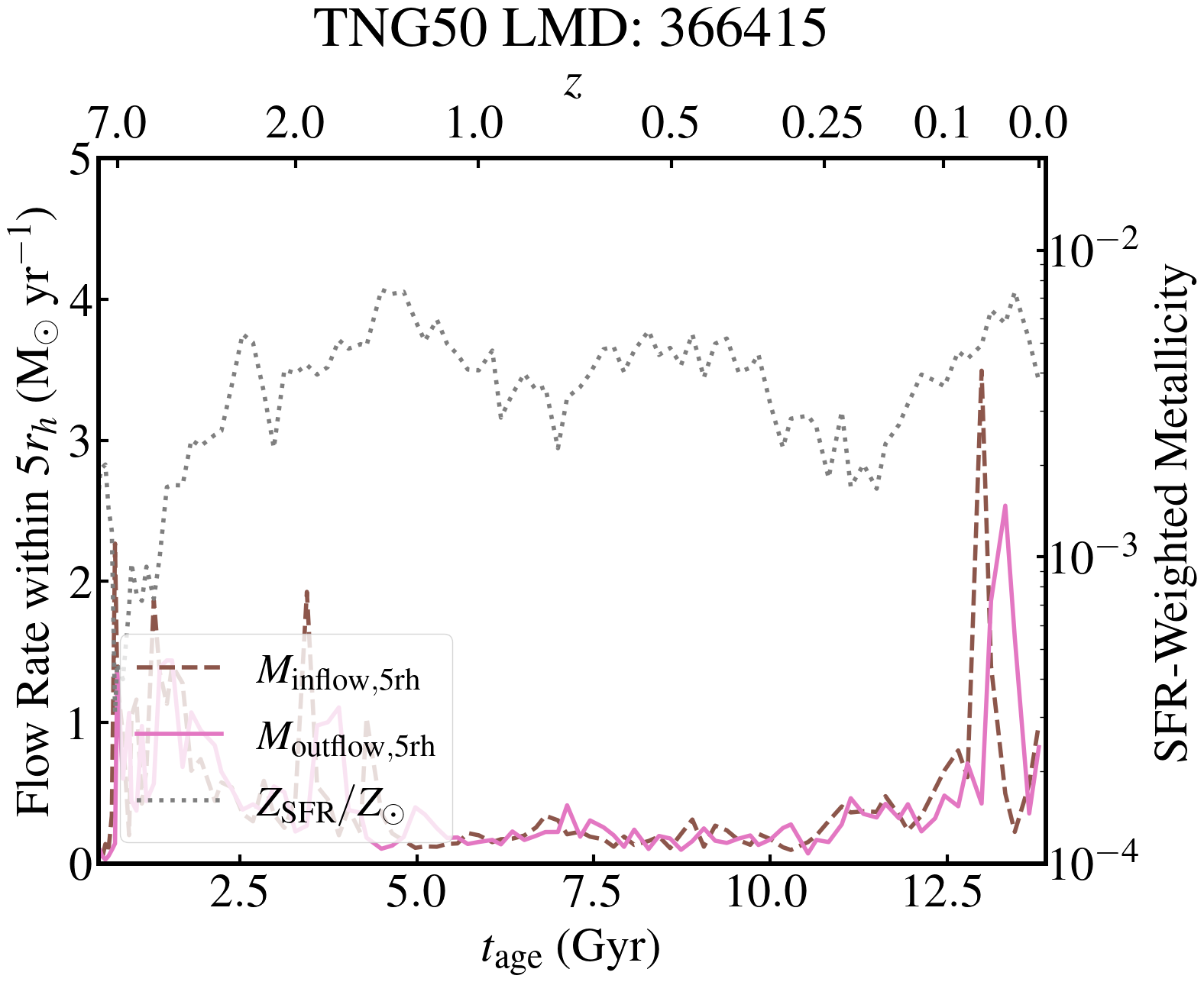} & 
    \includegraphics[width=.3\linewidth]{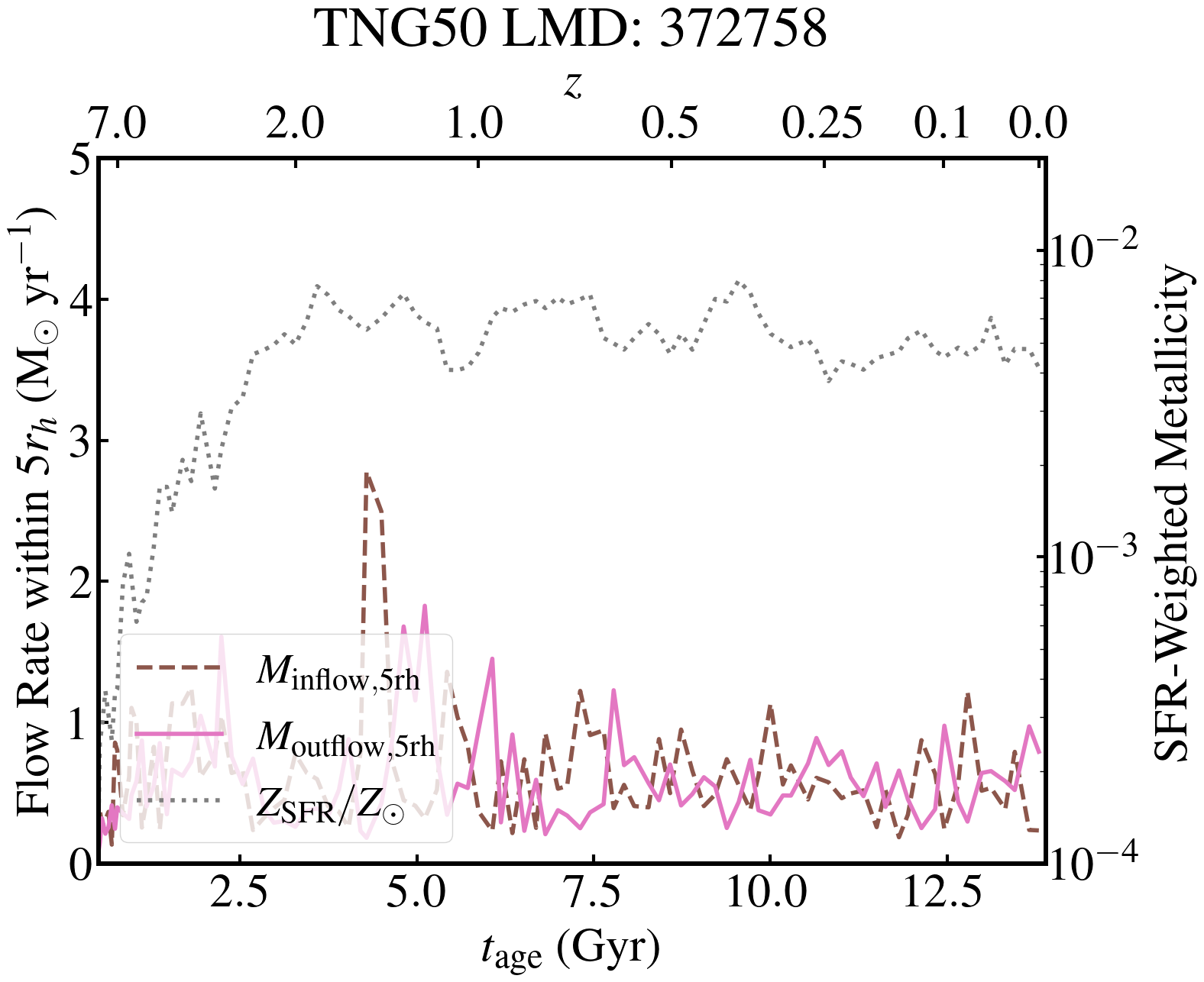}
    \end{tabular}
    \caption{Same as Fig~\ref{fig:flowxmd}, but for LMDs. }
    \label{fig:flowlmd}
\end{figure*}

\begin{figure*}
   \centering
   \begin{tabular}{ccc}
    \includegraphics[width=.3\linewidth]{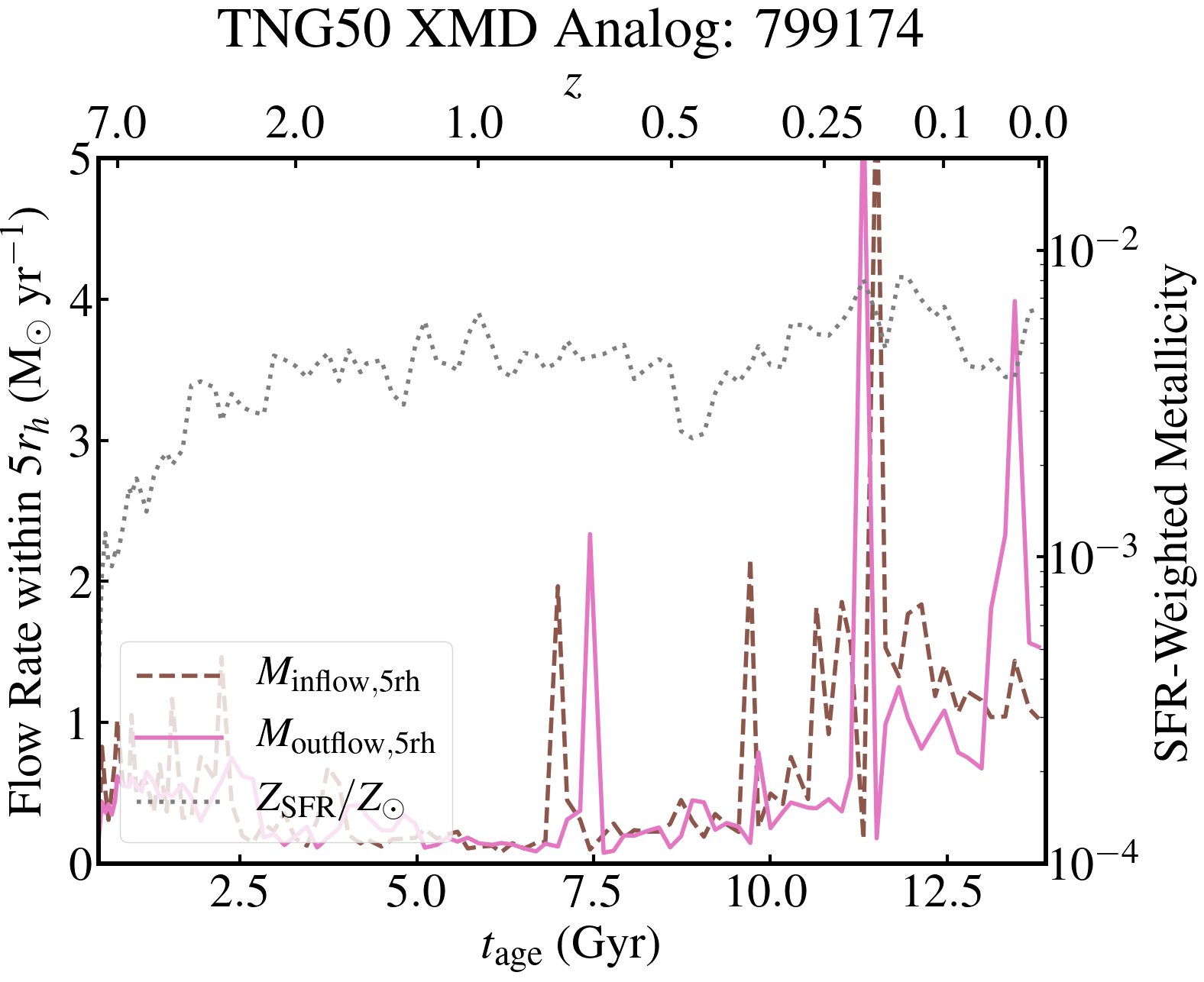} & 
    \includegraphics[width=.3\linewidth]{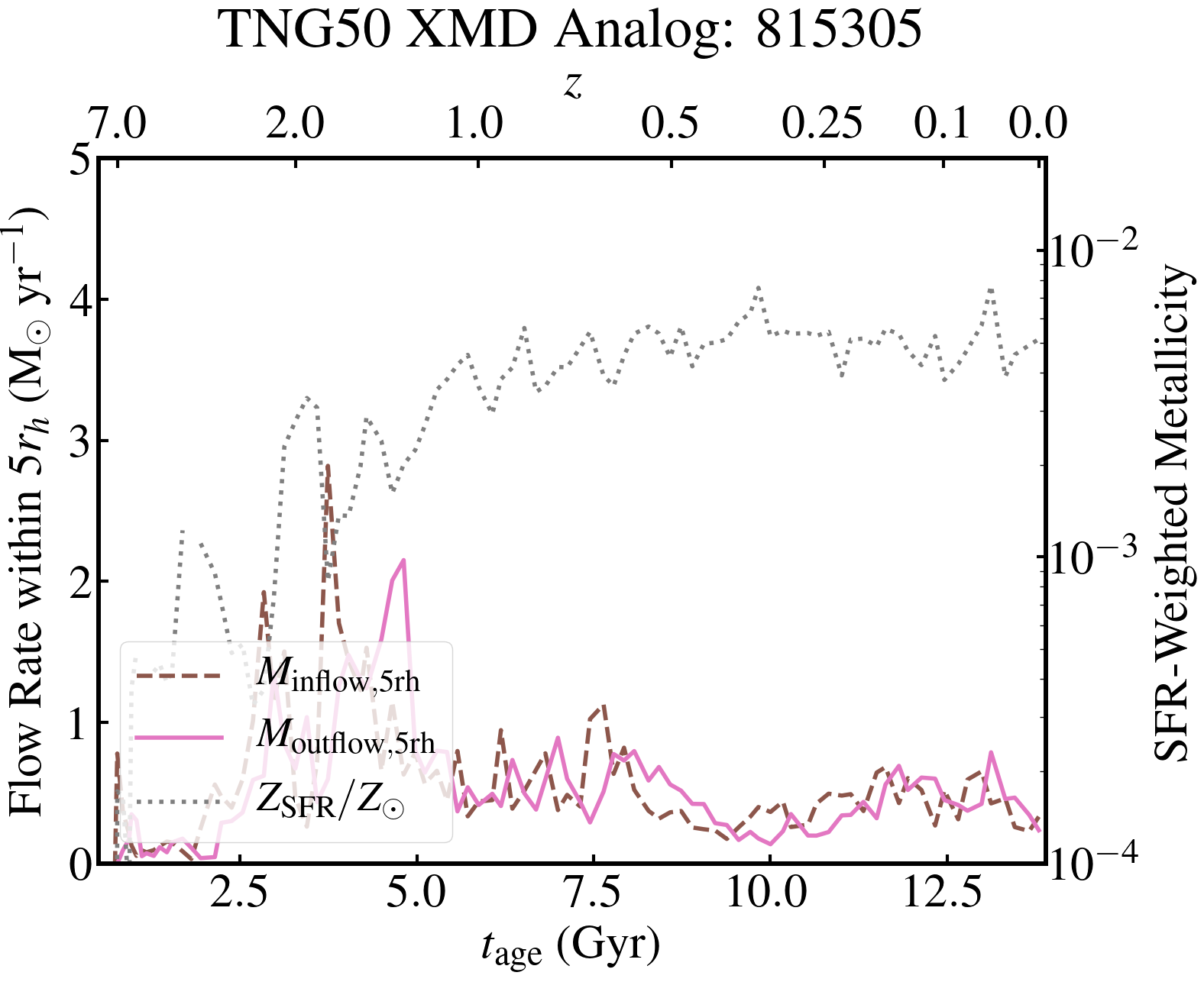} & 
    \includegraphics[width=.3\linewidth]{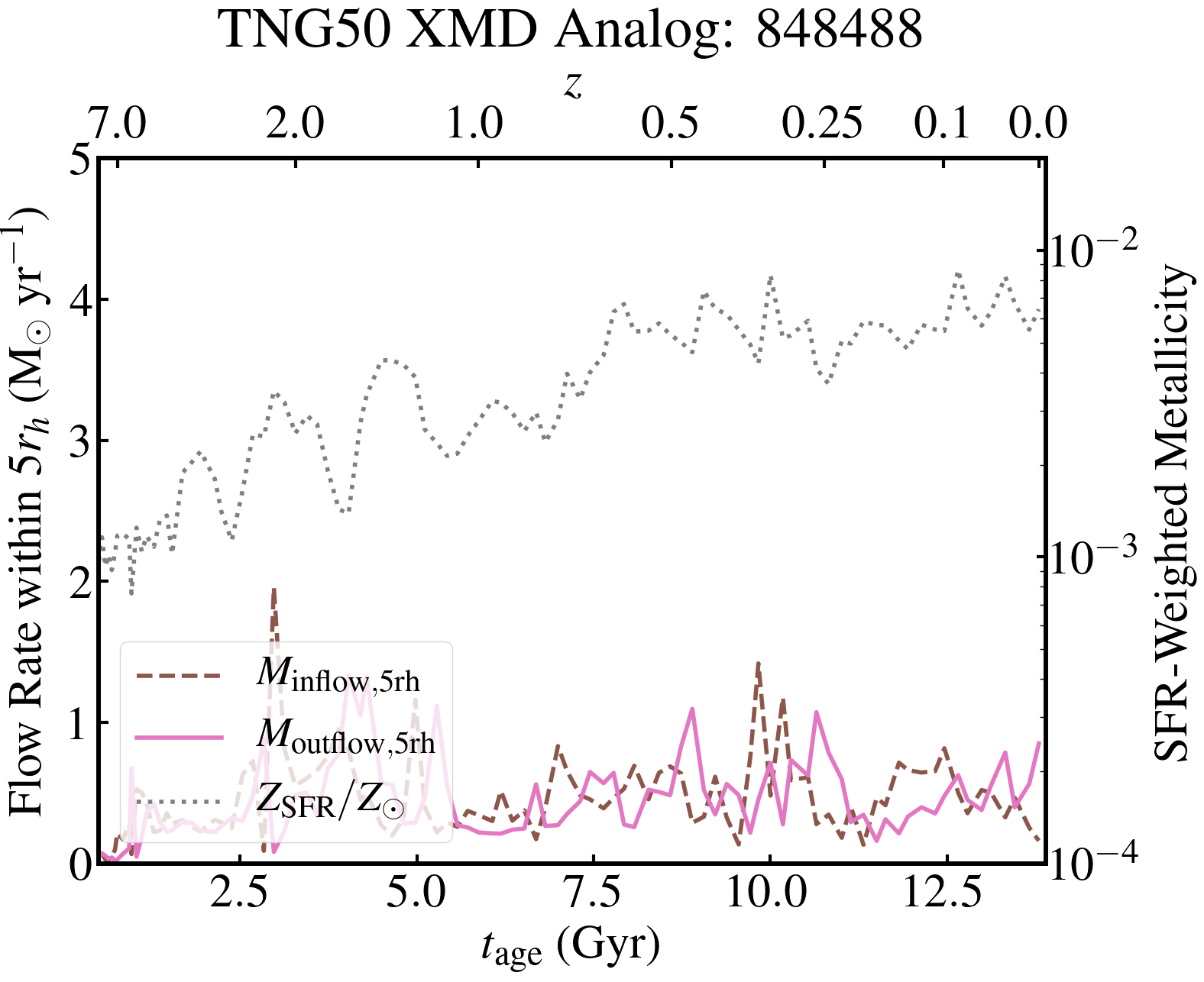}
    \end{tabular}
    \caption{Same as Fig~\ref{fig:flowxmd}, but for XMD analogs.}
    \label{fig:flowxmda}
\end{figure*}

\begin{figure*}
   \centering
   \begin{tabular}{ccc}
    \includegraphics[width=.3\linewidth]{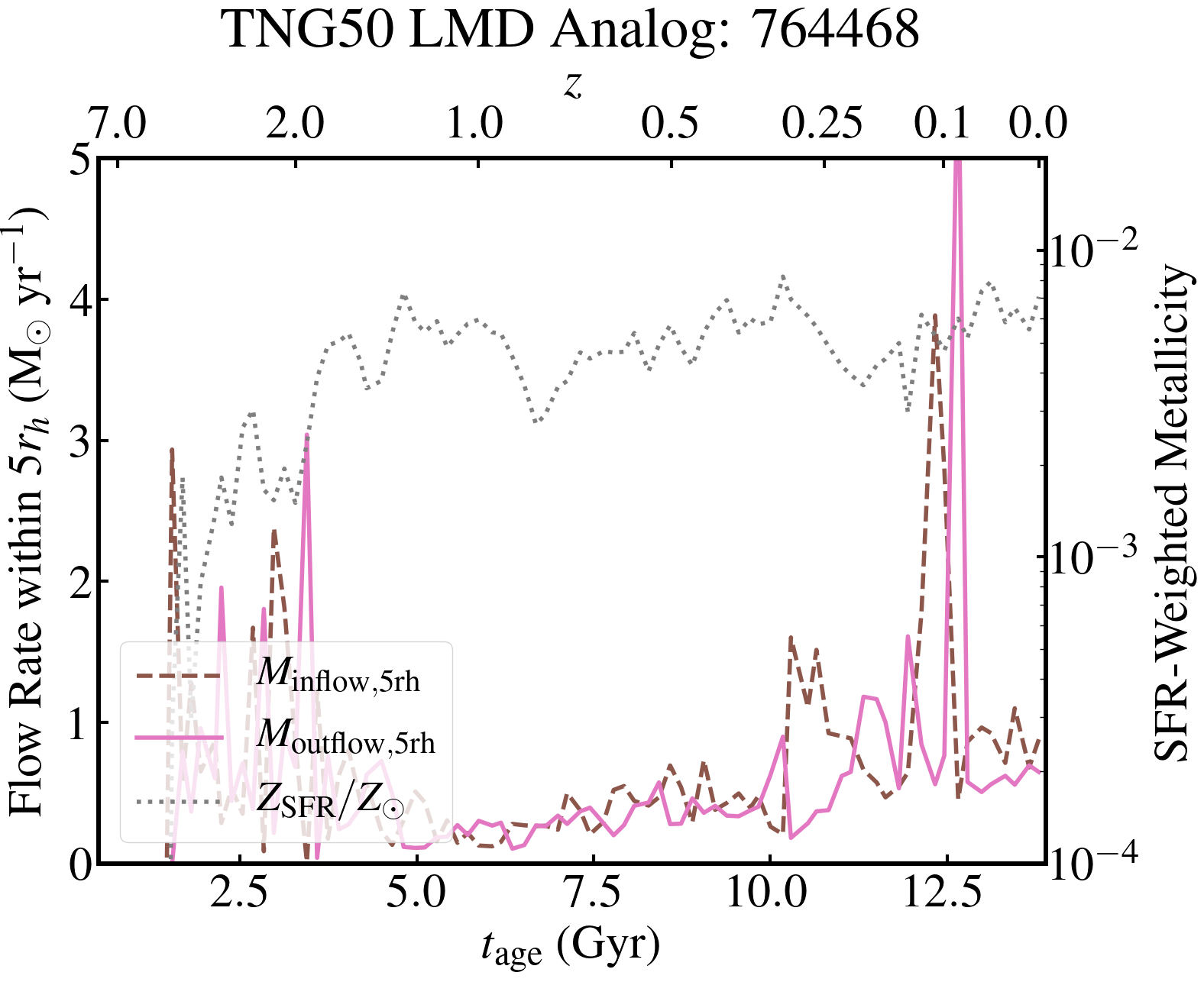} & 
    \includegraphics[width=.3\linewidth]{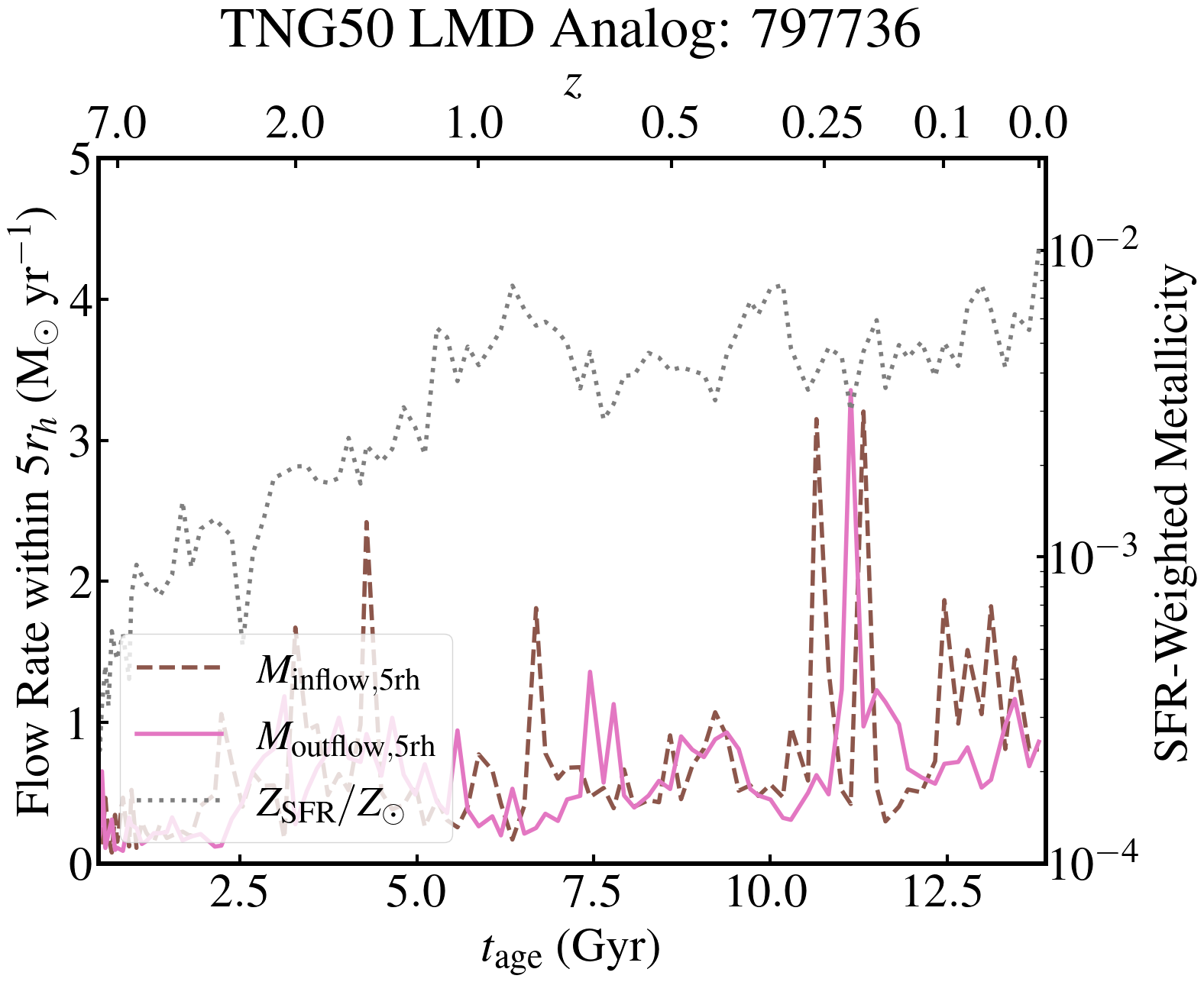} & 
    \includegraphics[width=.3\linewidth]{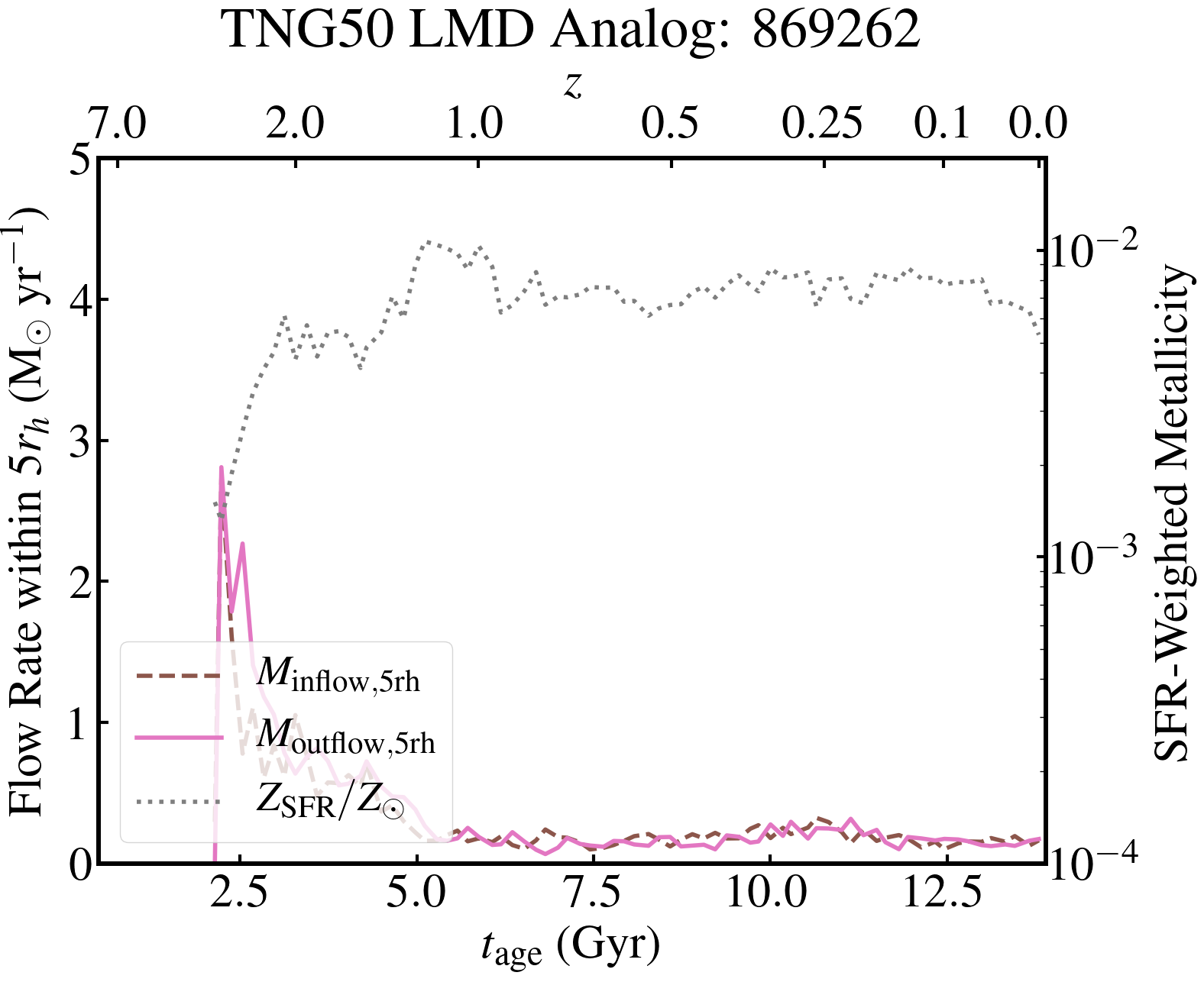}
    \end{tabular}
    \caption{Same as Fig~\ref{fig:flowxmd}, but for LMD analogs.}
    \label{fig:flowlmda}
\end{figure*}

\begin{figure*}
    \begin{tabular}{cc}
    \includegraphics[width=0.45\linewidth]{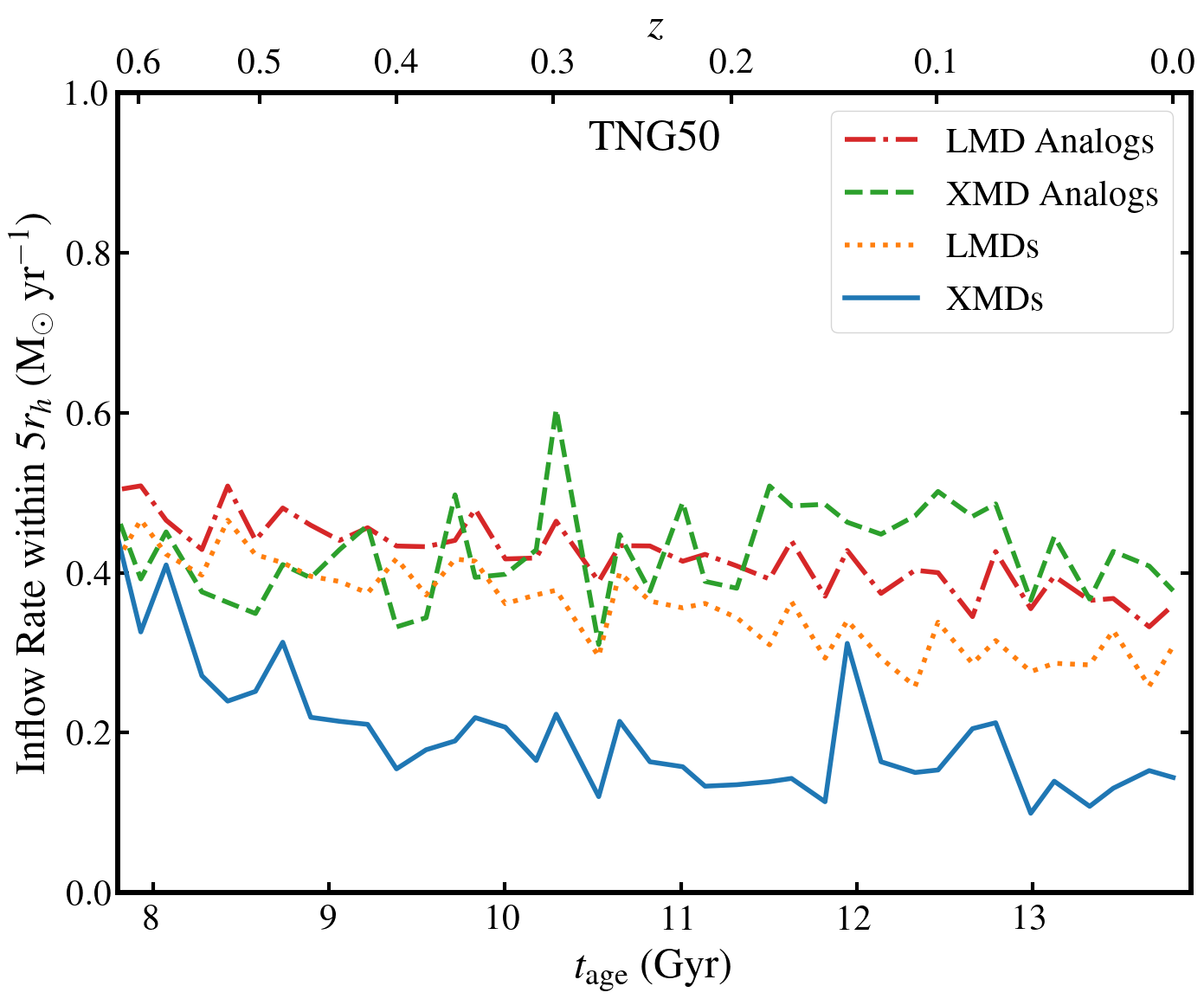} & \includegraphics[width=0.45\linewidth]{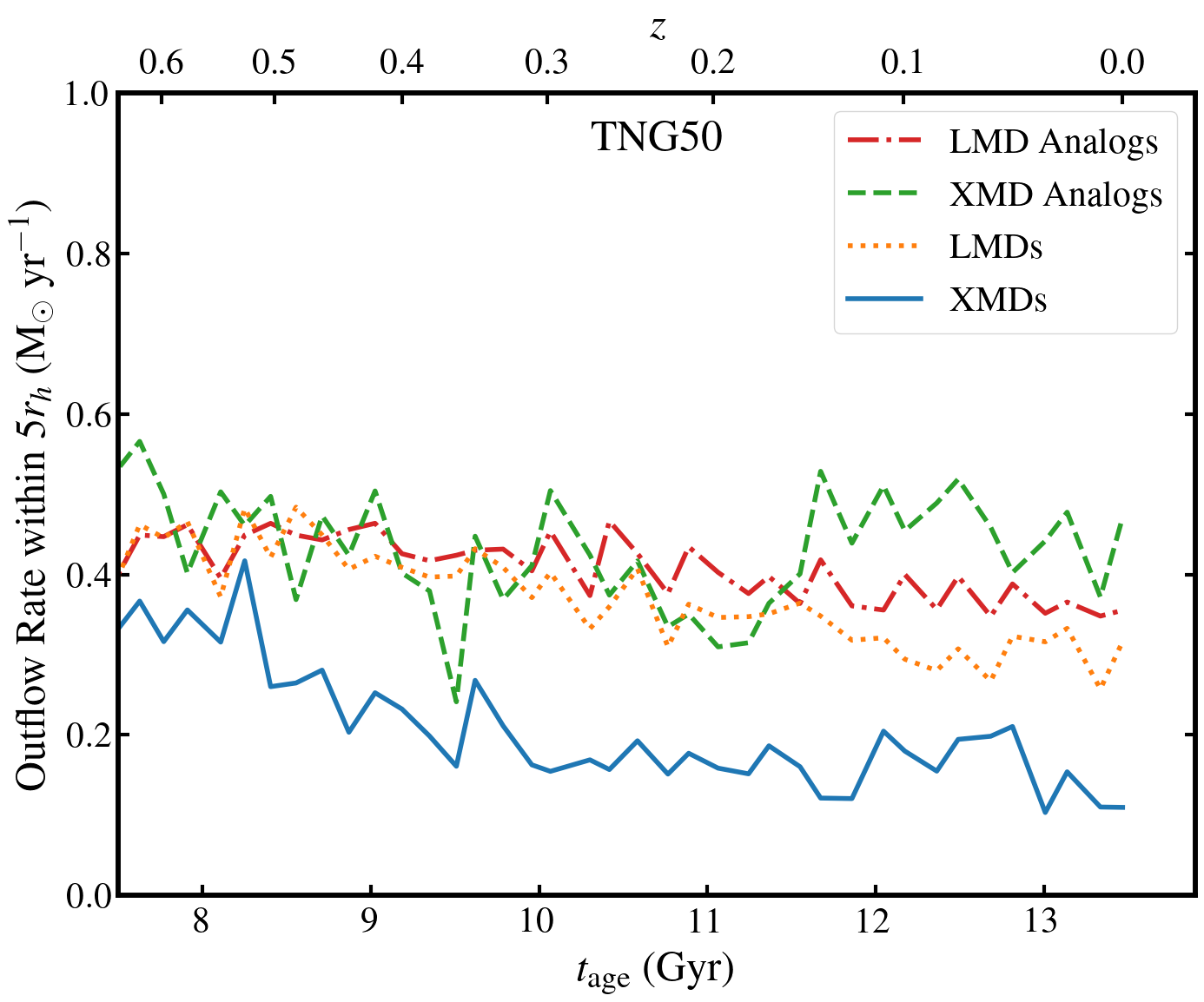}
    \end{tabular}
    \caption{The median inflow (left) and outflow (right) rates as a function of time for XMDs (blue solid line), LMDs (orange dashed line), XMD analogs (green dotted line) and LMD analogs (red dot-dashed line). }
    \label{fig:flowstack}
\end{figure*}

\begin{figure}
    \centering
    \includegraphics[width=1\linewidth]{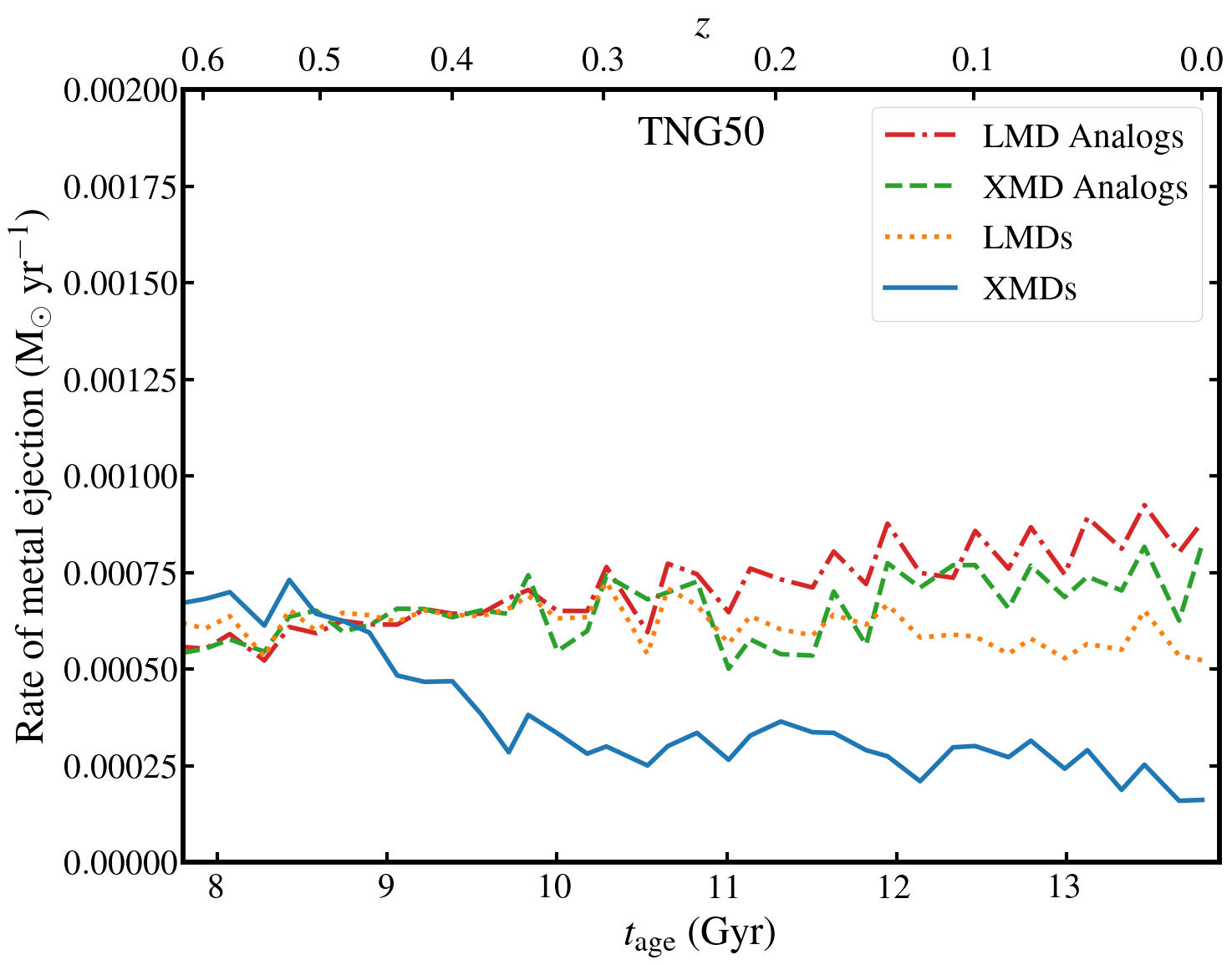}
    \caption{The rate of metal ejection for objects in TNG50, with the same linestyle as Fig.~\ref{fig:flowstack}. Contrary to expectations, XMDs actually eject fewer metals than similar-mass analogs, suggesting that particularly strong outflows are not the cause of their low metallicity in this simulation.}
    \label{fig:mzej}
\end{figure}

\section{What it's not: Proposed XMD Formation Mechanisms}
\label{sec:formation}
Having identified XMDs in the IllustrisTNG and TNG50 simulation, we are able to study their formation in more detail. In this section, we investigate the possibility that the low metallicities of XMDs and LMDs in IllustrisTNG are due to some previously proposed XMD formation mechanisms such as (a) XMDs being relics of primordial galaxy formation (b) particularly strong inflows or outflows, (c) high gas fractions or outflow efficiencies, based on simplified galaxy evolution models, or (d) particularly metal-poor inflows or isolated environments. Given the higher resolution of TNG50, we focus on the results from that simulation in this section.

Ultimately, we find that these processes are not sufficient to fully explain the formation of XMDs in IllustrisTNG. In Sec.~\ref{sec:nomix} we investigate the formation of these XMDs in more detail, and we show that because most gas in the ISM of dwarf galaxies in our sample was recently accreted, the difference between XMDs and non-XMDs can primarily be traced to differences in the degree of enrichment of this inflowing gas \citep{Williamson2016}.
\subsection{Not Primordial Relics}
\label{sec:relics}
First, we look to establish whether XMDs and LMDs have always had low metallicities as fossilized relics of primordial galaxy formation, or whether their low metallicities are recent. Figure~\ref{fig:metvstime} illustrates the time evolution of XMDs, LMDs, and their analogs. This shows that XMDs and LMDs have existed for all of cosmic time: they didn't form from IGM gas in a single burst of recent star formation. Rather, some internal or external processes have recently lowered the metallicity of their gas.
Additionally, it is apparent that the metallicities of individual objects vary substantially over the course of the simulation --- objects that are XMDs at $z=0$ are not necessarily metal-poor objects at any earlier time, and objects that would be considered XMDs at earlier times are not necessarily metal poor at $z=0$. Despite this scatter, the metallicity evolution of XMDs, and to a lesser extent LMDs, show a noticeable trend; in particular a dramatic drop in metallicity in the past $3-4$ Gyr before $z=0$. This drop can be substantial, with the metallicity of some objects decreasing by over $0.5$ dex. This is not a consequence of choosing SFR-weighted metallicity as opposed to mass-weighted metallicity for our measurement. The bottom panel of Figure~\ref{fig:metvstime} illustrates that the mass-weighted metallicity shows a similar pattern, albeit at a lower metallicity overall. This suggests that the low metallicities of XMDs and LMDs in IllustrisTNG are a consequence of some recent internal or external influence on their evolution.

\subsection{Not Particularly Strong Inflows or Outflows}
\label{sec:flows}
As gas inflows and outflows are considered likely candidates for XMD formation, we next study the inflows and outflows of XMDs in TNG50.
Inflows and outflows are identified at 5 times the stellar half-mass radius ($r_h$) of the galaxy (far enough away that it isn't too affected by internal gas motions, but close enough that infalling gas is expected to largely be bound to the galaxy). Figures~\ref{fig:flowxmd}, \ref{fig:flowlmd}, \ref{fig:flowxmda}, and \ref{fig:flowlmda} show the inflow and outflow rates of a sample of XMDs, LMDs, XMD analogs, and LMD analogs respectively, alongside the SFR-weighted metallicity for reference. All objects, regardless of metallicity, show ``bursts" of inflowing gas, followed by star formation and significant outflows. The bursts of outflows are substantial, such that the entire gas reservoir of the galaxy is recycled every few Gyr.

However, this pattern is not more pronounced among XMDs or LMDs compared to normal-metallicity analogs. In fact, the stacked inflow and outflow profiles of all objects (Fig.~\ref{fig:flowstack}; focusing on the last 50 timesteps or $5$ Gyr) show that XMDs and LMDs have slightly \emph{lower} inflow and outflow rates than their analogs. Additionally, Fig.~\ref{fig:mzej} highlights the rate of metal ejection from objects in TNG50. The metal ejection rate among XMDs is about $75\%$ \emph{lower} than that for similar-mass analogs. These plots show that, despite the strong outflows in the IllustrisTNG simulation, the presence of \emph{particularly} strong inflows or outflows are not responsible for the low metallicities of XMDs in the simulation.

We also note that explicit recycling of gas is not significant in our sample: only $\sim20\%$ of the gas particles within $5r_h$ have been within $5r_h$ before $z=0$ (for both XMDs and non-XMDs). We note, however, that two points make analyzing the effect of recycling difficult in this experiment.
First, much of the gas within the simulated galaxies at $z=0$ is composed of gas particles that have recently split as they become more dense during accretion. 
Gas cells that are created as part of this process are not technically ``recycled" because they have not been in the galaxy before. However, if the inflowing gas was recycled, those new gas particles should be considered as being affected by recycling. Since the newly-formed gas particles are not necessarily associated with a single progenitor, it is difficult, if not impossible, to include this effect in our analysis. The second effect is more physical: gas that is outflowing from the galaxy mixes with inflowing gas as it is ejected. Again, the inflowing gas would not be considered ``recycled" because it has not been in the galaxy before. However, this would cause inflowing gas to be more metal-rich in a way that largely resembles the effect of recycling. Given these limitations, we ignore the effect of explicit gas recycling on our analysis, though we reconsider the more general effect in Sec.~\ref{sec:conculsions}.

TNG100 presents a slightly different story than TNG50 regarding the importance of inflows and outflows. In TNG100, inflows and outflows are slightly more pronounced among XMDs than non-XMDs. However, this effect is still not substantial, with the strength of the maximum $\log M_{\rm outflow}$ in the last $3$~Gyr only responsible for $4\%$ of the variance in metallicity and the maximum $\log M_{\rm inflow}$ in the last $3$~Gyr only responsible for $7\%$ of the variance in metallicity.

All this is not to say that outflows are not important. Without their strong outflows, the dwarf galaxies in IllustrisTNG would certainly have higher metallicities than we observe in the simulation. However, \emph{within} the dwarf galaxy population, the relative strength of outflows does not appear to be a determining factor for the galaxies' metallicity.

\subsection{Mostly Not Current Gas Fraction or Outflow Efficiency}
\label{sec:testmod}
\begin{figure}
    \centering
    \includegraphics[width=1\linewidth]{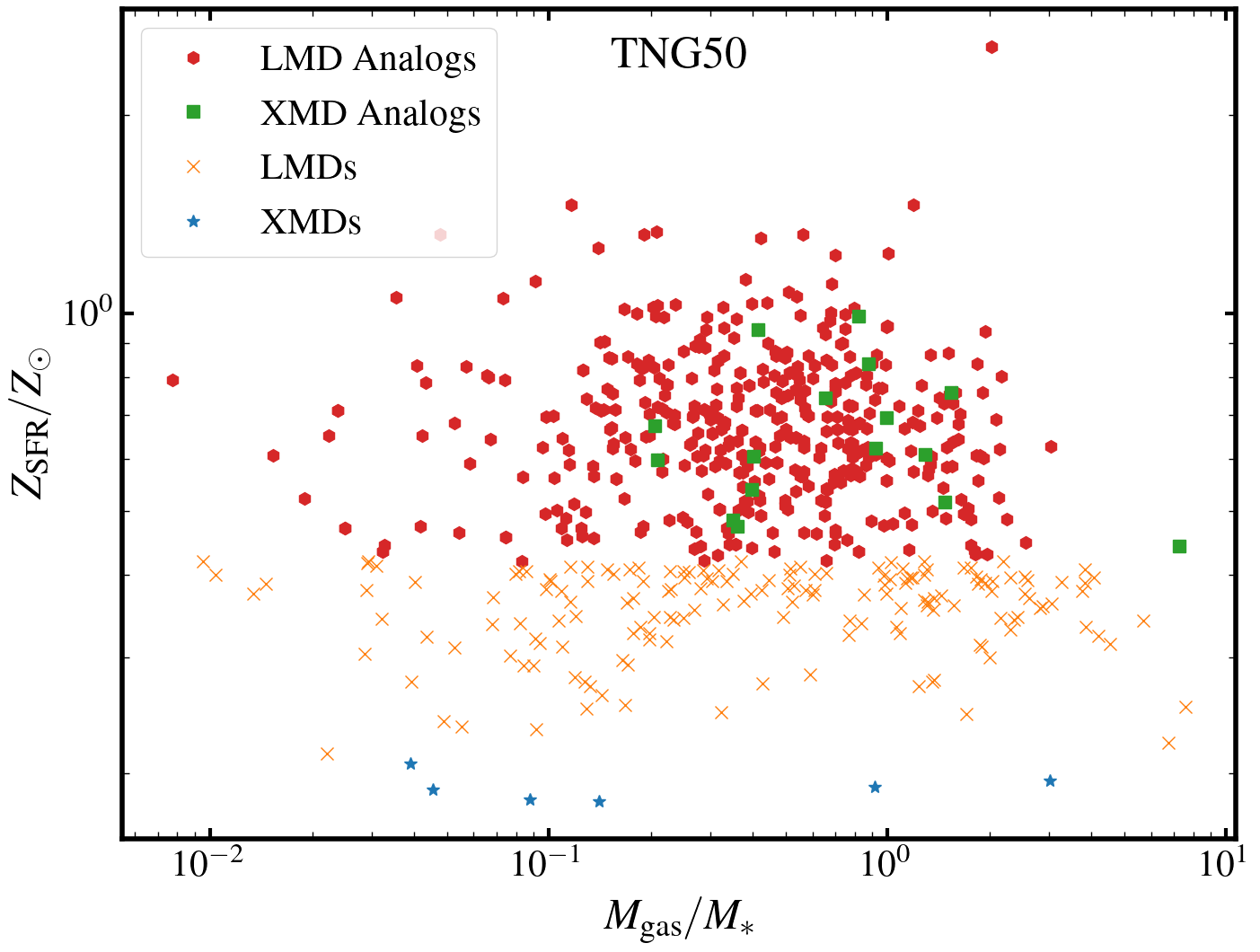}\\
    \includegraphics[width=1\linewidth]{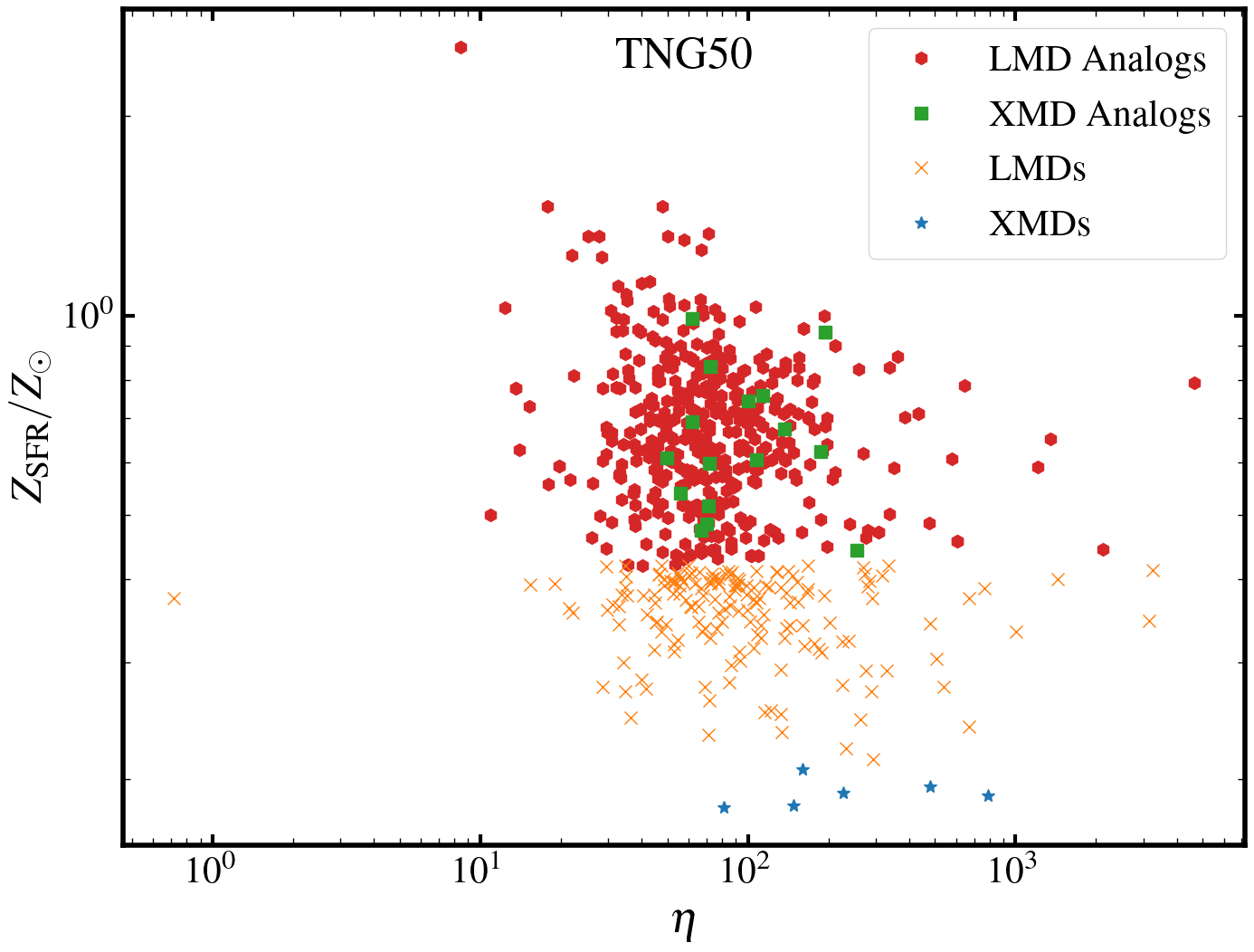}
    \caption{Metallicity as a function of gas fraction ({top}) and outflow efficiency ({bottom}), with points the same style as Fig.~\ref{fig:mzrplot}. Neither shows a strong correlation, suggesting that ``naive" expectations from galaxy evolution models do not explain the XMD population.}
    \label{fig:zvsgasfrac}
\end{figure}

Given that XMDs and LMDs do not appear to be associated with recent particularly strong inflows or outflows, we next examine the possible causes of low metallicity in the context of galaxy evolution models, such as the predictions of Eqn.~\ref{eqn:zmodel}. These models suggest that some internal properties of XMDs, such as their gas fraction or outflow efficiency, are associated with their low metallicities. On the other hand, Fig.~\ref{fig:zvsgasfrac} shows that neither gas fraction nor outflow efficiency strongly correlates with metallicity. Calculating the equilibrium metallicity from Eqn.~\ref{eqn:zmodel} (assuming $R=0.6$ and $y=0.008$) shows that the relationship between equilibrium metallicity and $Z_{\rm SFR}$ only accounts for $16\%$ of the variation in $Z_{\rm SFR}$\footnote{Throughout this work, we describe parameter $X$ as accounting for $p$ percentage of the variation of $Z_{\rm SFR}$. This is calculated as $1-\frac{\sigma_{\rm residual}^2}{\sigma_{\log Z}^2}$, where $\sigma_{\rm residual}^2$ is the variance in the residuals after fitting a first-order polynomial to the relationship between $X$ and $Z_{\rm SFR}$, and $\sigma_{\log Z}^2$ is the original variance of the metallicity distribution of our sample.}.

In the course of our analysis, we note two properties of XMDs that render the use of simple galaxy evolution models, such as Eqn.~\ref{eqn:zmodel}, challenging, if not completely ineffective, for XMDs. Firstly, as seen by Fig.~\ref{fig:metvstime}, XMDs have clearly not reached an equilibrium state, such that the equilibrium limit of Eqn.~\ref{eqn:zmodel} is not valid. Secondly, the second term in Eqn.~\ref{eqn:zmodel} reaches values on the order of $10^{-4}$ or lower, given the IllustrisTNG model yield of $0.008$, gas fractions $10$ or higher, and $\eta$ values $10-100$. The inflow metal fractions observed for dwarf galaxies in our sample are typically $\sim10^{-3}$, implying that ongoing star formation only causes minor alterations to the metallicity of the accreted gas.

\subsection{Not Environment}
\label{sec:env}
\begin{figure}
    \centering
    \includegraphics[width=1\linewidth]{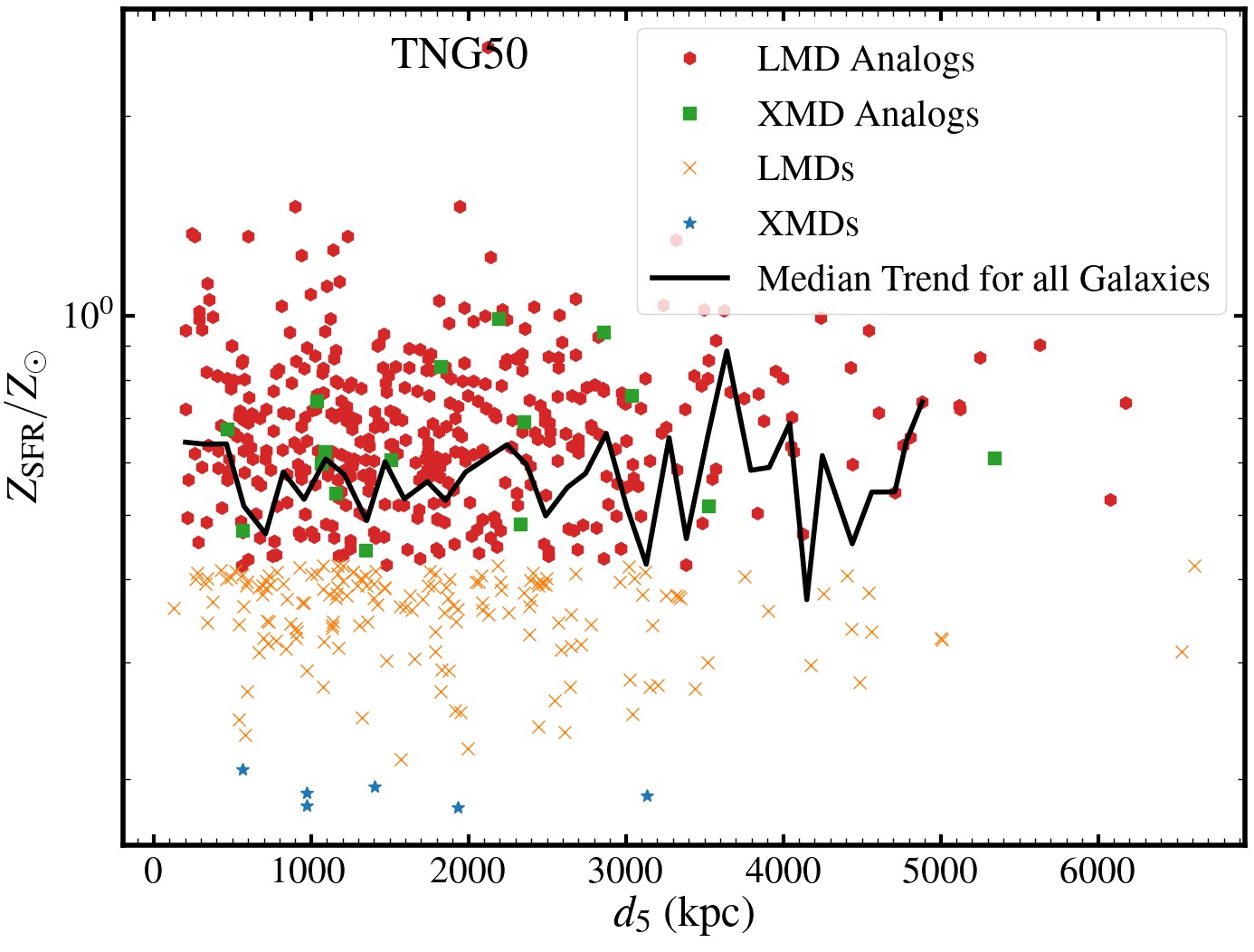}
    \caption{The relationship between environment and metallicity for XMDs, LMDs, and their analogs, with same symbols as Fig.~\ref{fig:mzrplot}. The black line represents a binned median of all points. Most star-forming dwarf galaxies live in relatively isolated environments where environment has no impact on metallicity.}
    \label{fig:dist5}
\end{figure}

Following the reasoning of Sec~\ref{sec:testmod}, it is natural to expect that external properties may influence the metallicities of XMDs and their analogs. 
Figure~\ref{fig:dist5} shows the relationship between metallicity and environment (as measured by the 3D 5th-nearest-neighbor distance, $d_5$). Among our sample, there is no significant correlation between metallicity and environment. This is likely due to our selection of low-mass dwarf galaxies that are (a) star forming and (b) not tidal debris. Given the short quenching times of low-mass dwarf galaxies \citep{Fillingham2015}, this primarily selects a sample of galaxies that have not experienced any interactions with more massive galaxies, and thus do not have very enriched inflows that could affect their current metallicity. This is in agreement with observational studies suggesting that XMDs are not preferentially found in voids \citep{Douglass2019}. 

\begin{figure}
    \includegraphics[width=\linewidth]{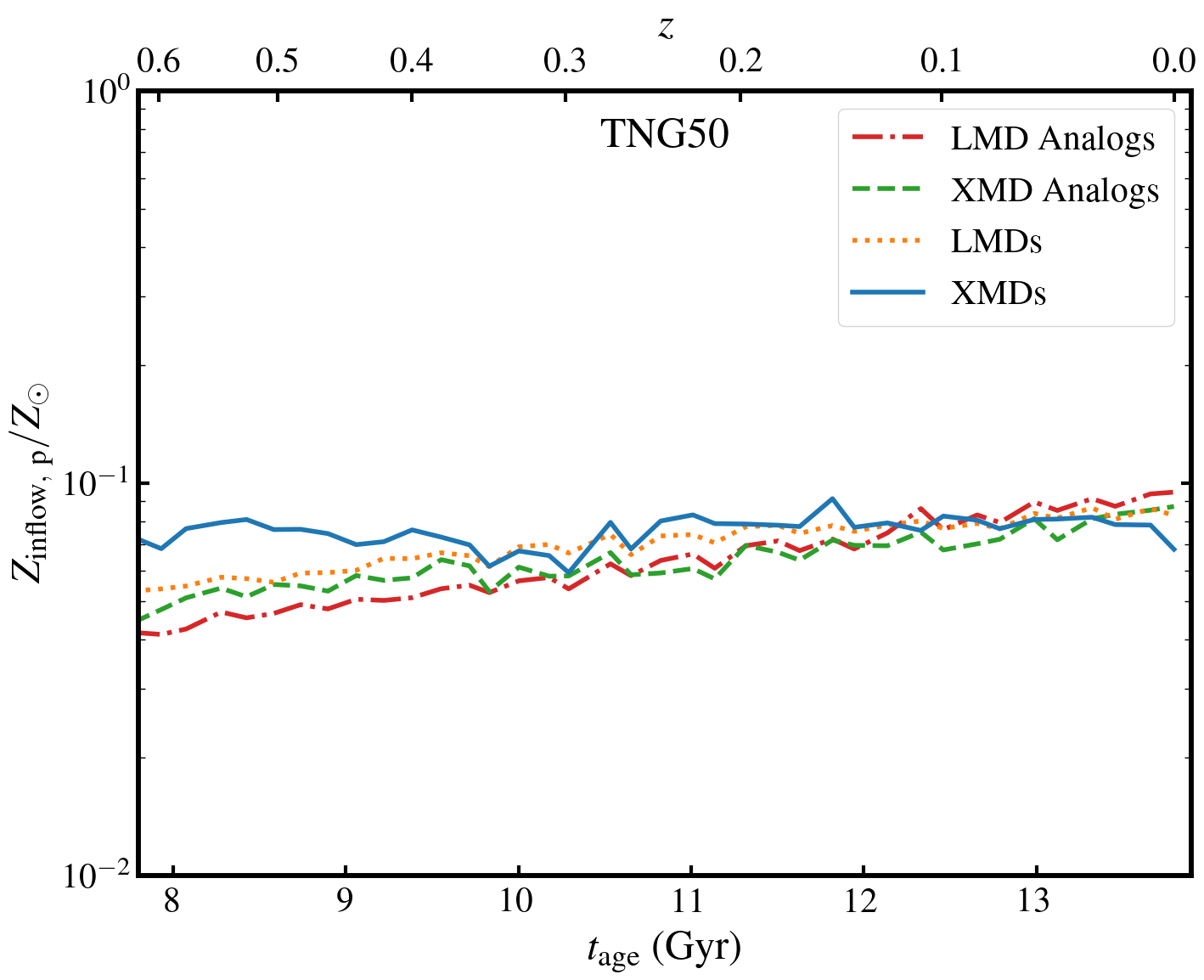}\\
    \caption{The median metallicity of inflows in our TNG50 sample (same colouring as Fig.~\ref{fig:flowstack}). The metallicity of inflowing gas is largely the same between low-metallicity galaxies and their analogs, further suggesting that environmental influences are not responsible for differences in metallicity among the populations. $Z_p$ refers to particle metallicity, as discussed in Footnote~\ref{foot:part}. }
    \label{fig:zflowstack}
\end{figure}

Perhaps more convincingly, Fig.~\ref{fig:zflowstack} shows the metallicity of accreted gas at $R_{\rm vir}$ as a function of time for our TNG50 sample. The metallicity of inflowing gas is nearly identical for XMDs, LMDs and their analogs, suggesting that most of the variation in observed metallicity is due to differences in enrichment of that inflowing gas.

\section{What it is: XMD Formation by Poor Metal Enrichment of Inflowing Gas}
\label{sec:nomix}
\begin{figure}
    \centering
    \includegraphics[width=1\linewidth]{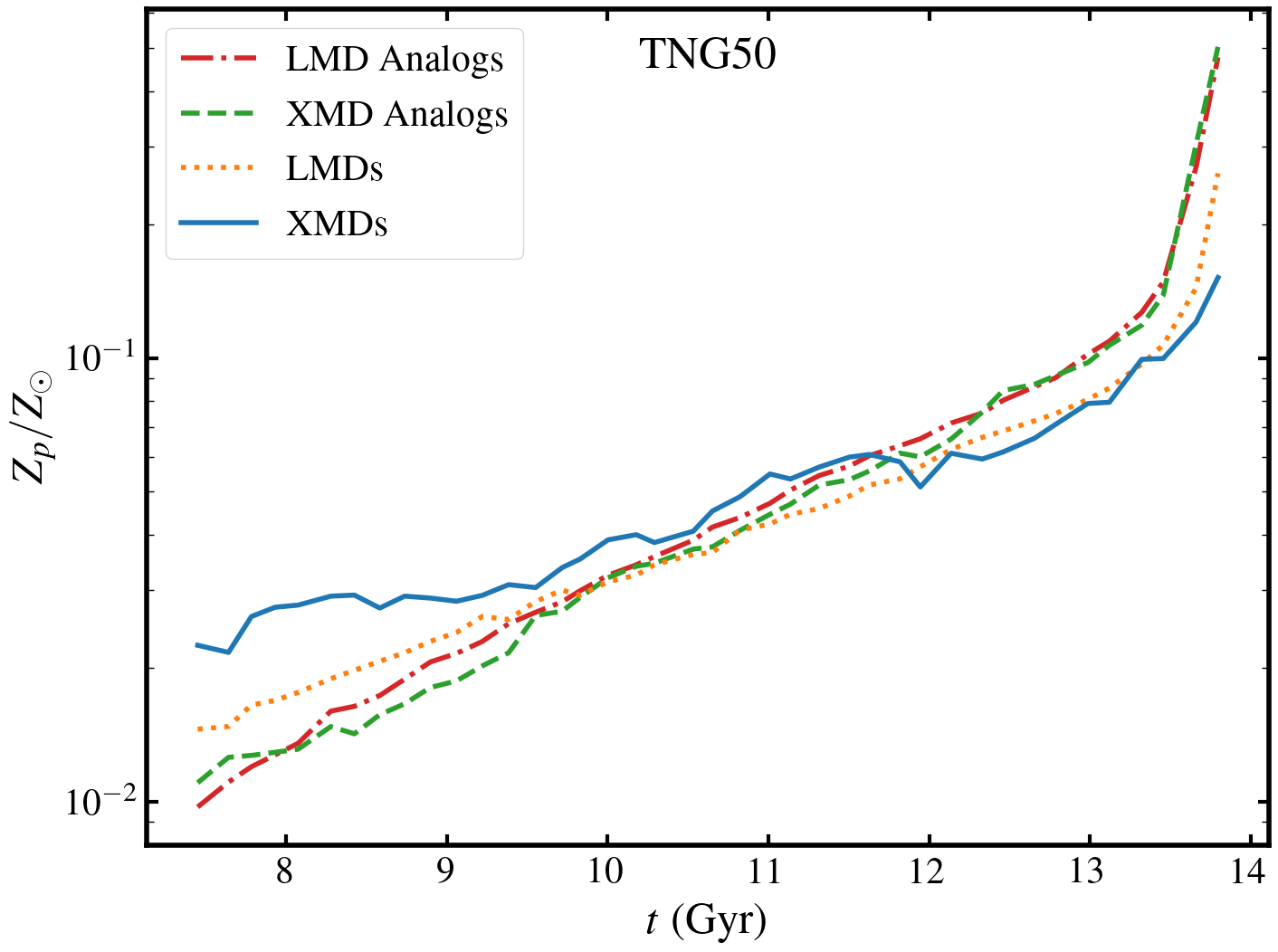}\\
    \caption{Evolution of the metallicity of gas particles that are star forming at $z=0$ (i.e. contributing to the measured metallicity), with the same style as Fig.~\ref{fig:mzrplot}. Gas in XMDs starts out with a similar metallicity as that in XMD analogs. The difference is that in the last $\sim1-2$~Gyr gas in XMD analogs becomes much more enriched, whereas gas in XMDs remains at its low metallicity. $Z_p$ refers to particle metallicity, as discussed in Footnote~\ref{foot:part}.}
    \label{fig:particleenrichment}
\end{figure}

Given the inability of ``naive" expectations from existing theories to fully explain the XMD population in the IllustrisTNG simulation, we look to the detailed evolution of particle metallicities within our sample to understand the metallicity evolution of the galaxy. If external effects were responsible for the low metallicity of XMDs, we would expect the metallicity of star-forming gas to start with lower metallicities than non-XMDs. If internal effects were responsible, we would expect the initial metallicities to be similar, but change less when accreted onto the galaxy. Fig.~\ref{fig:particleenrichment}, in which we select gas particles that are star forming at $z=0$ and trace their evolution back in time, shows that indeed internal effects are principally responsible. The initial metallicities of star-forming particles in non-XMDs are similar to those of XMDs, but change more when accreted (by mixing with the outflows of recent star formation). However, it is also apparent in Fig.~\ref{fig:particleenrichment} that the star-forming gas particles have only been in the galaxy for a small amount of time. Gas inflowing into XMDs mixes less with gas that has recently been enriched by star formation during the short time since its accretion than non-XMD analogs. 

There are a few relationships that demonstrate that inefficient enrichment of inflowing gas is the primary mechanism responsible for metallicity variations among our sample.

\subsection{Inflow Star Formation Efficiency}
\label{sec:lowisfe}
\begin{figure}
    \centering
    \includegraphics[width=1\linewidth]{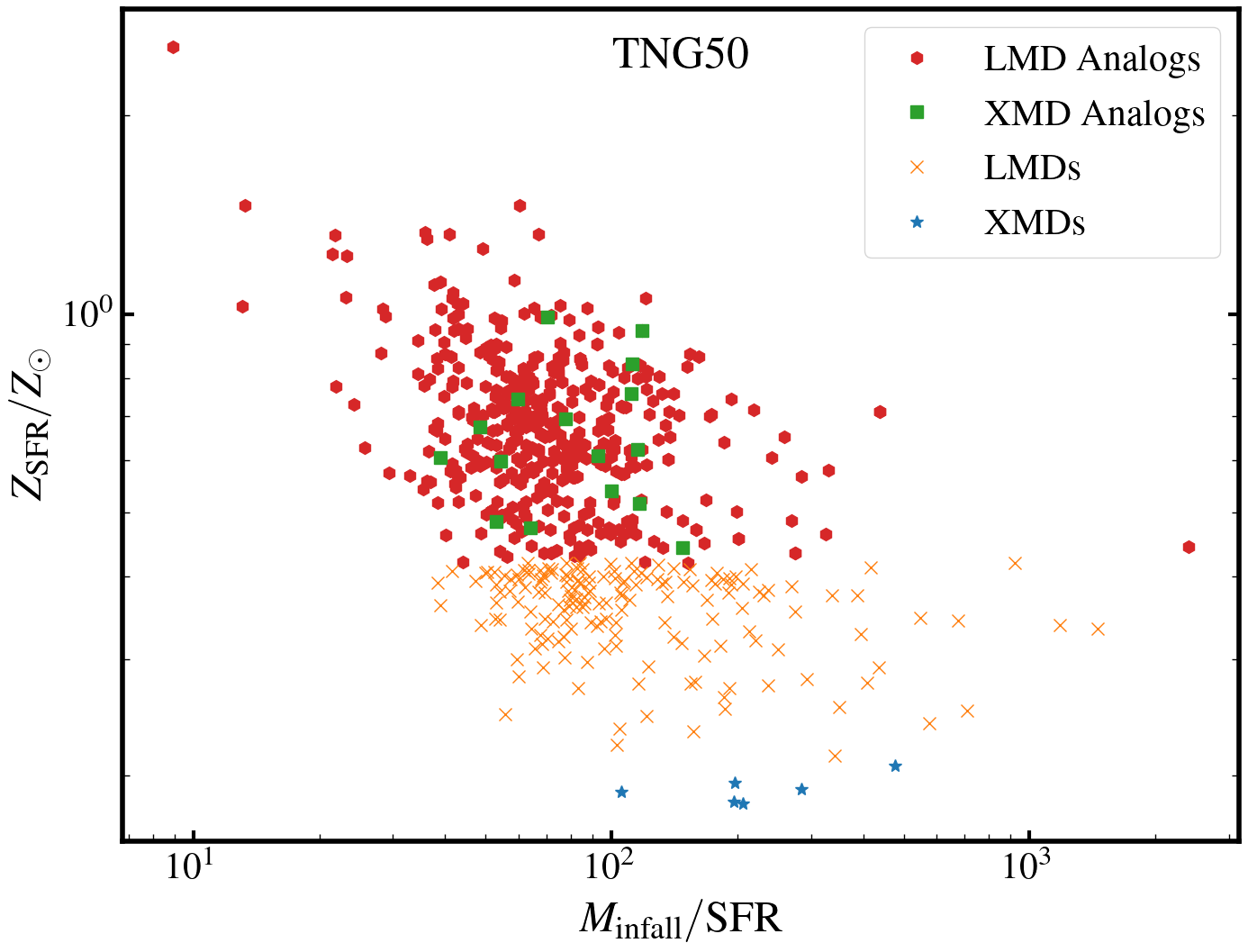}
    \caption{The relationship between metallicity and inflow star formation efficiency (the average inflow rate over the past 676 Myr divided by the average SFR within that time). In this space, XMDs and LMDs distinguish themselves from their normal-metallicity counterparts, suggesting that inflowing gas primarily contains gas that has been accreted but not been polluted by metals ejected from star formation.}
    \label{fig:sfeinf}
\end{figure}

Figure~\ref{fig:sfeinf} shows the relationship between inflowing star formation efficiency and current gas-phase metallicity. There is a significant negative correlation in this space (explaining $23\%$ of the metallicity variance) - similar to the negative correlation between metallicity and total star formation efficiency; i.e. SFR/total halo gas mass. In this Figure, XMDs are galaxies that have accreted a significant amount of gas, but have not converted that gas to stars, and thus have not enriched the star-forming gas with metals.

\subsection{Uniform Gas Metallicity}
\label{sec:uniformgas}
\begin{figure}
    \centering
    \includegraphics[width=1\linewidth]{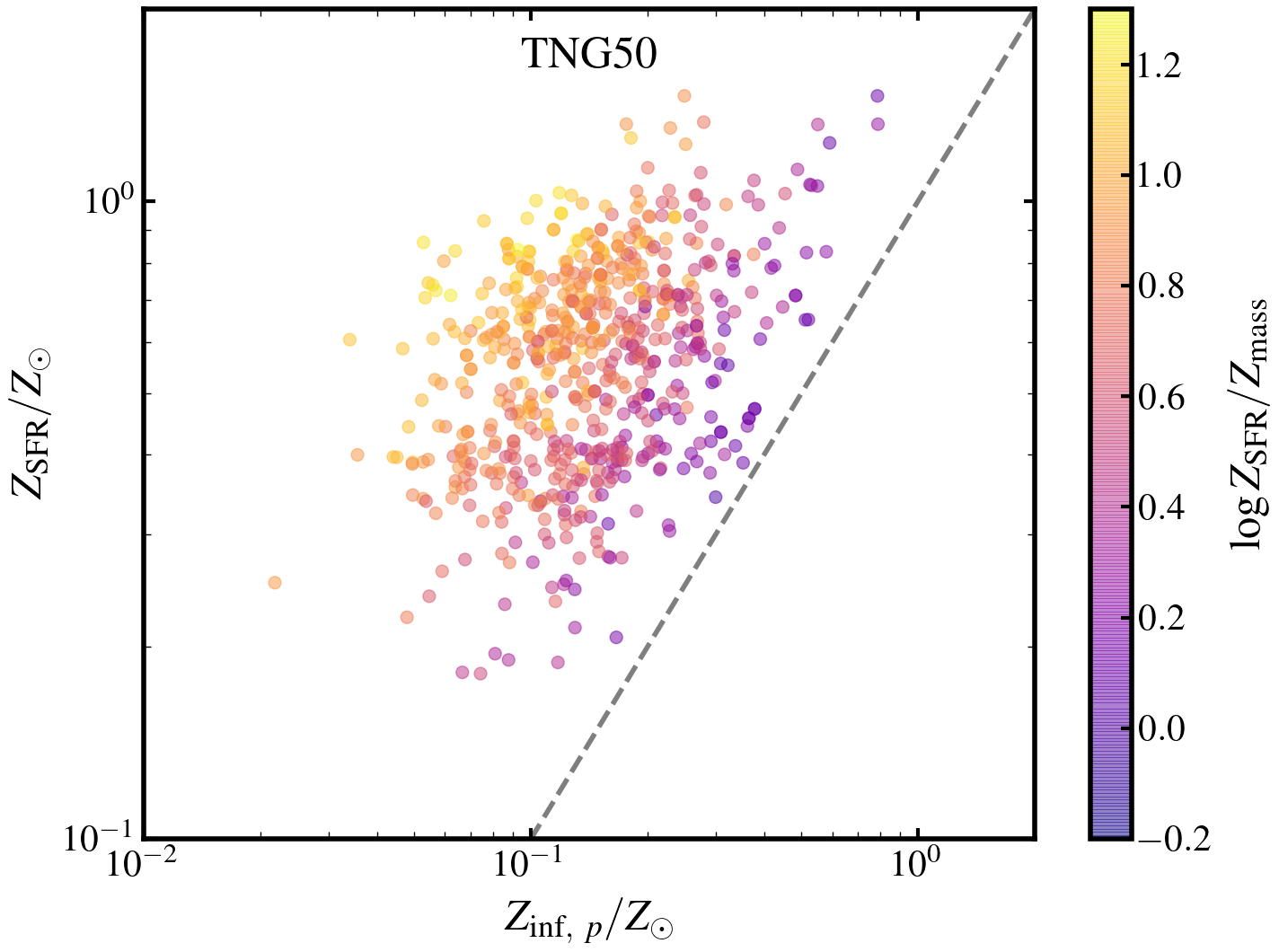}
    \caption{SFR-weighted gas-phase metallicity of gas in the galaxy as a function of inflow metallicity for all objects in our TNG50 sample (XMDs, LMDs, XMD analogs, and LMD analogs). Points are colour-coded by the log of the ratio of SFR-weighted metallicity to mass-weighted metallicity. Given the relatively narrow range of inflow metallicities, the low metallicities of XMDs are primarily due to the fact that the metallicity of star-forming gas more resembles that of inflowing gas compared to non-XMDs. The dashed line shows $Z_{\rm inf,~p}=Z_{\rm SFR}$, the limit expected in the case of no enrichment of accreted gas. $Z_p$ refers to particle metallicity, as discussed in Footnote~\ref{foot:part}.}
    \label{fig:zinfvszsfr}
\end{figure}

Figure~\ref{fig:zinfvszsfr} shows the relationship between $Z_{\rm SFR}$ and inflow metallicity, colour-coded by the log of the ratio of the metallicity of star-forming gas to that of the metallicity of the overall gas reservoir. With a small number of exceptions, the metallicity of inflowing gas occupies a small range, consistent with the results of Fig.~\ref{fig:zflowstack}. Furthermore, the mass-weighted metallicity of the gas in the galaxy largely reflects the metallicity of inflowing gas, as expected given the high relative inflow rates in our sample. Except for these rare exceptions where inflowing gas has been pre-enriched, and in agreement with Fig.~\ref{fig:particleenrichment}, XMDs are distinguished by the fact that the metallicity of their star-forming gas is closer to that of the total gas reservoir than for non-XMDs, further suggesting that the star-forming gas in XMDs is more likely to be unenriched by nearby star formation once accreted.

\subsection{Low Nearby Star Formation}
\label{sec:nearbysf}
\begin{figure}
    \centering
    \includegraphics[width=1\linewidth]{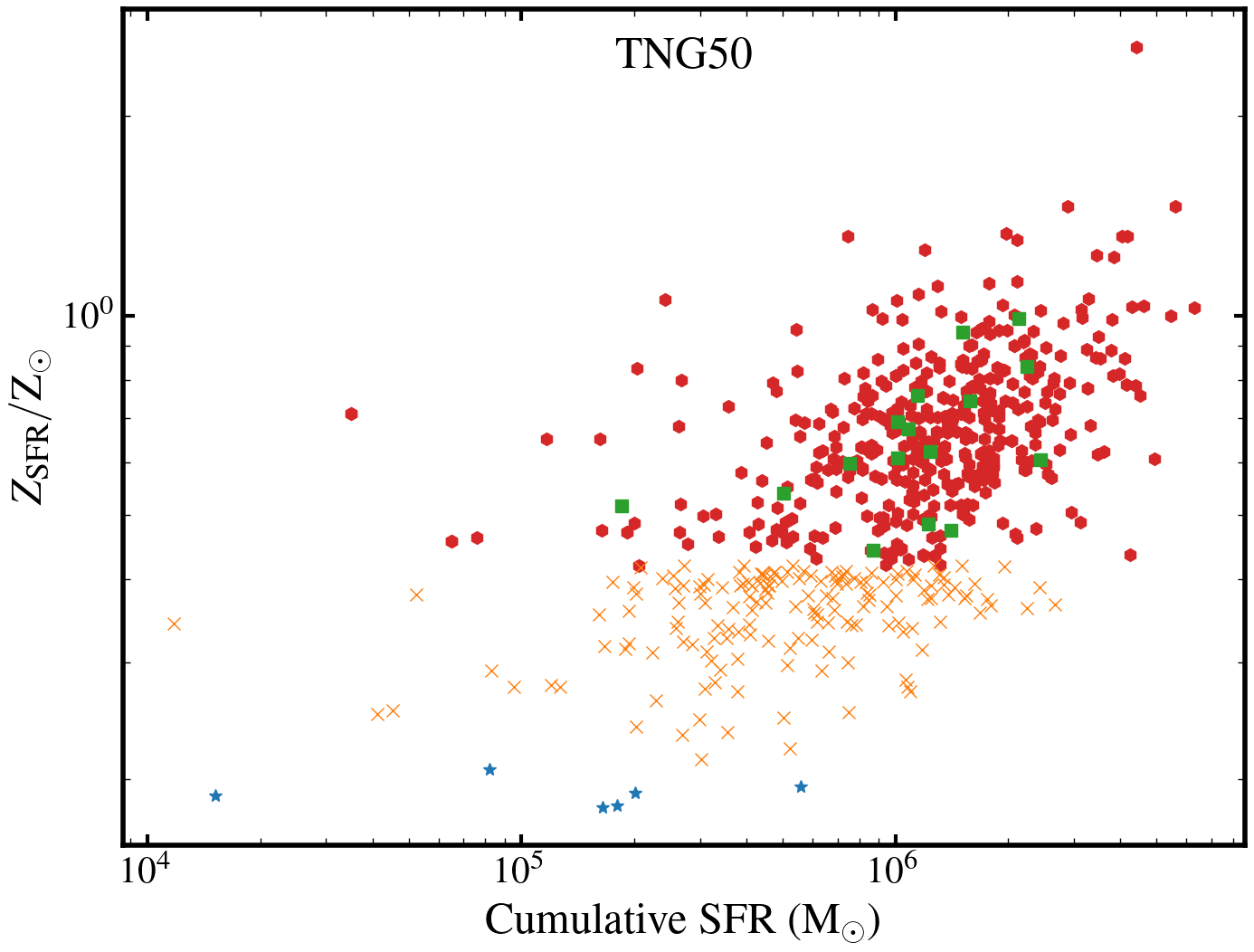}
    \caption{Galaxy metallicity as a function of the cumulative nearby (within $500$~pc of the gas particle) star formation experienced by gas particles which are star forming at $z=0$. The strong correlation in this space suggests that regional star formation is the primary parameter responsible for influencing dwarf galaxy metallicities.}
    \label{fig:regionalsfr}
\end{figure}

To most directly probe if a lack of enrichment of inflowing gas from star formation outflows is responsible for the low metallicities of XMDs, we plot metallicity as a function of the total cumulative regional star formation rate of star-forming gas particles in Fig.~\ref{fig:regionalsfr}. To do this, we track the evolution of particles that are star forming at $z=0$ back in time, noting the total amount of star formation within $500$~pc (i.e. nearby) as a function of time. While this is an imperfect measurement due to the continued formation of gas particles from condensing gas, the strong correlation observed in this space strongly suggests that low levels of regional star formation is a key factor for forming XMDs. This relation accounts for $38\%$ of the variation in metallicity among our sample, suggesting that it is the dominant parameter determining the metallicity of a dwarf galaxy in IllustrisTNG. Notably, we do not find that star-forming gas in XMDs was more recently accreted. The time that star-forming gas is within $r_h$ is similar between XMDs and non-XMDs (less than $\sim200$~Myr), consistent with the results of Sec.~\ref{sec:flows}. While this timescale is an imperfect measurement due to the same condensation effects discussed above, the short timescales point to a fundamental difference between the metallicity evolution of dwarf and massive galaxies in IllustrisTNG. In dwarf galaxies in IllustrisTNG, enriched gas does not stay in the galaxy long enough to mix with newly accreted gas and come to equilibrium, so the accreted gas has to get enriched directly from star formation ejecta. As seen in Fig.~\ref{fig:regionalsfr}, the difference is primarily in the levels of recent nearby star formation among XMDs and non-XMDs. 

\vspace{1em}

The exact details of this scenario differ slightly between the TNG50 and TNG100 simulations. In the TNG50 simulation, the difference in metal enrichment is most apparent close to the disk of the galaxy. However, in the TNG100 simulation, the difference in metal enrichment extends significantly into the CGM. The difference in metallicity between XMDs and non-XMDs at a distance of 10$r_h$ is $\log Z_{\rm XMD}/Z_{\rm XMD~Analog}=-0.30$~dex in TNG100, whereas it is $+0.12$~dex in TNG50. While the higher resolution of the TNG50 simulation would generally favor the former case, the fact that the TNG100 simulation more closely matches the observed mass-metallicity relation suggests that a possible correlation between the metallicity of CGM gas around dwarf galaxies and the metallicity of ISM gas within dwarf galaxy disks should be considered \citep{Muratov2017}.

This fits into galaxy evolution models slightly differently on different scales. On small scales (approximately the scale of star-forming regions), the lower metallicities of XMDs can be thought of as a consequence of lower metallicity inflows (i.e. gas flowing into star-forming regions of XMDs has metallicity closer to a cosmological value than the pre-enriched value of non-XMDs). On larger scales (the galaxy/halo-wide scale, where the inflow metallicity among both XMDs and non-XMDs is similar) this can be viewed partly as a consequence of the sparsity of star-forming gas. The mass-weighted metallicity, considering all of the gas in the halo, largely resembles that of the inflowing gas. This is not surprising given that $M_{\rm gas}/M_*$ ratio is $\sim20$ and $\eta$ is $\sim100$ for our objects (even considering outflows at the virial radius rather than $5r_h$). Even a relatively substantial burst of star formation producing $\sim10^7$~\msun{} only produces $\sim 10^5~$\msun{} of metals. This is quickly lost to the IGM and is not able to substantially affect the metallicity of the halo, which has a typical gas mass of nearly $10^9$~\msun{} and a metal mass of $\sim10^6$~\msun{}, even if composed primarily of recently accreted gas. \textbf{Whether a galaxy is observed to be low or high metallicity is primarily determined by whether or not the star-forming gas has been enriched during the short time since being accreted or not.}

\subsection{Decreased Recent Star Formation}
\label{sec:sfh}
\begin{figure}
    \centering
    \includegraphics[width=1\linewidth]{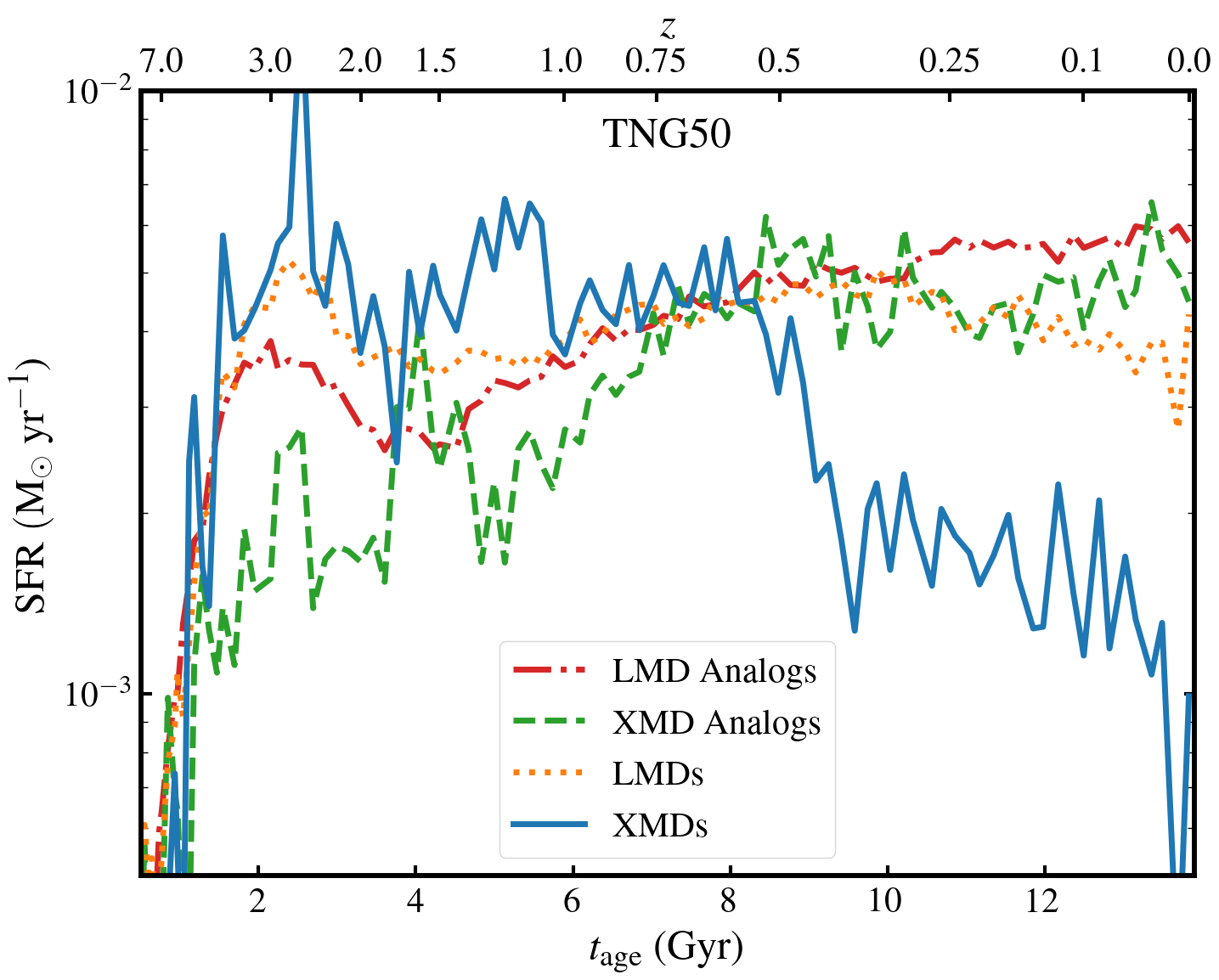} \\
    \caption{The star formation histories of our sample. A dramatic difference between XMDs, LMDs, and normal dwarfs is observed here. The star formation rates of metal-poor galaxies drops substantially between $z=0.5$ and $z=0.1$.}
    \label{fig:sfrhist}
\end{figure}

While objects were selected to have non-zero star formation at $z=0$, the lack of metal enrichment suggests that XMDs have less recent star formation than non-XMDs. Figure~\ref{fig:sfrhist} shows that XMDs, and to a lesser extent LMDs, show a dramatic drop in their star formation rates in the past $1-6$ Gyr. It is also apparent that XMDs and LMDs had higher levels of early star formation to reach the same stellar mass as their analogs given their low levels of recent star formation. We note that the correlation between metallicity and recent star formation has significantly more scatter than in Fig.~\ref{fig:regionalsfr}, implying that geometric effects are also at play for Fig.~\ref{fig:regionalsfr}.

\section{Properties}
\label{sec:properties}
Given the XMD formation scenario predicted by IllustrisTNG and outlined in Sec.~\ref{sec:nomix}, here we describe additional observational properties of XMDs in IllustrisTNG.
These can all be explained as associated with the inefficient enrichment of gas infalling into XMDs, and most can be explained by the particular rising-then-decreasing star formation histories of XMDs shown in Fig.~\ref{fig:sfrhist}.
Given that this physical cause can be observed in TNG50 and TNG100, we primarily examine XMD properties among TNG100 galaxies given the larger sample size. Again, all conclusions are generally consistent between TNG50 and TNG100 unless otherwise noted.

\subsection{XMDs/LMDs: Cored Metallicity Profiles}
\label{sec:zinf}
\begin{figure}
    \centering
    \includegraphics[width=1\linewidth]{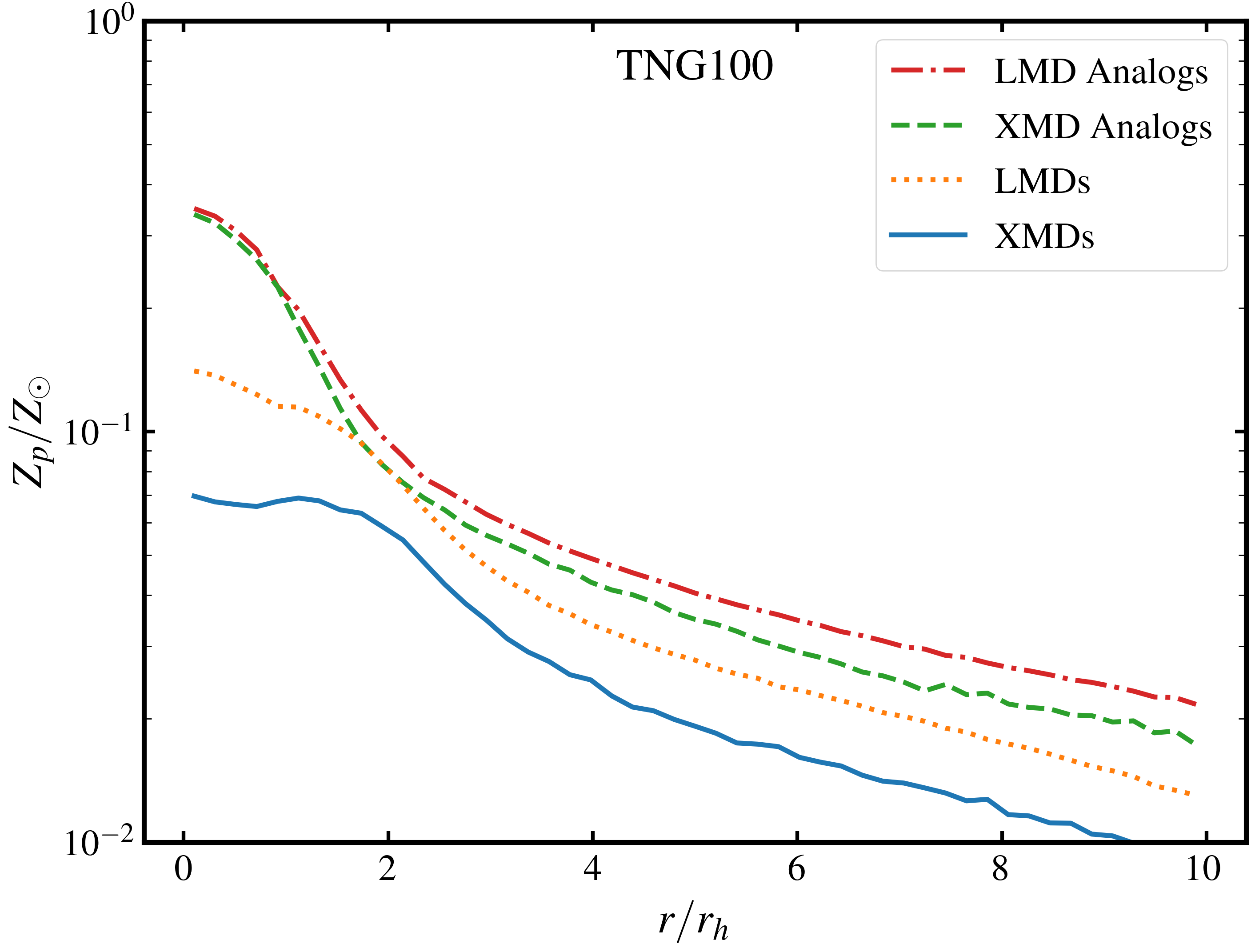}
    \caption{Stacked mass-weighted gas-phase metallicity as a function of galaxy-centric radius, normalized by the stellar half-mass radius, for XMDs, LMDs, and their analogs in the IllustrisTNG100 simulation. The relatively weak metallicity gradients among XMDs are a direct consequence of the inefficient enrichment of inflowing gas. $Z_p$ refers to particle metallicity, as discussed in Footnote~\ref{foot:part}.}
    \label{fig:metgrad}
\end{figure}

A direct consequence of the low enrichment efficiency of XMDs and LMDs is that the metallicity of star-forming gas more closely resembles that of the overall halo gas reservoir. Fig.~\ref{fig:metgrad}, which shows the mass-weighted average gas-phase metallicity as a function of galaxy-centric radius, normalized by the stellar half-mass radius, illustrates this more specifically. The central metallicity of gas in XMDs is only 4 times higher than that at $5r_h$, whereas the central metallicity of XMD analogs is 8 times that at 5$r_h$. This is in contrast with expectations from an outflow-driven model in which metal-rich outflows occupy the outskirts of low-metallicity galaxies. On the other hand, this matches the expectations of a model in which differences in level of central star formation are responsible for metallicity differences. Observations of the metallicity of diffuse gas in the galaxy outskirts and CGM should be able to test this prediction.

The metallicity gradients observed in XMD and LMD analogs are consistent with \cite{Garcia2023} using TNG50-1 and TNG50-2, who identify a change in the slope of the metallicity gradient (seen in Fig.~\ref{fig:metgrad} at $\sim 2r_h$). The shallow outer slope is explained as due to the fact that gas-mixing is driving metallicity evolution in the outer regions, whereas enrichment from stellar evolution is driving metallicity evolution in the inner regions. Given the low current SFRs of XMDs in IllustrisTNG, they are expected to have longer enrichment timescales. On the other hand, our results are in contrast with \cite{Acharyya2024} using FOGGIE, who find that metallicity inflows and outflows have substantial impacts on metallicity gradients.

\begin{figure}
    \centering
    \includegraphics[width=1\linewidth]{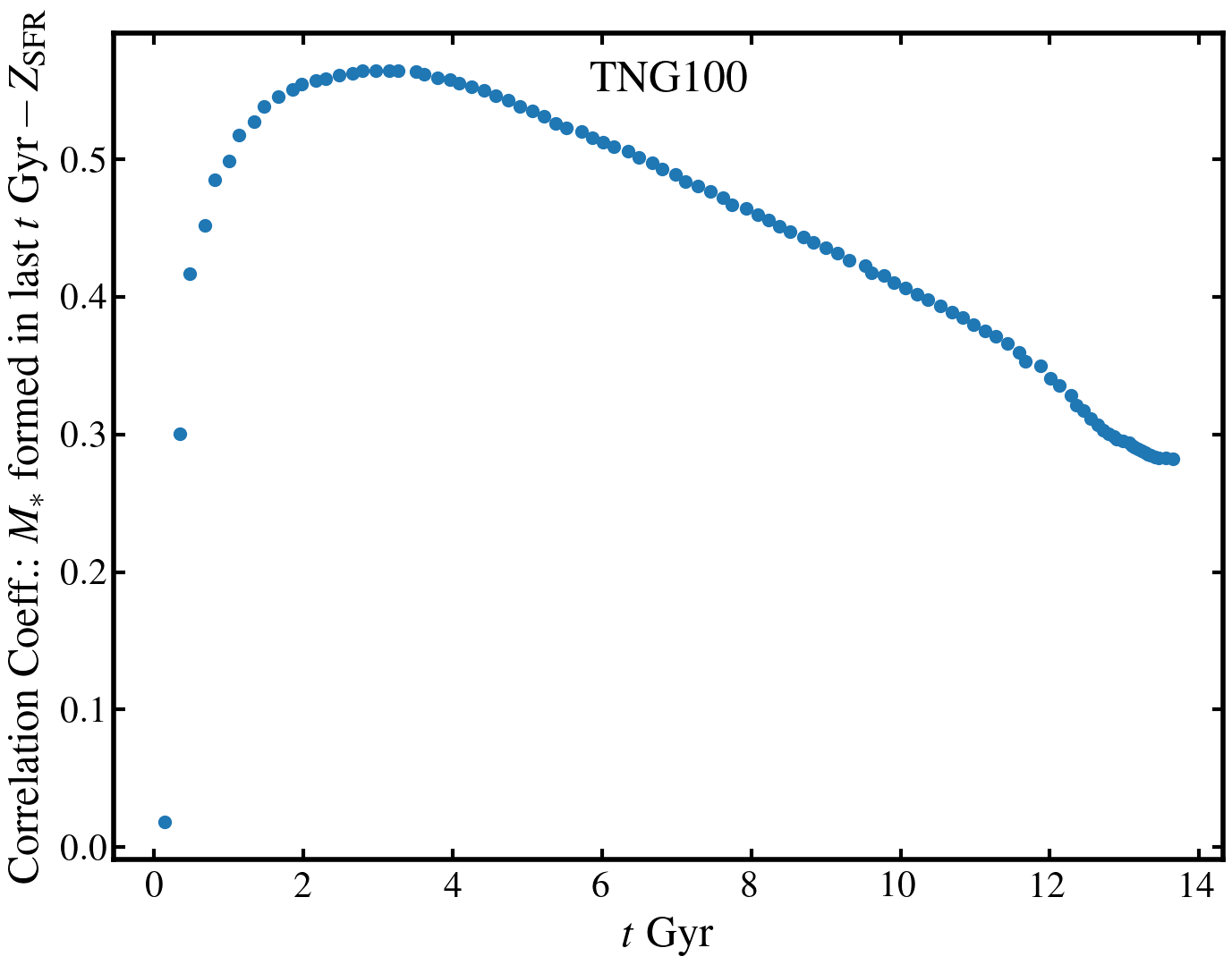}
    \caption{The correlation coefficient between metallicity and stellar mass formed in the past $t$ Gyr as a function of $t$. The Gyr-long enrichment of inflowing gas means that metallicity is most dependent on the amount of stars formed in the past $\sim3$~Gyr, rather than the total stellar mass.}
    \label{fig:stagecorr}
\end{figure}

\subsection{Star Formation Histories}
\label{sec:sfhist}
Here, we again refer to Fig.~\ref{fig:sfrhist} illustrating the star formation rate evolution of XMDs, LMDs, and their analogs. Our results paint a distinct picture regarding XMDs' star formation histories. Their star formation rates decrease significantly from a peak at $z>1$ to today, whereas the SFRs of normal metallicity analogs continue to increase on average until $z=0$.
On average, the fraction of stellar mass in XMDs with ages of $0-730$ Myr is $2\%$, the fraction with ages $730$ Myr to $5.1$~Gyr is $17\%$, and the fraction with ages $>5.1$~Gyr is $81\%$. The corresponding fractions for XMD analogs are $8\%$, $32\%$, and $60\%$. These differences should be observable with detailed spectroscopic observations (the degenerate effects of Gyr-scale age and metallicity make it difficult to study this effect photometrically: XMDs and XMD analogs have similar optical colours).

We note that this result is in apparent disagreement with observations finding XMDs with \emph{higher} than average star formation rates \citep{Isobe2021, McQuinn2020}, suggesting that some details of XMD formation could be different between observed galaxies and those in Illustris-TNG.

We note that the timescale of the decreased star formation activity present in XMDs is on the order of many Gyr in IllustrisTNG, as shown in Fig~\ref{fig:stagecorr}. This shows the Spearman correlation coefficient between the stellar mass formed in the last $t$~Gyr and metallicity, as a function of $t$. The peak at $\sim 3$~Gyr is indicative of the timescale in which accreting gas can be enriched by star formation. The low correlation coefficients as $t$ approaches $0$ and $14$~Gyr are indicative that (a) enrichment is not happening instantaneously, and (b) that enriched gas eventually leaves the halo completely and does not influence the $z=0$ metallicity. In TNG50, this trend is flatter between $t=1$~Gyr and $8$~Gyr, possibly as a result of differences in large-scale enrichment discussed in Sec.~\ref{sec:nearbysf}, or simply a result of the smaller sample size. More detailed analysis is required to fully understand the origin of this timescale, though it is likely associated with the duration that gas is within the galaxy before it begins star formation. This means that short-timescale tracers such as H$\alpha$ or UV emission should not necessarily correlate with metallicity, but that longer-timescale tracers such as H$\delta$ absorption and D$_{\rm 4000}$ may work better. 

\subsection{Stellar Size}
\label{sec:size}
\begin{figure}
    \centering
    \includegraphics[width=1\linewidth]{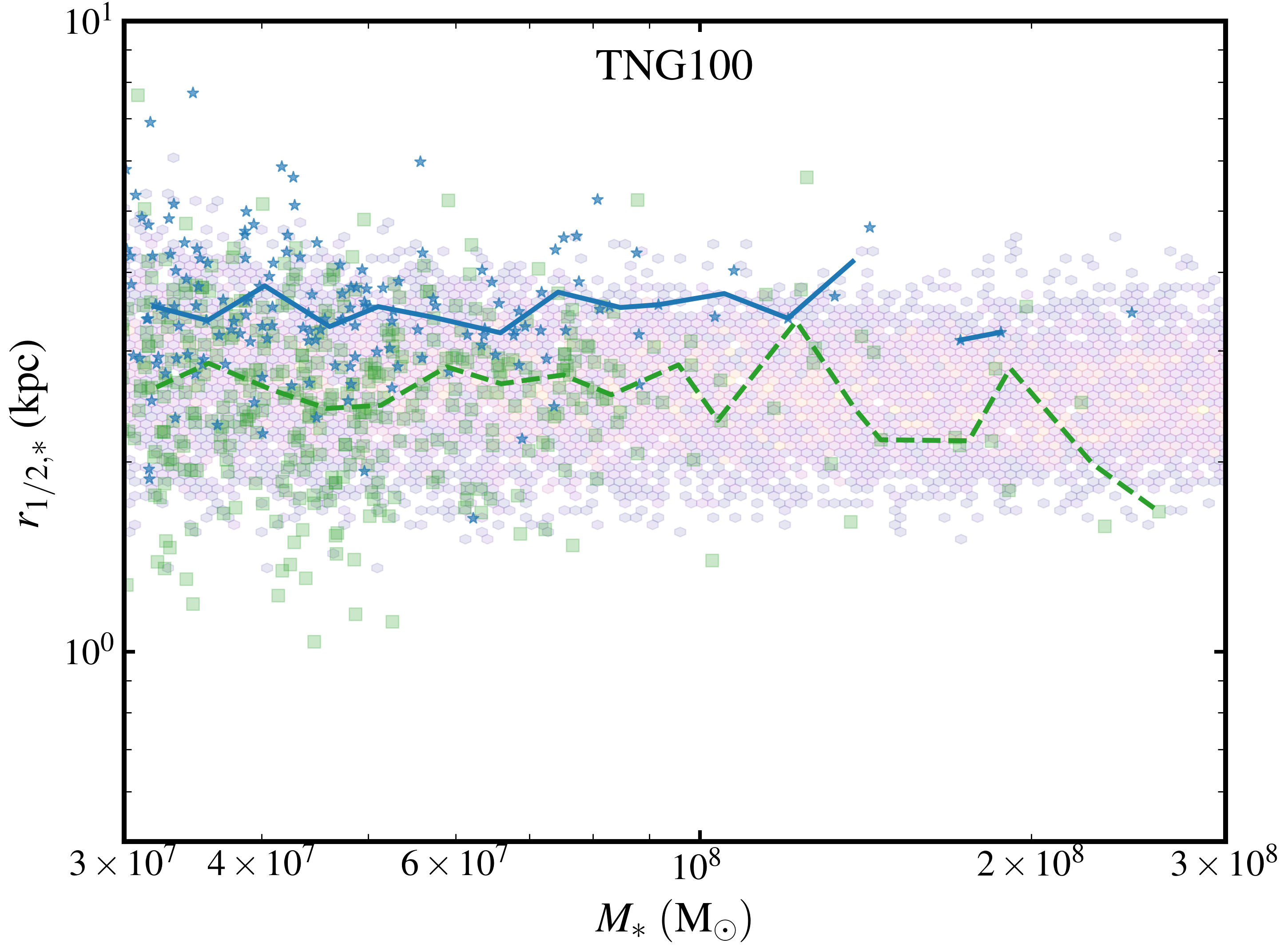}
    \caption{The stellar size-mass relation of our sample (where ``size" refers to the galaxy's 3D stellar half-mass radius; colouring the same as in Fig~\ref{fig:mzrplot}). Solid blue and dashed green lines show the median binned values of the XMD and XMD-analog points respectively. XMDs tend to have slightly larger sizes than their analogs, which is related to both the lower star formation efficiency of XMDs and the lowered efficiency of that star formation enriching inflowing gas.}
    \label{fig:masssize}
\end{figure}

Figure~\ref{fig:masssize} shows the stellar half-mass radius as a function of total stellar mass for our sample. It is apparent that the structure of dwarf galaxies in our sample is significantly correlated with their metallicity. This correlation stems from two points. First, galaxies with large sizes are less efficient at forming stars, consistent with observations of local galaxies \citep{Kennicutt2012}. The galaxies that are less efficient at forming stars from inflowing gas become less enriched, as in Fig.~\ref{fig:sfeinf}. Secondly, in galaxies with more extended star formation, the chance of inflowing gas interacting with that star formation is lower, further decreasing the enrichment of the galaxy's gas, as in Fig.~\ref{fig:regionalsfr}. This suggests an interesting connection between XMDs and isolated H{\sc i}-rich ultra-diffuse galaxies.

These effects are further demonstrated by the fact that the larger stellar sizes of XMDs apparent in Fig.~\ref{fig:masssize} are largely driven by the larger size of the young stellar population. For XMD analogs, the young stellar population is $0.9$~kpc smaller than the overall galaxy. For XMDs, the young (age$<730$~Myr) stellar population is $3$~kpc \emph{larger} than the overall galaxy size.

\subsection{Stellar Metallicity and Abundance Patterns}
\label{sec:stellarmet}
\begin{figure}
    \centering
    \includegraphics[width=1\linewidth]{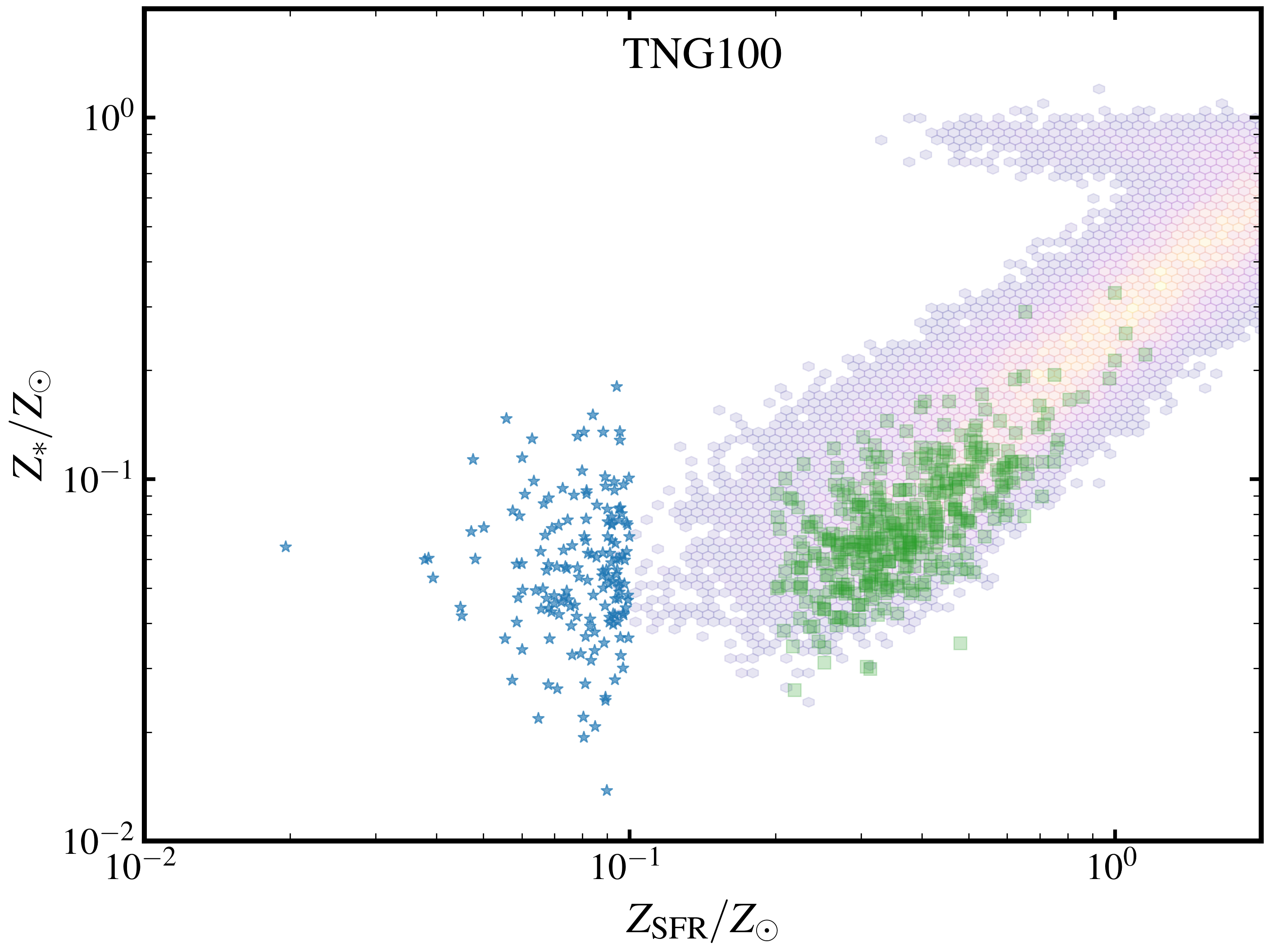}
    \caption{The correlation between gas-phase and stellar metallicity among our sample. While normal-metallicity dwarfs show a strong correlation between gas and stellar metallicity driven by cycles of star formation, gas enrichment, and star formation from enriched gas, lower metallicity dwarfs have a plateau in stellar metallicity. This is because most of stars in dwarf galaxies formed when XMDs and their analogs had a similar metallicity.}
    \label{fig:gasstellarmet}
\end{figure}

One would expect that the distinct star formation histories of XMDs should have an imprint on their stellar metallicity and abundance patterns. However, the effect is subtle. First, we note in Fig~\ref{fig:metvstime} that up until $z\sim0.5$, the metallicity of XMDs, LMDs, and their analogs was similar, and that on average $68\%$ of the stellar mass of these galaxies was formed by that point. The consequence is that the stellar metallicity of XMDs, LMDs, and their analogs are similar. Figure~\ref{fig:gasstellarmet} shows the correlation between gas and stellar-phase metallicity for XMDs and XMD analogs, alongside the other dwarf galaxies in TNG100. While there is a strong correlation at higher metallicities, stellar metallicity plateaus for low-gas-phase-metallicity objects. This is also a consequence of the early star formation disconnected with the $z=0$ gas-phase metallicities. Some observations agree with this prediction \citep{SanchezAlmeida2018}, but more are required to build up a statistical sample.

\begin{figure}
    \centering
    \includegraphics[width=1\linewidth]{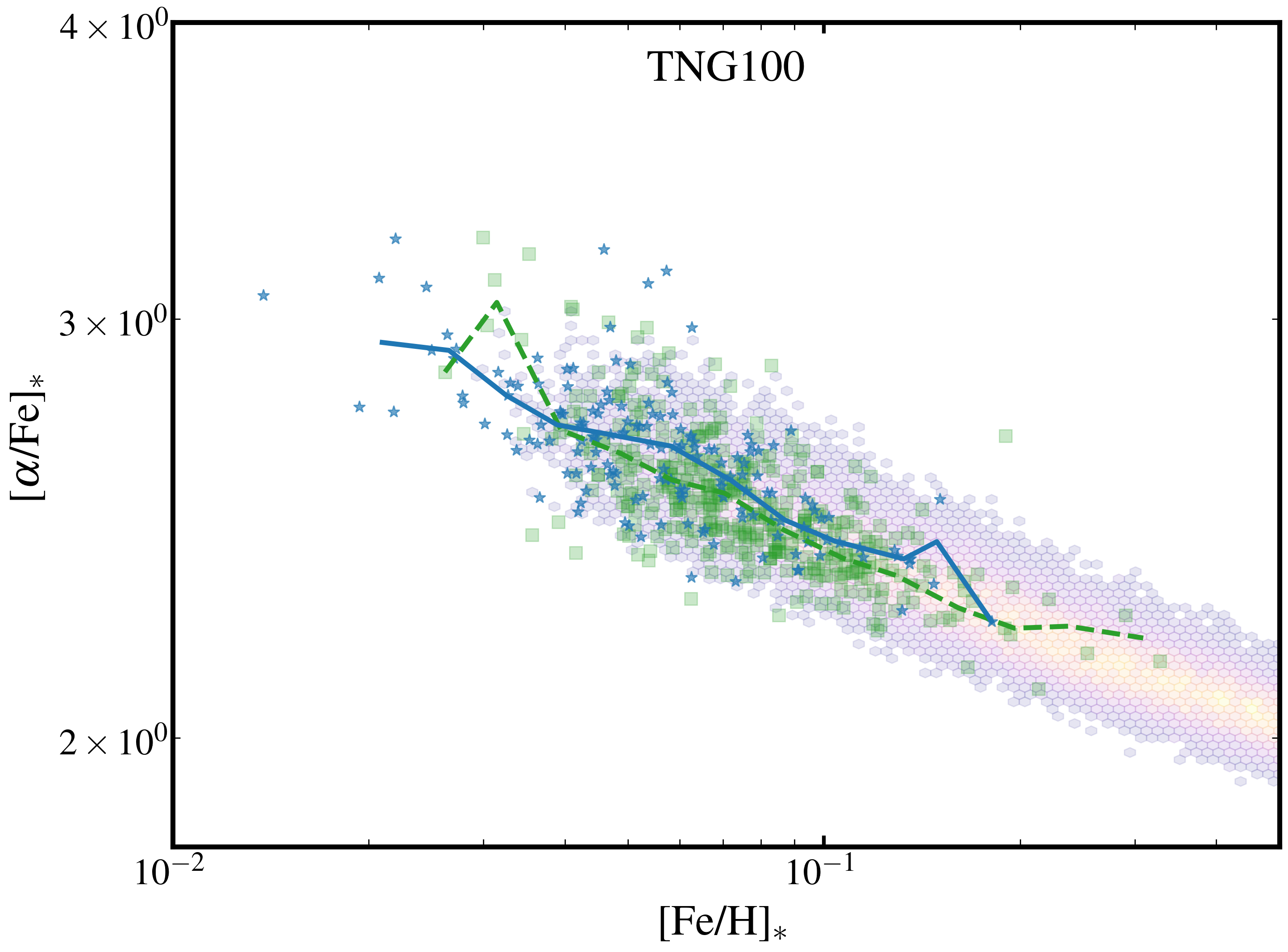}
    \caption{The $\alpha$ over Fe ratio as a function of stellar metallicity (colours the same as in Fig.~\ref{fig:mzrplot}, with green dashed and blue solid lines showing the median binned values for XMD analogs and XMDs respectively). Ultimately, the differences in star formation history are not significant enough to generate a significant difference in $\alpha/$Fe between XMD and non-XMD objects.}
    \label{fig:alphairon}
\end{figure}

Furthermore, when examining the stellar $\alpha/$Fe (including C, O, Ne, Mg, and Si as $\alpha$ elements) abundance patterns of XMDs compared with other objects (seen in Figure~\ref{fig:alphairon}), XMDs appear similar to non-XMDs for the same reason.

\subsection{Halo Properties}
\label{sec:halos}
\begin{figure}
    \centering
    \includegraphics[width=1\linewidth]{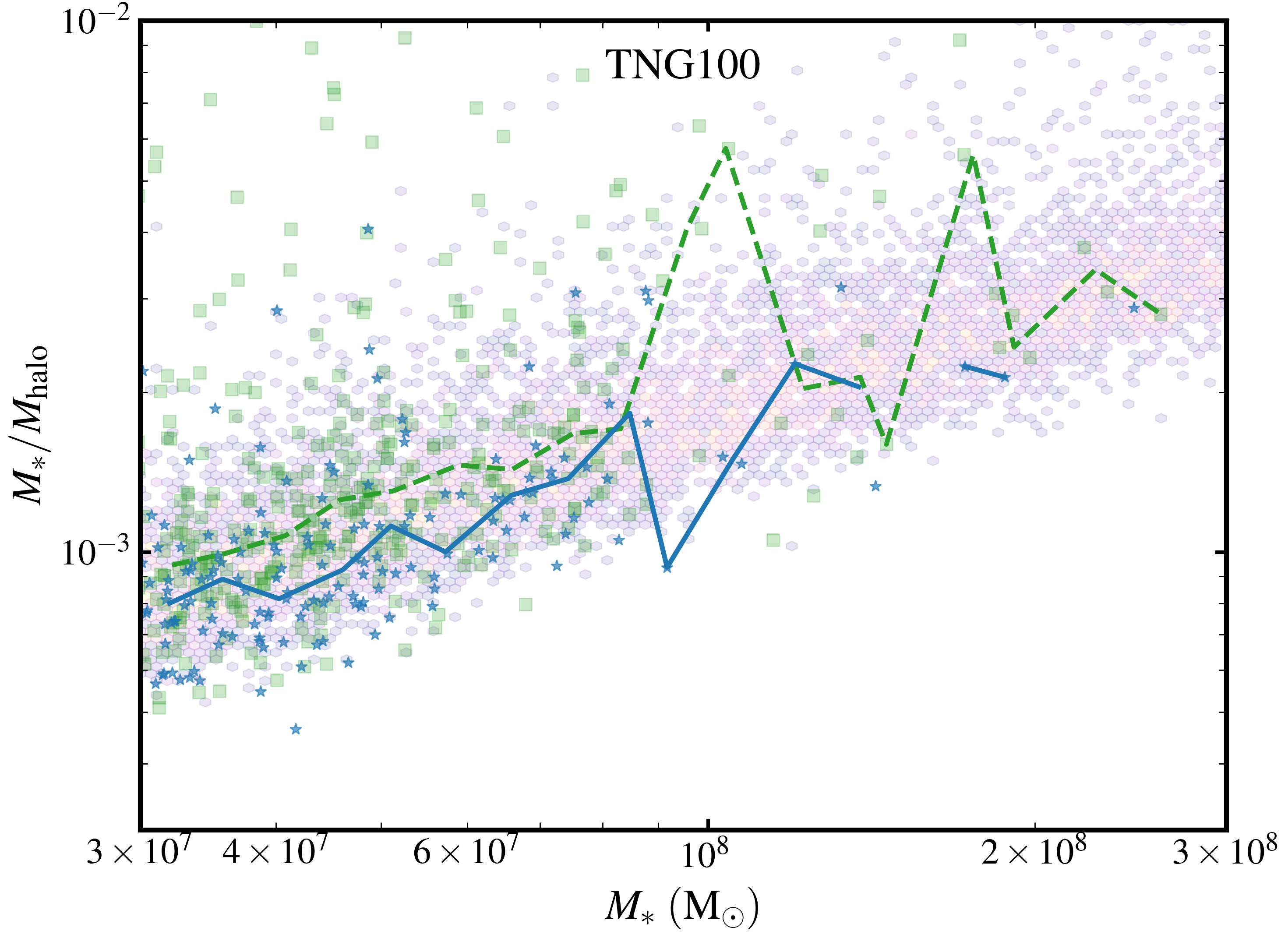}\\
    \includegraphics[width=1\linewidth]{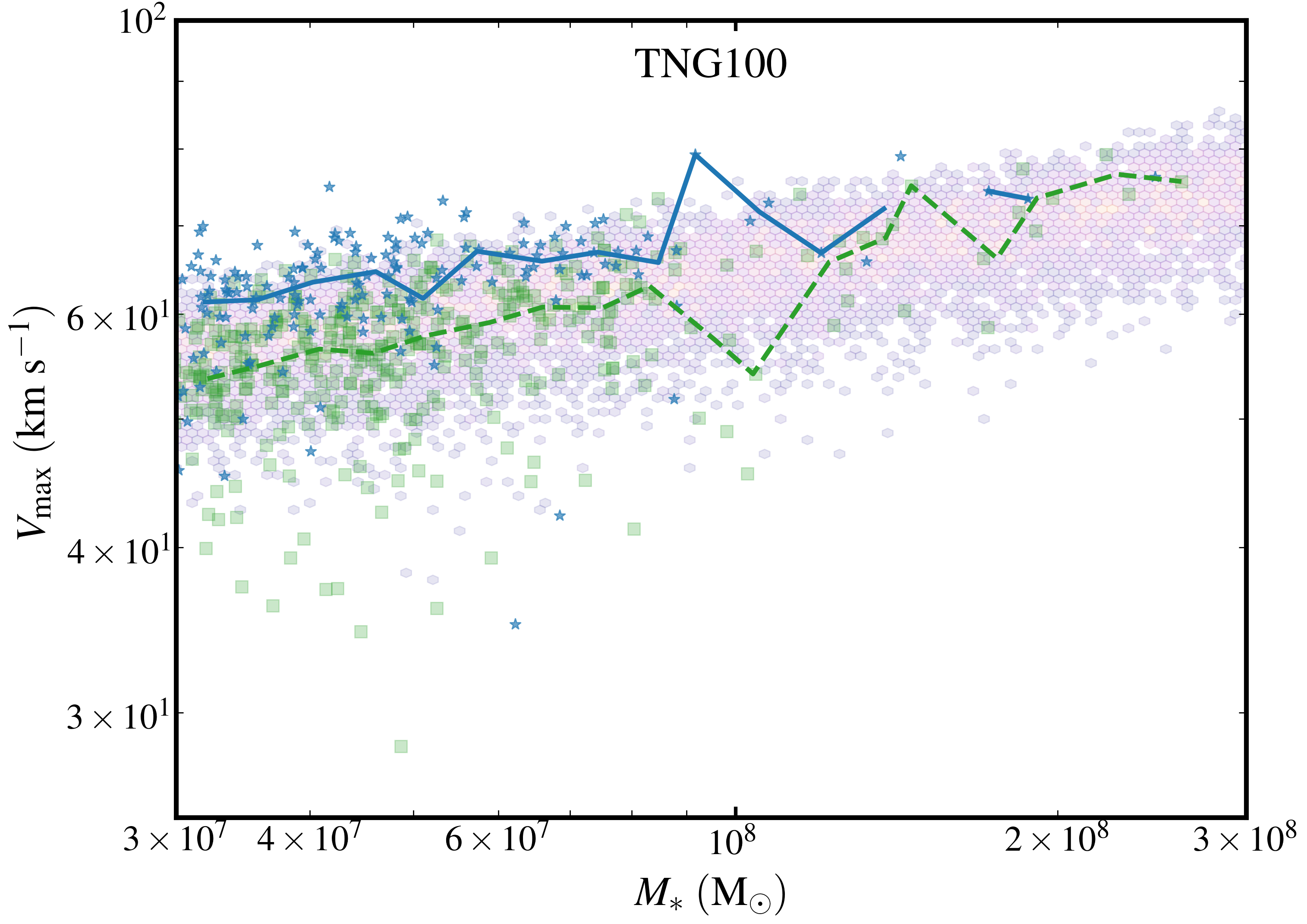}
    \caption{{\bf Top:} The stellar mass/halo mass ratio as a function of halo mass of objects in our sample (colours the same as in Fig.~\ref{fig:mzrplot}). The decreased star formation rates of XMDs at late times result in decreased stellar masses at a given halo mass.
    {\bf Bottom:} The correlation between $M_*$ and $V_{\rm max}$. The halos of XMDs are not more or less concentrated than XMD analogs, so the elevated halo masses correspond to higher $V_{\rm max}$ values, which should be observable with H{\sc i} observations.}
    \label{fig:mstmh}
\end{figure}

Given the influence of halo properties on galaxy evolution and the range of halo properties observed among dwarf galaxies, we compare the dark-matter halo properties of XMDs and their analogs. The top panel of Figure~\ref{fig:mstmh} illustrates the location of the TNG100 XMDs in $M_{*}-M_*/M_{\rm halo}$ space and the bottom panel of Figure~\ref{fig:mstmh} shows them in $M*-v_{\rm max}$ space. Notably, we find that XMDs live in dark-matter halos that are \emph{overmassive} for their stellar mass. Again, this effect is consistent with being caused by the lower recent SFR of XMDs. The halo growth histories of XMDs and non-XMDs are similar. For XMD analogs, stellar mass continues to grow alongside halo mass until $z=0$, but for XMDs, the stellar mass growth stalls, resulting in a galaxy with an overmassive halo. This effect is less apparent in TNG50, in which the halo masses of XMDs and XMD analogs are similar. In TNG50, the halo mass growth more closely mirrors the stellar mass growth, peaking in early times and stalling in later times.

\subsection{Redshift Evolution}
\label{sec:zevolution}
\begin{figure}
    \centering
    \includegraphics[width=1\linewidth]{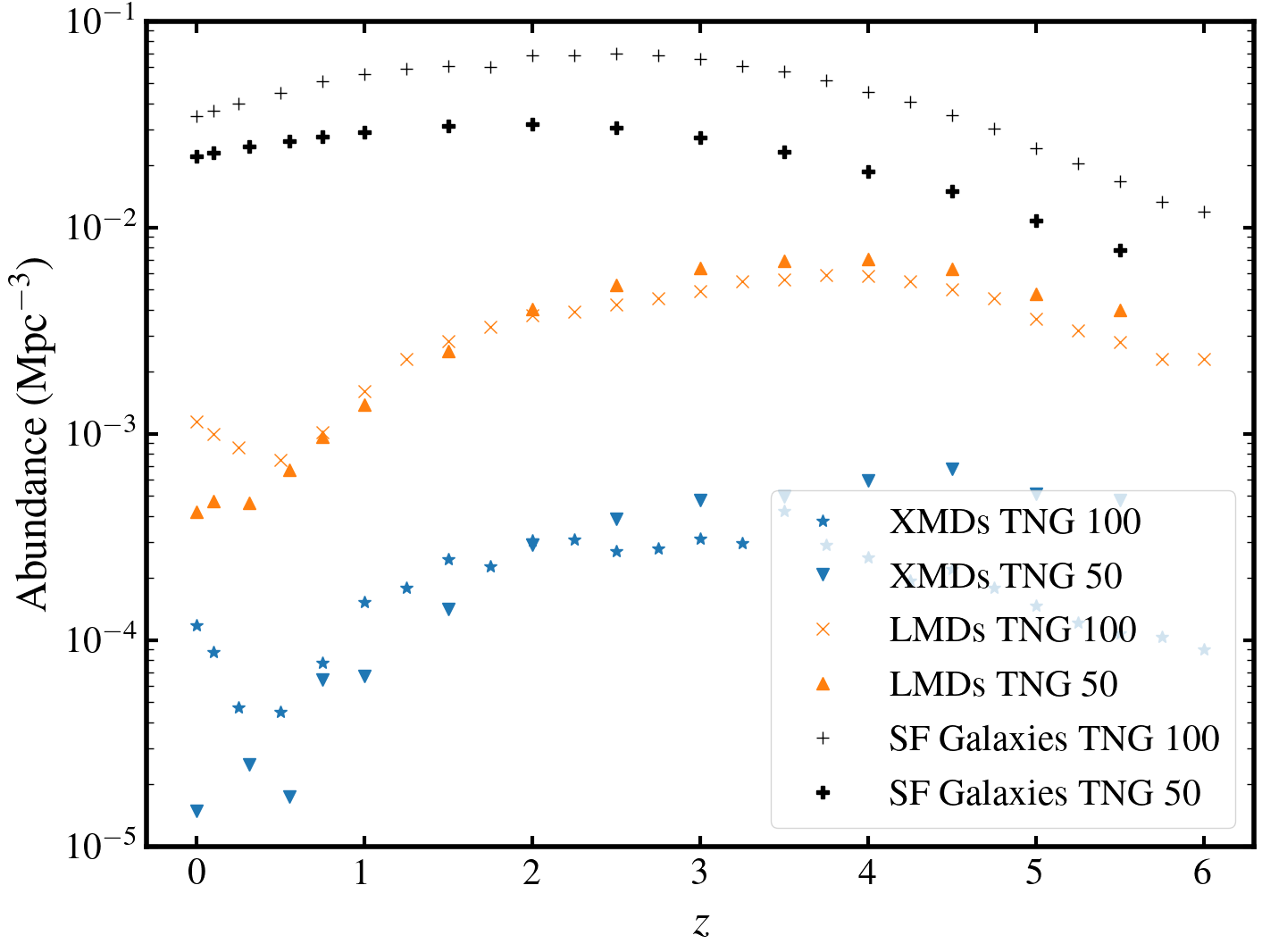}
    \caption{The comoving volume density of XMDs in TNG100 (blue stars) and TNG50 (blue down-facing triangles), LMDs in TNG100 (orange ``x"s) and TNG50 (orange up-facing traingles), and star-forming objects in TNG100 (grey thin ``+"s) and TNG50 (thick black ``+"s) as function of redshift. Before $z=0.75$, XMD and LMD abundance decreases with time as ISM gas is continually enriched. However, after $z\sim0.75$, the XMD and LMD abundance flattens out (or in the case of XMDs in TNG100, begins to increase with time). This suggests that XMD formation through inefficient star formation begins to dominate after $z\sim0.75$.}
    \label{fig:abvsz}
\end{figure}

Additionally, we note that a consequence of this formation is that the evolution of XMD abundance with redshift (illustrated in Fig.~\ref{fig:abvsz}) is non-trivial. XMD and LMD abundance decreases with time from early redshifts as gas is enriched in most galaxies; however, after $z\sim0.75$, XMD and LMD abundance flattens out.

Notably, the XMD and LMD abundance are similar among TNG50 and TNG100 up until this point. This is consistent with a scenario in which XMDs and LMDs at early times (through $z\sim0.75$) form through an external mechanism (such as the accretion of unenriched gas). At later times, the more complex mechanisms described above become dominant.

\section{Conclusions}
\label{sec:conculsions}
We have presented an investigation into the origin and nature of XMDs in the IllustrisTNG simulation. While these objects are likely different from observed XMDs, they provide valuable insight into the detailed mechanisms affecting dwarf galaxy metallicity evolution.

We find what distinguishes XMDs from the broader dwarf galaxy population in TNG is not primordial formation, extreme outflows or inflows, or a metal-poor environment, but rather inefficient enrichment of gas during the short time between its accretion and $z=0$.

We find that different issues affect galaxy evolution models on different scales, hindering their ability to predict dwarf galaxy metallicities. In general, the large gas fractions and outflow efficiencies of dwarf galaxies mean that the observed metallicity is strongly correlated with the metallicity of inflowing gas (i.e. the first term in Eqn.~\ref{eqn:zmodel} has the most variance). On small (galaxy-disk) scales, the metallicity of this inflowing gas has the most variance, but is only weakly correlated with galaxy properties.
On larger (halo-wide) scales, the metallicity of (cosmological) accreting gas is similar within the dwarf galaxy population, but only a small percentage of this gas is traced by observations.
As this scenario does not fit in nicely with metallicity evolution scenarios the authors are aware of, more work will need to be done to develop a theory of metallicity evolution driven by the enrichment of inflowing gas that is more applicable for dwarf galaxies. Although explicit recycling of gas is not considered important in our analysis (see Sec.~\ref{sec:testmod}), this mixing of outflowing and inflowing gas somewhat resembles recycling, so a recycling-dominant model may be necessary.

Observationally, this is largely characterized by a decrease in the SFR of XMDs in recent times. The lower recent star formation rates and larger sizes of XMDs compared with other dwarf galaxies means that they are less efficient at enriching this inflowing gas, although the specific geometric details of the recent star formation history and gas accretion geometry make pinning down the exact nature of metallicity evolution difficult.

We have identified a number of observable parameters to compare the IllustrisTNG XMDs to in order to test this theory.
\begin{itemize}
    \item Normal outflow indicators
    \item Large $M_{\rm HI}/{\rm SFR}$ ratios
    \item Decreasing star formation histories
    \item Stellar metallicities consistent with other dwarfs (both in terms of $[{\rm Fe}/{\rm H}]$ and $[{\rm \alpha}/{\rm Fe}]$)
    \item Isolated environments, consistent with other star-forming dwarf galaxies
    \item Large halo masses and normal concentrations, resulting in high circular velocities
    \item Large sizes, with the young stellar population more extended than the old stellar population
\end{itemize}

Lastly, we note that our analysis suggests that XMDs are not a ``special class" of galaxies, but an extension of the scatter in the mass-metallicity relation to the extreme cases. Furthermore, while the gas reservoirs of XMDs are largely composed of unenriched gas, similar to high-$z$ proto-galaxies, many XMD properties are likely different from high-$z$ counterparts.

\vspace{1em}
\noindent{\bf Acknowledgements:}

We are extremely grateful for the comments from the anonymous referee, which significantly improved the manuscript. The authors thank K. Finlator, A. Pillepich, and S. Borthakur for their insightful comments and discussion. J. Monkiewicz is supported by an NSF Astronomy and Astrophysics Postdoctoral Fellowship under award AST-1903944. T. Carleton acknowledges support from the Beus Center for Cosmic Foundations.

We also acknowledge the indigenous peoples of Arizona, including the Akimel O’odham (Pima) and Pee Posh (Maricopa) Indian Communities, whose care and keeping of the land has enabled us to be at ASU’s Tempe campus in the Salt River Valley, where much of our work was conducted. 

\section{Data Availability}
The outputs from the IllustrisTNG simulation used for this analysis are available here: \url{https://www.tng-project.org/data/}. The code used to create the plots from that data is here: \url{https://github.com/timcarleton/Illustris_XMD}.

\bibliography{main}

\appendix
\section{Resolution Comparison}
In various places in this work, we have noted that the different resolution of TNG50 vs. TNG100 causes differences in the resulting correlations. To test that this is really due to differences in resolution, we have these parts of our analysis using data from TNG50-2, which has a similar resolution as TNG100, but in a 35~Mpc/h volume.

\begin{figure*}
    \centering
    \begin{tabular}{ccc}
       \includegraphics[width=0.3\linewidth]{sampleplots/mzrnodup.png}  & 
       \includegraphics[width=0.3\linewidth]{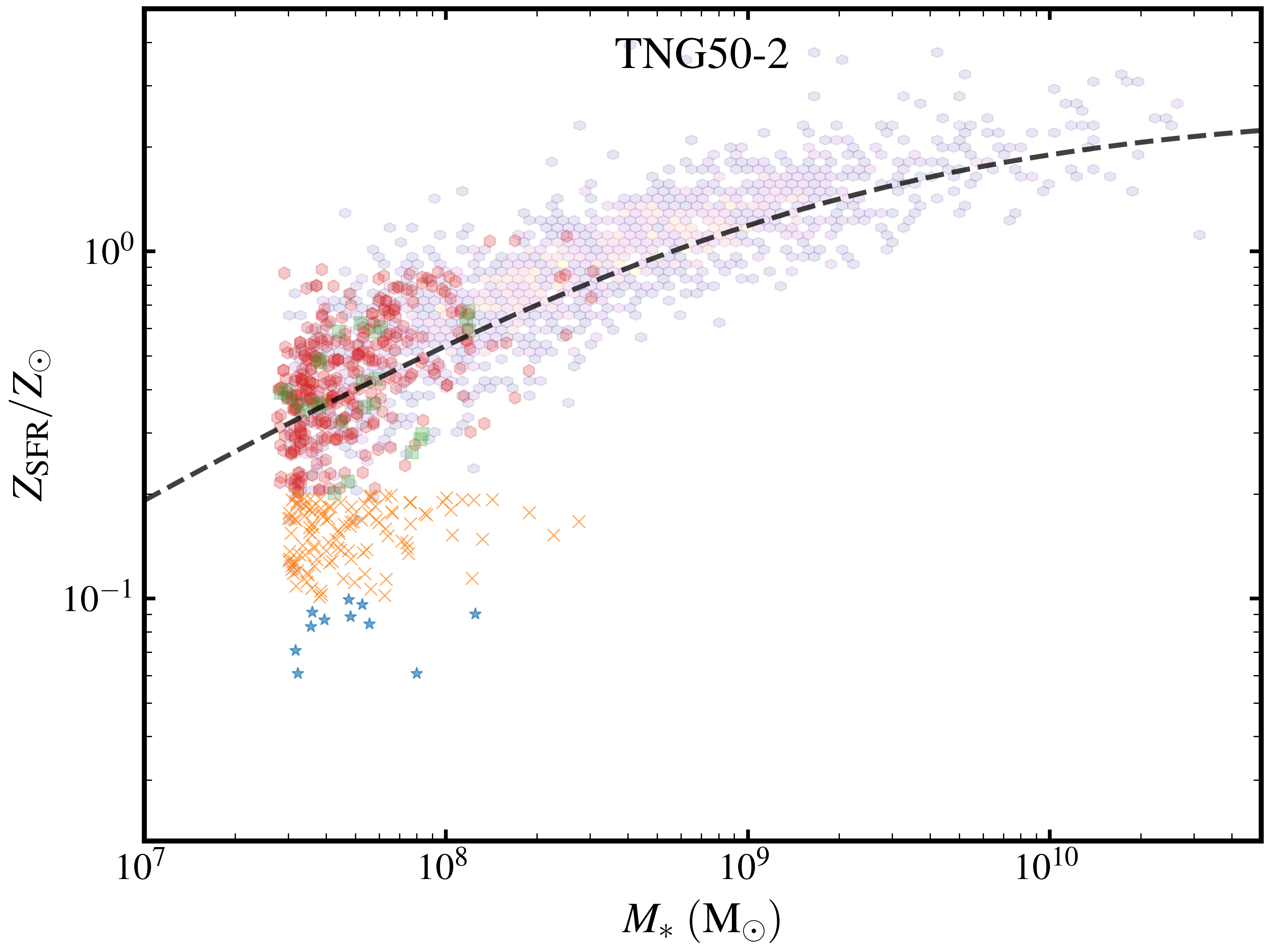} &
       \includegraphics[width=0.3\linewidth]{sampleplots/mzrtng50nodup.png}
    \end{tabular}
    \caption{The mass-metallicity relation (MZR) for the Illustris TNG100-1 (left), TNG50-2 (middle), and TNG50-1 (right) simulations (as in Fig.~\ref{fig:mzrplot}) The similarity of the TNG50-2 and TNG100-1 MZRs suggests that the higher resolution of TNG50-1 is responsible for the difference between it and TNG100-1.}
    \label{fig:mzr502}
\end{figure*}

\begin{figure*}
    \centering
    \begin{tabular}{ccc}
       \includegraphics[width=0.3\linewidth]{metgrad_noscale.png}  & 
       \includegraphics[width=0.3\linewidth]{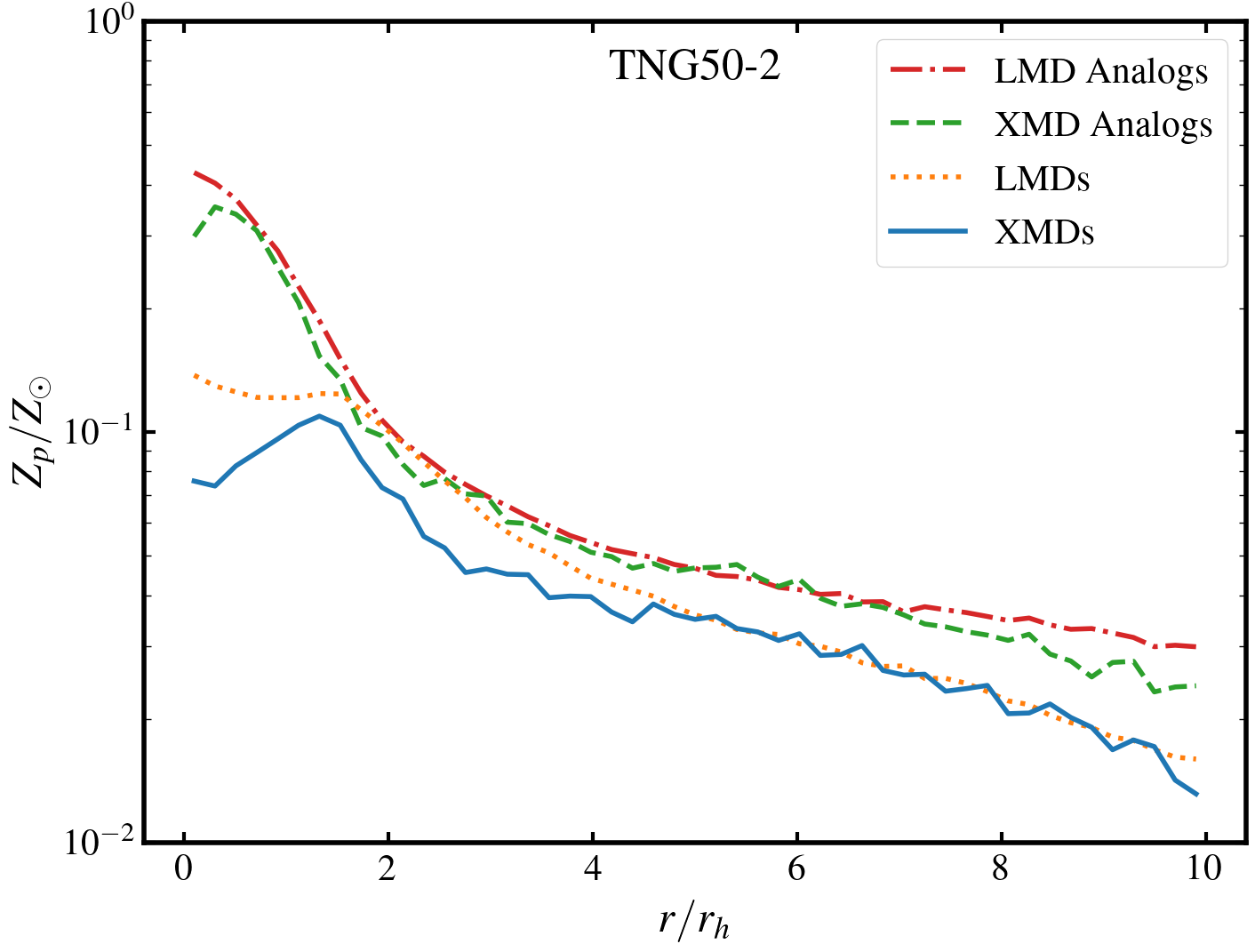} &
       \includegraphics[width=0.3\linewidth]{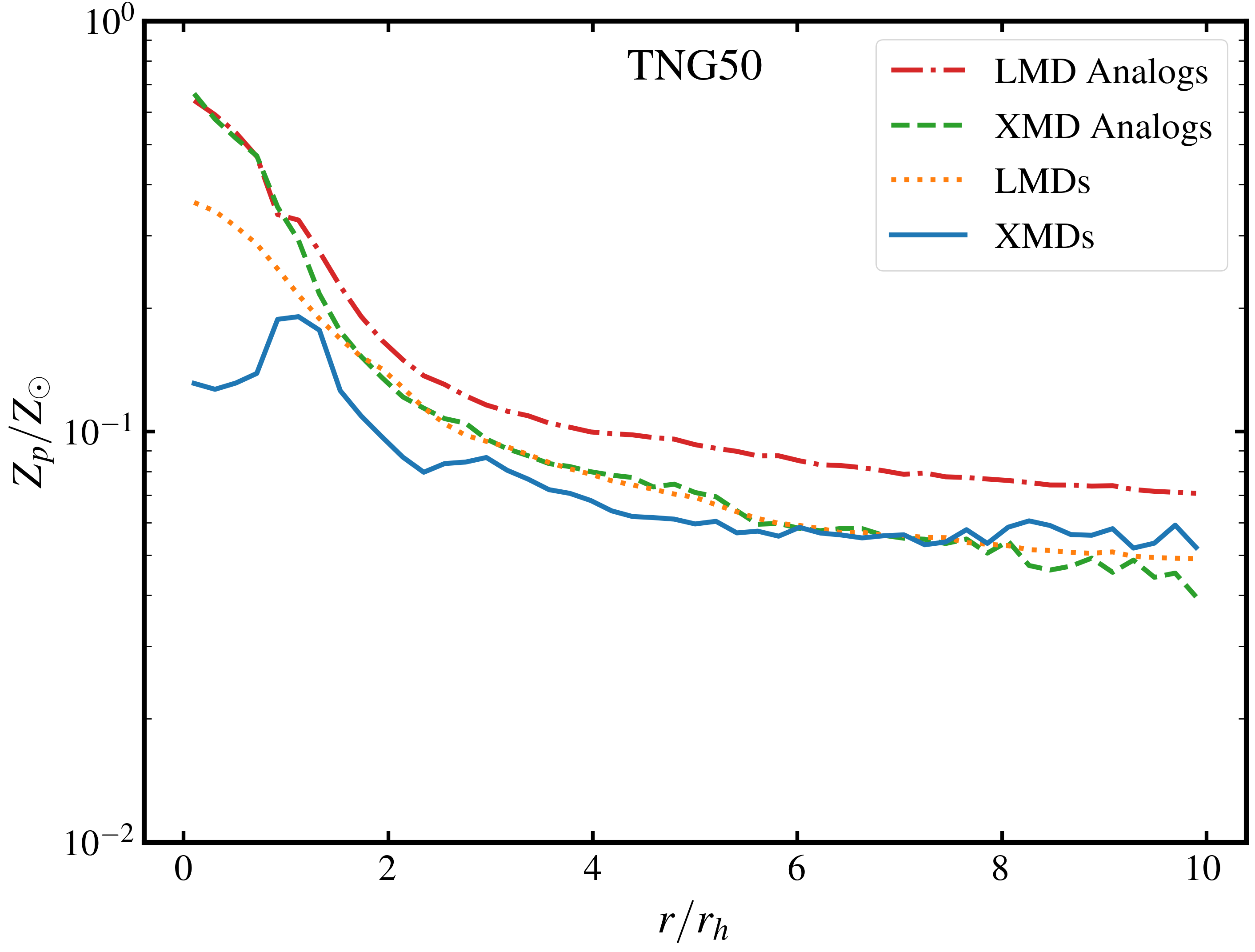}
    \end{tabular}
    \caption{The stacked radial metallicity profiles of XMDs, LMDs, and their analogs (as in Fig.~\ref{fig:metgrad}) for TNG100-1 (left), TNG50-2 (middle) and TNG50-1 (right). The results from TNG50-2 are in between those of TNG100-1 and TNG50-1, suggesting that resolution plays a role in the differences between TNG100 and TNG50. Selection effects and small number statistics likely account for the remaining differences between TNG50-2 and TNG100-1.}
    \label{fig:zgrad502}
\end{figure*}

\begin{figure*}
    \centering
    \begin{tabular}{ccc}
       \includegraphics[width=0.3\linewidth]{sampleplots/mstarhalonodup.png}  & 
       \includegraphics[width=0.3\linewidth]{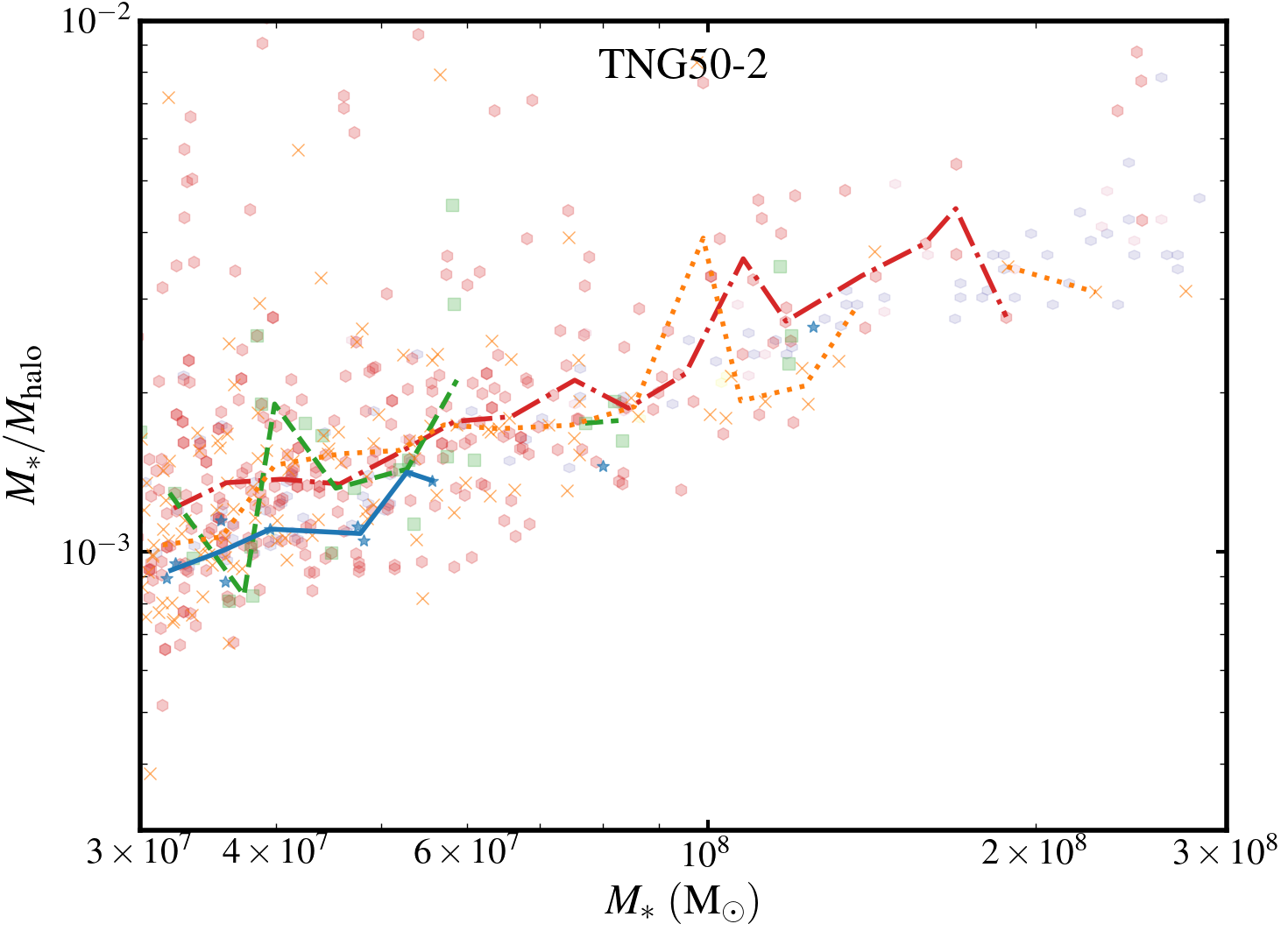} &
       \includegraphics[width=0.3\linewidth]{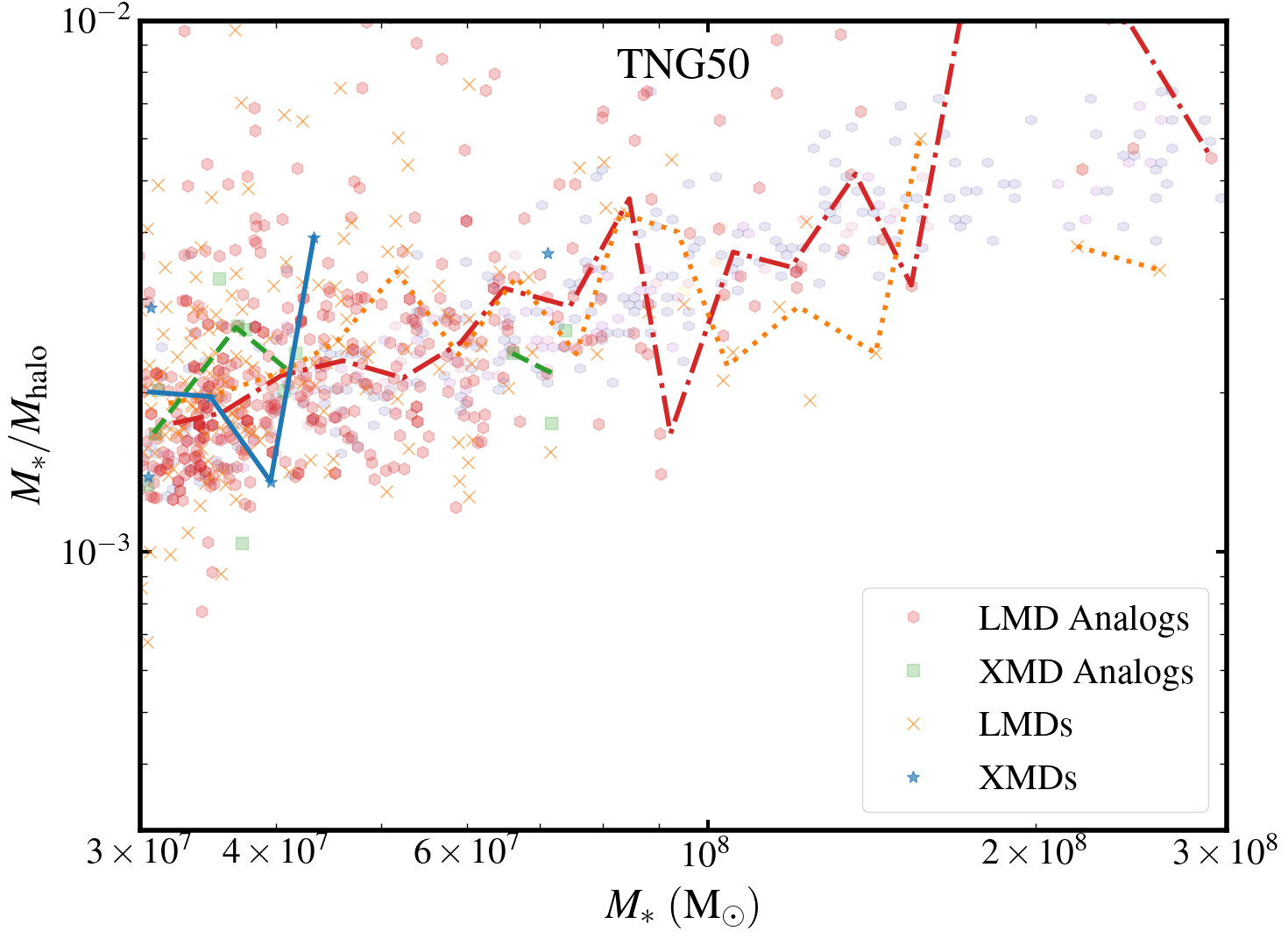}
    \end{tabular}
    \caption{The halo mass/stellar mass ratio as a function of stellar mass for our sample (as in Fig.~\ref{fig:mstmh}) for TNG100-1 (left), TNG50-2 (middle) and TNG50-1 (right). LMDs and their analogs are included in TNG50-2 and TNG50-1 plots due to the scarcity of points, whereas they are not included in the TNG100-1 plot due to the large number of galaxies. In TNG50-2, there is a hint that the XMDs have lower stellar mass-halo mass ratios than XMD analogs, although the limited number of objects makes the conclusion difficult.}
    \label{fig:mhalo502}
\end{figure*}

\begin{figure*}
    \centering
    \begin{tabular}{ccc}
       \includegraphics[width=0.3\linewidth]{corretimescaleall_nodup.png}  & 
       \includegraphics[width=0.3\linewidth]{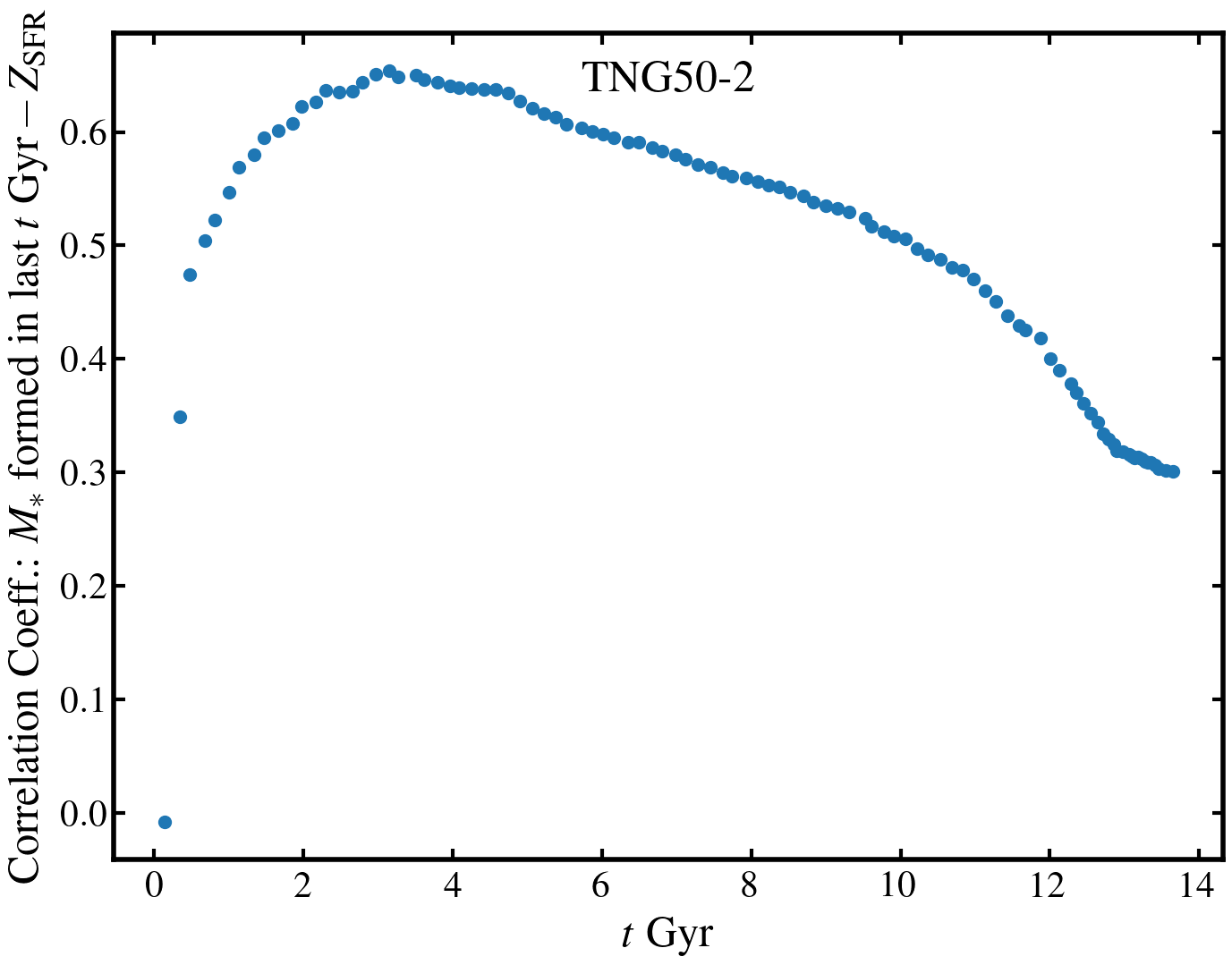} &
       \includegraphics[width=0.3\linewidth]{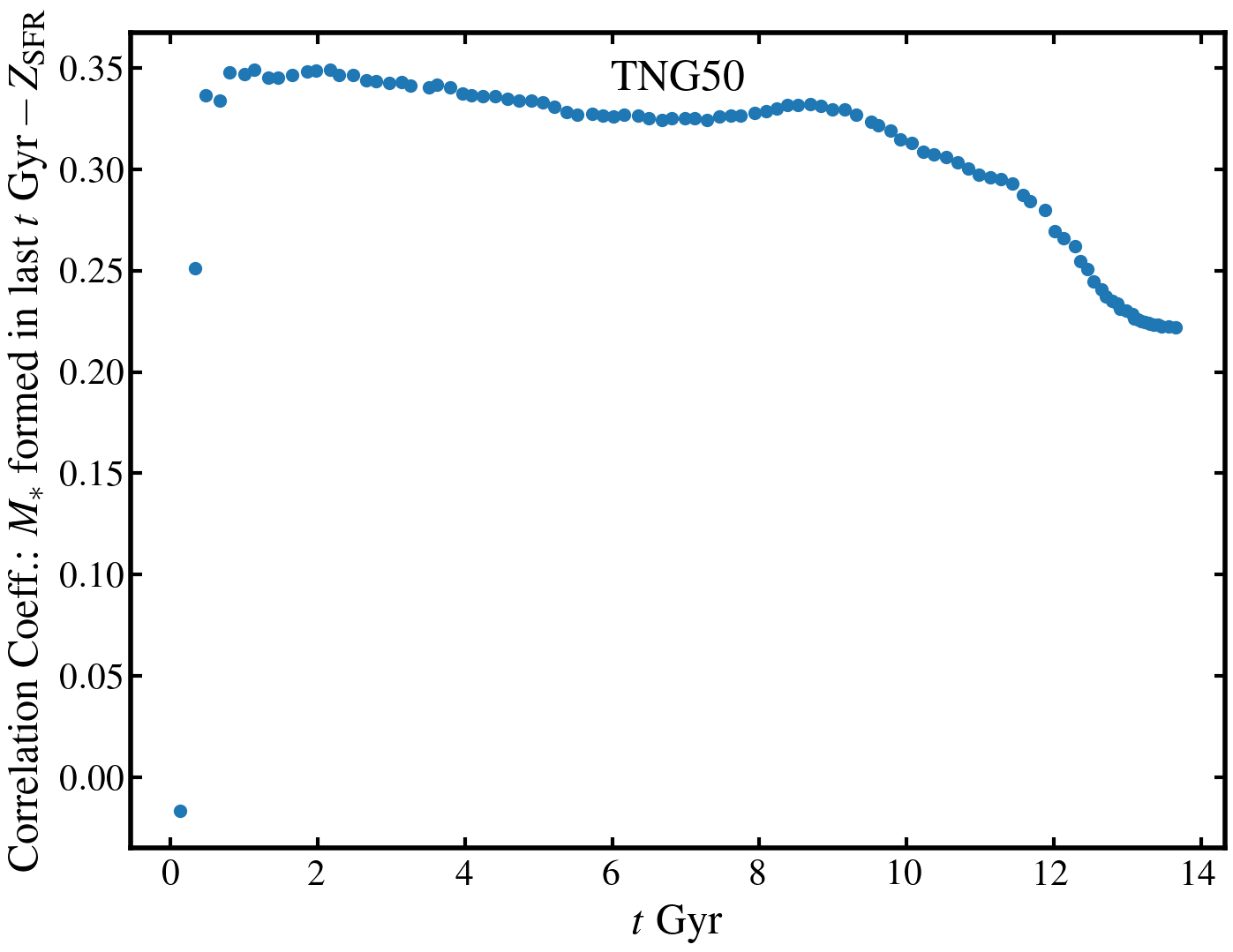}
    \end{tabular}
    \caption{The metallicity-mass formed correlation as a function of time (as in Fig.~\ref{fig:stagecorr}) for TNG100-1 (left), TNG50-2 (middle) and TNG50-1 (right). As with Fig.~\ref{fig:mzr502}, the TNG50-2 result matches the TNG100 result closely, suggesting that resolution is the primary reason for the different results.}
    \label{fig:corr502}
\end{figure*}

Figures~\ref{fig:mzr502},~\ref{fig:zgrad502},~\ref{fig:mhalo502}, and \ref{fig:corr502} show the mass-metallicity relation, radial metallicity profiles, halo mass-stellar mass ratios, and correlation timescales using data from TNG100, TNG50, and TNG50-2. As seen in Fig.~\ref{fig:mzr502}, the mass-metallicity relation in TNG50-2 matches the TNG100 relation closely, so XMDs are selected in the same way as TNG100. For the mass-metallicity relation and correlation timescale plots, the TNG50-2 data matches the TNG100-1 data well, suggesting that the differences in resolution are the cause of the differences between TNG100-1 and TNG50-1 results. For the radial metallicity profiles, more of a difference between TNG100-1 and TNG50-2 is apparent. This suggests that, although resolution plays a part, selection effects and small-number statistics may also play a role in the TNG50 vs. TNG100 differences. The results considering the stellar mass/halo mass ratio are less conclusive. The TNG50-2 results appear similar to the TNG100-1 results, but are limited by the small numbers of XMDs in the TNG50-2 volume.

\end{document}